\documentclass[prx, aps,
 twocolumn,
 superscriptaddress,
 amsmath,amssymb,10pt,
]{revtex4-2}

\usepackage{times}
\usepackage{color}
\usepackage{graphicx}
\usepackage[dvipsnames]{xcolor}
\usepackage{physics}
\usepackage{bm}
\usepackage{mathtools}
\usepackage{upgreek}
\usepackage{footmisc}
\usepackage{makecell}
\usepackage{soul}
\usepackage{wasysym}
\usepackage[colorlinks=true,allcolors=blue]{hyperref}

\usepackage{booktabs}
\newcommand{\wdrv}{\omega_\text{d}}
\newcommand{\EJ}{E_\text{J}}
\newcommand{\EC}{E_\text{C}}

\begin{document}

\title{Strong-Drive Limits in Josephson Circuits: From Chaos to an Unbound-Resonance Threshold}

\author{Xinyuan You}
\email{xinyuan@fnal.gov}
\affiliation{Superconducting Quantum Materials and Systems Division, Fermi National Accelerator Laboratory (FNAL), Batavia, IL 60510, USA}

\author{Aniket Maiti}
\affiliation{Departments of Applied Physics and Physics, Yale University, New Haven, Connecticut 06511, USA}
\affiliation{Yale Quantum Institute, Yale University, New Haven, Connecticut 06511, USA}

\author{Yao Lu}
\email{yaolu@fnal.gov}
\affiliation{Superconducting Quantum Materials and Systems Division, Fermi National Accelerator Laboratory (FNAL), Batavia, IL 60510, USA}

\begin{abstract}
Strong microwave drives enable fast measurement and parametric control in superconducting circuits but can induce transitions out of the intended low-energy manifold.
We develop a unified description of strong-drive limits in flux- and charge-driven Josephson circuits across drive frequency and dc flux bias.
Using classical phase-space analysis and Floquet--Markov simulations, we identify distinct low- and high-frequency mechanisms.
At low frequency, we characterize bound-state resonances and separatrix chaos and find that the flux-drive chaos threshold depends strongly on dc flux bias.
At high frequency, these mechanisms are suppressed, and the dissipative steady state transfers from the central bound-state sector to outer resonances formed by above-barrier running trajectories.
The resulting unbound-resonance threshold is nearly independent of drive frequency and circuit parameters over the regime studied and is controlled primarily by dc flux bias.
Coherent simulations show that parametric operation persists beyond this threshold, but at a reduced rate, setting an effective upper bound on the achievable operation speed.
We derive analytical criteria for both thresholds, validate them numerically, and experimentally confirm the predicted dc-bias dependence of the low-frequency threshold in a flux-driven SQUID.
We also determine the timescales for transfer into the unbound-resonance regime and relaxation back to the bound-state manifold after the drive is removed. Finally, we relate the stability limits to a complementary picture based on the junction critical current and extend the framework to multitone drives and inductively shunted circuits. Together, these results identify the mechanisms limiting strong driving and suggest routes to extend the stable operating range of Josephson circuits.
\end{abstract}

\maketitle


\section{Introduction}

High-fidelity quantum operations require gates and measurements to be completed on timescales short compared with the relevant coherence times~\cite{Krantz2019}.
Although coherence times of superconducting qubits have improved substantially~\cite{Place2021,Bland2025}, they remain finite, making faster operation a complementary route to reducing errors.
Strong microwave driving through charge or flux modulation is therefore a central resource for fast measurement and control.
Such driving can increase dispersive-readout rates~\cite{Gambetta2007Readout,Kurilovich2025} and parametrically activate interactions used for longitudinal readout, entangling gates, frequency conversion, tunable coupling, and bosonic operations~\cite{Didier2015,Caldwell2018,Gao2018Interference,Sete2021ParametricCoupler,Chapman2023HighOnOffRatio,Lu2023HighFidelityParametricBeamsplitting,Noh2023ParametricDispersive,PhysRevApplied.19.044003,Maiti2025LINC,Baskov2025ExactAmplitudes}.
These advantages, however, are ultimately limited by unwanted transitions and leakage out of the intended operating manifold at sufficiently large drive amplitudes~\cite{Sank2016,Shillito2022,Cohen2023,Dumas2024,Xia2025,Dai2026,Repicky2026Scaling,Beaulieu2026JunctionReadout}.
Understanding these limitations is essential for extending the regime of reliable operation in driven Josephson devices.

The mechanisms underlying these strong-drive limitations have been studied most extensively in the context of dispersive readout, where the coherent field of a strongly populated resonator provides an effective charge drive on a transmon.
Strong driving can activate multiphoton resonances that connect low-energy states to highly excited transmon states, producing measurement-induced state transitions (MIST) and, in some cases, population transfer to states above the Josephson potential barrier, a process commonly referred to as transmon ionization~\cite{Sank2016,Lescanne2019Escape,Shillito2022,Khezri2023,Dumas2024,Wang2025,Fechant2025}.
Near the primary separatrix, periodic driving can generate a chaotic layer, leading to broad hybridization and loss of phase-space localization~\cite{Cohen2023,Dumas2024,Xia2025,1zff-vp5q}.
Related strong-drive transitions and instabilities have also been studied in fluxonium and inductively shunted circuits~\cite{Verney2019,PhysRevApplied.18.064044,Nesterov2024FluxoniumMIST,Singh2025FluxoniumArrayModes,Maiti2025LINC,Bista2026FluxoniumLeakage,Zobrist2026, Chapple2026FluxoniumMIST,Zwanenburg2026FluxoniumMIST}.
More broadly, experimentally observed drive-induced unwanted state transitions can arise from resonant exchange with parasitic two-level systems or from inelastic scattering involving the electromagnetic environment or two-level-system defects~\cite{Connolly2025Transitions,Dai2026}.

Taken together, these studies identify several routes to strong-drive breakdown, but the intrinsic limiting mechanism and its threshold remain incompletely understood across drive channels and frequencies.
Flux-driven circuits, in particular, remain less systematically explored.
Recent studies of inductively coupled readout schemes whose semiclassical dynamics map onto flux modulation find suppressed multiphoton ionization and reduced separatrix chaos at zero dc flux bias, while finite bias activates otherwise symmetry-forbidden MIST channels~\cite{Chapple2025Longitudinal,Mori2026CosPhi}. 
These readout-focused studies, however, do not establish how dc flux bias shifts the strong-drive threshold under flux modulation more generally.
This dependence is particularly relevant for parametric operation, where the dc flux bias is often an intentional control parameter used to set the circuit operating point~\cite{Caldwell2018,Sete2021ParametricCoupler,Chapman2023HighOnOffRatio,Lu2023HighFidelityParametricBeamsplitting,Maiti2025LINC}.
Drive frequency presents a complementary gap.
High-frequency driving can suppress the bound-state resonances and separatrix chaos that dominate at lower frequencies~\cite{Cohen2023,Dumas2024,Xia2025} and has enabled robust high-power readout~\cite{Kurilovich2025,Dixit2026}.
At still stronger flux drive, the collapse and inversion of the effective Josephson potential has also recently been observed~\cite{Deve2026Collapse}. However, how the driven-dissipative steady state evolves in this high-frequency regime, the characteristic thresholds associated with these changes, and their implications for useful parametric operation remain incompletely understood.

In this work, we develop a unified description of strong-drive limits in flux- and charge-driven Josephson circuits.
A Jacobi--Anger expansion reveals the distinct harmonic structures generated by the two drive channels, while classical phase-space analysis and Floquet--Markov simulations relate these structures to the resulting driven-dissipative states.
At low frequency, we characterize bound-state resonances and separatrix chaos under flux drive, compare them with charge drive, and determine the pronounced dc-flux-bias dependence of the flux-drive chaos threshold.
At high frequency, these mechanisms are suppressed, and we identify an unbound-resonance transition in which the dissipative steady state transfers from the central bound-state sector to outer resonances of above-barrier running motion.
Its threshold is nearly independent of drive frequency and circuit parameters over the regime studied and is controlled primarily by dc flux bias.
We derive analytical criteria for the low- and high-frequency thresholds, validate them numerically, and experimentally verify the predicted dc-bias dependence of the low-frequency threshold in a flux-driven SQUID.
Coherent simulations show that parametric operation persists beyond the unbound-resonance transition, but at a reduced rate in the outer-resonance sector, setting an effective upper bound on the achievable operation speed.
We also determine the timescales for transfer into this sector and for relaxation back to the bound-state manifold after the drive is removed.
Finally, we apply the framework to multitone flux driving, relate the thresholds to the junction critical current, and examine the effect of inductive confinement.

The remainder of this paper is organized as follows. Section~\ref{sec:charg_and_flux} introduces the flux- and charge-driven Josephson Hamiltonians and compares their harmonic structures through the Jacobi--Anger expansion. Section~\ref{sec:flux_0dc} analyzes the low- and high-frequency strong-drive regimes for flux modulation at zero dc flux bias. Section~\ref{sec:dc} derives the dc-flux-bias dependence of both thresholds and presents the experimental verification of the low-frequency result. Section~\ref{sec:parametric} studies parametric operation across the unbound-resonance threshold, and Sec.~\ref{sec:timescale} characterizes the associated transfer and relaxation dynamics. Sections~\ref{sec:charge_drive}, \ref{sec:multitone}, \ref{sec:current}, and \ref{sec:linc} extend the analysis to charge drive, multitone flux drive, the junction-current picture, and inductively shunted circuits, respectively.
Section~\ref{sec:conclusion} summarizes the main conclusions and their implications for strongly driven superconducting devices.

\section{Flux- and charge-driven Josephson circuits}\label{sec:charg_and_flux}

To compare flux and charge driving within a common setting, we consider a superconducting quantum interference device (SQUID)~\cite{Tesche1977SQUID} as a representative Josephson circuit.
The SQUID can be driven capacitively through a charge-bias line or by modulating the magnetic flux through its loop~\cite{Lu2023HighFidelityParametricBeamsplitting}.
Below, we derive both drive channels and use a Jacobi--Anger expansion to expose their structural differences, which will be useful for interpreting the strong-drive thresholds discussed in later sections.

\subsection{Flux- and charge-driven Hamiltonians}
\label{subsec:flux_charge_ham}

We begin from the Hamiltonian of a driven SQUID,
\begin{align}
    \hat{H} ={}& 4 E_\text{C} (\hat{n}-n_\text{g})^2 +\epsilon(t) \hat{n} \notag\\
    &- E_\text{J1}\cos\left[\hat{\theta} + \frac{\Phi_\text{ext}(t)}{2\phi_0}\right]
    - E_\text{J2}\cos\left[\hat{\theta} - \frac{\Phi_\text{ext}(t)}{2\phi_0}\right].
\end{align}
In an appropriate gauge, this general form describes both the modulation by a time-varying magnetic flux $\Phi_\text{ext}(t)$ and an effective charge drive $\epsilon(t)$.
The latter is often referred to as a common-mode drive and contains contributions from both an externally applied charge bias and the electromotive force induced by the time-dependent magnetic flux~\cite{You2019TimeDependentFlux,Riwar2022TimeDependentFields,Bryon2023TimeDependentFlux,Lu2023HighFidelityParametricBeamsplitting,Lu2025SystematicConstruction}.
Here, $\hat{n}$ and $\hat{\theta}$ are the conjugate Cooper-pair-number and phase operators, $E_\text{C}$ is the charging energy, $E_\text{J1}$ and $E_\text{J2}$ are the Josephson energies of the two junctions, $n_\text{g}$ is the offset charge, and $\phi_0=\Phi_0/(2\pi)$ is the reduced flux quantum.
In the following, we consider identical junctions, i.e., $E_\text{J1}=E_\text{J2}=E_\text{J}/2$.
The general case in the presence of junction asymmetry is discussed in Appendix~\ref{app:asymmetry}.
We set $n_\text{g}=0$ to keep the presentation compact and restore its contribution where quantitatively relevant, with the effects of finite offset charge discussed in Appendix~\ref{app:offset_charge}.

To isolate the distinct roles of each drive channel, we analyze the flux- and charge-driven cases independently.
The more general case with simultaneous flux and charge drives, including the use of charge drive to compensate junction-asymmetry-induced effects, is discussed in Appendix~\ref{app:asymmetry}.
In the absence of common-mode drive, i.e., $\epsilon(t)=0$, the Hamiltonian simplifies to
\begin{equation}
    \hat{H}_\text{flux}
    = 4 E_\text{C} \hat{n}^2
    - E_\text{J}\cos\hat{\theta}
    \cos\left[\frac{\Phi_\text{ext}(t)}{2\phi_0}\right].
\end{equation}
We consider a monochromatic flux modulation,
\begin{equation}
    \frac{\Phi_\text{ext}(t)}{2\phi_0}
    = \phi_\text{dc} + \phi_\text{ac}\sin(\omega_\text{d} t).
\end{equation}
The generalization to multitone driving is discussed in Sec.~\ref{sec:multitone}.
The resulting Hamiltonian is
\begin{equation}
    \hat{H}_\text{flux}
    = 4 E_\text{C} \hat{n}^2
    - E_\text{J}\cos\hat{\theta}
    \cos\left[\phi_\text{dc} + \phi_\text{ac}\sin(\omega_\text{d} t)\right].
\end{equation}

For charge drive only, we first set $\Phi_\text{ext}(t)=0$, so that the Hamiltonian reduces to that of a charge-driven transmon,
\begin{equation}
    \hat{H}_\text{charge}
    = 4 E_\text{C} \hat{n}^2
    - E_\text{J}\cos\hat{\theta} + \epsilon(t)\hat{n}.
\end{equation}
For a symmetric SQUID, retaining a finite static flux bias simply replaces $E_\text{J}$ by $E_\text{J}\cos\phi_\text{dc}$ and does not change the charge-drive structure discussed below.
To cast the charge-driven Hamiltonian in a form that enables direct structural comparison with the flux case, we perform a gauge transformation,
\begin{align}
    \hat{H}_\text{charge}' ={}&
    \hat{U}^\dagger \hat{H}_\text{charge} \hat{U}
    - i\hat{U}^\dagger \partial_t\hat{U} \notag\\
    ={}&
    4 E_\text{C} \hat{n}^2
    - E_\text{J}\cos\left[
    \hat{\theta}+\int \epsilon(t)\mathrm{d}t
    \right],
\end{align}
with the unitary
$\hat{U}=\exp[-i\hat{n}\int\epsilon(t)\mathrm{d}t]$.
For a monochromatic charge drive $\epsilon(t)=\epsilon_\text{c}\cos(\omega_\text{c}t)$, this yields
\begin{equation}\label{eq:charge_ham}
    \hat{H}_\text{charge}'
    = 4 E_\text{C}\hat{n}^2
    - E_\text{J}\cos\left[
    \hat{\theta}+\phi_\text{c}\sin(\omega_\text{c}t)
    \right],
\end{equation}
where $\phi_\text{c}=\epsilon_\text{c}/\omega_\text{c}$.

The same two effective drive structures also arise in resonator-based readout.
In architectures where the readout-mode phase enters the Josephson potential, treating a strongly populated resonator semiclassically yields a modulation with the same form as the flux-driven Hamiltonian~\cite{Chapple2025Longitudinal,Mori2026CosPhi,Hazra2026NonlinearReadout}.
By contrast, conventional capacitive coupling acts through $\hat n$, corresponding to charge drive and, after the gauge transformation above, to Eq.~\eqref{eq:charge_ham}~\cite{Dumas2024}.

\subsection{Jacobi--Anger expansion of the driven Hamiltonian}
\label{subsec:jacobi_anger}

A Jacobi--Anger expansion of the drive terms separates the Hamiltonian into contributions at distinct temporal harmonics, with amplitudes determined by Bessel functions.
To expose the harmonic components that play a central role below, we write the $J_0$, $J_1$, and $J_2$ terms explicitly and leave higher Bessel orders implicit.
For flux drive,
\begin{align}
    \hat{H}_\text{flux} ={}&
    4 E_\text{C} \hat{n}^2
    - E_\text{J}\cos\hat{\theta}
    \cos\left[
    \phi_\text{dc}+\phi_\text{ac}\sin(\omega_\text{d}t)
    \right] \notag\\
    ={}&
    4 E_\text{C} \hat{n}^2
    - E_\text{J}\cos\hat{\theta}\cos\phi_\text{dc}
    \cos\left[\phi_\text{ac}\sin(\omega_\text{d}t)\right] \notag\\
    &+
    E_\text{J}\cos\hat{\theta}\sin\phi_\text{dc}
    \sin\left[\phi_\text{ac}\sin(\omega_\text{d}t)\right] \notag\\
    ={}&
    4 E_\text{C} \hat{n}^2
    - E_\text{J}\cos\hat{\theta}\cos\phi_\text{dc}
    \left[
    J_0(\phi_\text{ac}) \right. \notag\\
    &\left.\hspace{2.5cm}
    +2J_2(\phi_\text{ac})\cos(2\omega_\text{d}t)
    +\cdots
    \right] \notag\\
    &+
    E_\text{J}\cos\hat{\theta}\sin\phi_\text{dc}
    \left[
    2J_1(\phi_\text{ac})\sin(\omega_\text{d}t)
    +\cdots
    \right].
    \label{eq:jacobi_flux}
\end{align}
The same expansion applied to charge drive yields a structurally similar but distinct decomposition,
\begin{align}
    \hat{H}_\text{charge}' ={}&
    4 E_\text{C} \hat{n}^2
    - E_\text{J}\cos\left[
    \hat{\theta}+\phi_\text{c}\sin(\omega_\text{c}t)
    \right] \notag\\
    ={}&
    4 E_\text{C} \hat{n}^2
    - E_\text{J}\cos\hat{\theta}
    \cos\left[\phi_\text{c}\sin(\omega_\text{c}t)\right] \notag\\
    &+
    E_\text{J}\sin\hat{\theta}
    \sin\left[\phi_\text{c}\sin(\omega_\text{c}t)\right] \notag\\
    ={}&
    4 E_\text{C} \hat{n}^2
    - E_\text{J}\cos\hat{\theta}
    \left[
    J_0(\phi_\text{c}) \right. \notag\\
    &\left.\hspace{2.5cm}
    +2J_2(\phi_\text{c})\cos(2\omega_\text{c}t)
    +\cdots
    \right] \notag\\
    &+
    E_\text{J}\sin\hat{\theta}
    \left[
    2J_1(\phi_\text{c})\sin(\omega_\text{c}t)
    +\cdots
    \right].
    \label{eq:charge_jacobi}
\end{align}

The expanded forms reveal three structural differences between flux and charge drive.
(1) In charge drive, both even- and odd-order Bessel sectors are generically present.
In flux drive, their relative weight is controlled by $\phi_\text{dc}$.
At $\phi_\text{dc}=0$ ($\phi_\text{dc}=\pi/2$), only the even (odd) contributions survive, while at $\phi_\text{dc}=\pi/4$ the dc-bias prefactors of the even- and odd-order sectors are equal in magnitude.
(2) In flux drive, both even and odd Bessel contributions couple to $\cos\hat{\theta}$, while in charge drive the even and odd contributions couple to $\cos\hat{\theta}$ and $\sin\hat{\theta}$, respectively.
Since $\sin\hat{\theta}$ is odd under $\hat{\theta}\rightarrow-\hat{\theta}$, whereas $\cos\hat{\theta}$ is even, charge drive contains both odd- and even-parity operators in $\hat{\theta}$.
By contrast, flux drive couples through $\cos\hat{\theta}$ at any dc flux bias, so the Hamiltonian remains even in $\hat{\theta}$.
A finite $\phi_\text{dc}$ activates odd temporal harmonics, such as the $J_1$ term, but does not change this phase-parity structure.
(3) In flux drive, the argument of the Bessel functions is the ac amplitude $\phi_\text{ac}$, which also equals the displacement of the differential mode in the SQUID at zero dc offset~\cite{Lu2023HighFidelityParametricBeamsplitting}.
In charge drive, the Bessel argument $\phi_\text{c}=\epsilon_\text{c}/\omega_\text{c}$ is not generally equal to the linearized phase displacement.
Within the harmonic approximation, the latter is
\begin{equation}
    \phi_\text{disp}
    = \frac{\epsilon_\text{c}\omega_\text{c}}
    {\omega_\text{c}^2-\omega_0^2},
\end{equation}
where $\omega_0$ is the small-amplitude oscillation frequency, and $\phi_\text{disp}$ approaches $\phi_\text{c}$ only in the high-frequency limit $\omega_\text{c}\gg\omega_0$.
Since the strength of parametric processes is typically characterized by this displacement~\cite{Rosenblum2018FaultTolerantDetection,Chapman2023HighOnOffRatio}, this distinction must be taken into account when comparing strong-drive thresholds and parametric rates under flux and charge drive.

\section{Flux drive with zero dc flux bias}
\label{sec:flux_0dc}

Strong microwave drives can increase parametric interaction rates and readout speeds, motivating operation at large drive amplitude~\cite{Lu2023HighFidelityParametricBeamsplitting,Xia2025,Khezri2023,Kurilovich2025,Repicky2026Scaling}.
At sufficiently strong drive, however, unwanted transitions and strong hybridization can qualitatively change the driven-state structure and degrade operation fidelity~\cite{Sank2016,Shillito2022,Cohen2023,Dumas2024,Xia2025,Dai2026,Huang2026Sideband,Xia2026Subharmonic}.
We first study this strong-drive limit for flux drive at zero dc flux bias, $\phi_\text{dc}=0$, by varying the ac drive amplitude $\phi_\text{ac}$ and the drive frequency $\omega_\text{d}$.

A common diagnostic of strong-drive hybridization is the overlap between driven Floquet eigenstates and the static eigenstates of the system~\cite{Sank2016,Shillito2022,Khezri2023,Cohen2023,Dumas2024,Wang2025,Fechant2025,Xia2025,Kurilovich2025,Dai2026}.
A complementary open-system diagnostic is the impurity of the steady-state density matrix of a driven dissipative system~\cite{Verney2019,Lu2023HighFidelityParametricBeamsplitting,Maiti2025LINC}.
The impurity quantifies how broadly the dissipative steady state is distributed over multiple Floquet states.
Throughout this work, we use impurity as the primary diagnostic and compare it with the overlap metric where the two provide complementary information.

For the Floquet--Markov calculation, we model the dominant dissipative channel as charge-coupled Ohmic noise, with spectral density $J(\omega>0)=\eta\omega/(2\pi)$. Additional noise channels can be incorporated through their corresponding coupling operators and are discussed briefly below.
The steady-state density matrix $\rho_\text{s}$ is obtained as the stationary solution of the Floquet--Markov equation from the null space of the Floquet--Markov generator~\cite{Grifoni1998DrivenQuantumTunneling,Verney2019,You2025}.
The impurity is obtained from $1-\mathrm{Tr}(\rho_\text{s}^2)$.
For simplicity, we take the bath temperature to be zero. In the strong-drive regime considered here, the steady-state excitation is dominated by drive-induced transitions rather than thermal excitation.
For a unique steady state, the overall scale $\eta$ in the spectral density sets the relaxation timescale but does not affect the stationary Floquet populations or the steady-state impurity within this bath model.

\begin{figure}[t]
    \centering
    \includegraphics[width=\columnwidth]{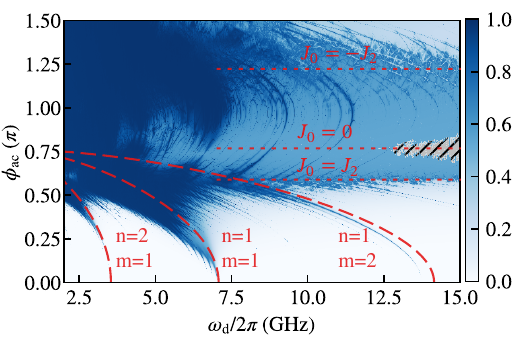}
    \caption{\textbf{Impurity at zero dc flux bias as a function of ac flux-drive amplitude and frequency.}
    The color scale shows the steady-state impurity obtained from Floquet--Markov simulations.
    Red dashed curves indicate resonance conditions extracted from the eigenenergies of the reduced static Hamiltonian Eq.~\eqref{eq:flux_static}.
    Red dotted lines mark drive amplitudes related to various Bessel functions.
    The hatched region denotes parameters for which the Floquet--Markov equation admits multiple stationary states.
    Parameters: $E_\text{C}/2\pi = 0.13$~GHz, $E_\text{J}/2\pi = 50$~GHz, $n_\text{g}=0$.}
    \label{fig:impurity}
\end{figure}

Figure~\ref{fig:impurity} shows the impurity as a function of drive frequency and amplitude.
Overall, the impurity tends to increase with drive amplitude, although its dependence is strongly structured and nonmonotonic.
Two distinct regimes emerge.
First, when the drive frequency is comparable to or below the characteristic frequency of the circuit, around $7$~GHz for the parameters of Fig.~\ref{fig:impurity}, narrow high-impurity stripes appear already at small drive amplitude.
These stripes arise from resonances involving low-lying bound states.
At larger amplitude, a broad high-impurity region emerges that we show below is associated with the primary separatrix chaos.

Second, at drive frequencies well above the characteristic circuit frequency, the low-energy bound-state resonances and the broad chaotic region are suppressed.
Instead, a sharp increase in impurity occurs near $\phi_\text{ac}=0.6\pi$, where it rises to values close to $1/2$. The location of this onset depends only weakly on drive frequency.
As shown below, this feature marks the high-frequency unbound-resonance transition.
At still larger amplitude, around $\phi_\text{ac}=1.2\pi$, the impurity decreases again.
The origin of this recovery is discussed below.
Around $\phi_\text{ac}=0.8\pi$ at high drive frequency, the Floquet--Markov equation admits multiple stationary states, as indicated by the hatched region and discussed in Appendix~\ref{app:bistability}.

The results in Fig.~\ref{fig:impurity} are obtained at zero offset charge.
A comparison with $n_\text{g}=0.25$ is presented in Appendix~\ref{app:offset_charge}.
For the parameters considered here, offset charge does not qualitatively modify the low-frequency regime or the location of the high-frequency transition.
It can, however, modify the steady-state impurity reached after the high-frequency transition by breaking the reflection symmetry in the charge degree of freedom.
Since this does not qualitatively change the strong-drive regimes, we focus on $n_\text{g}=0$ and include offset-charge averaging only where it is quantitatively relevant.

We next examine these two regimes separately, first identifying the bound-state resonances and separatrix chaos governing the low-frequency response and then analyzing the unbound-resonance transition that emerges at high frequency.

\subsection{Bound-state resonances and separatrix chaos at low frequency}
\label{subsec:bound}

\begin{figure*}[htbp]
    \centering
    \includegraphics[width=\textwidth]{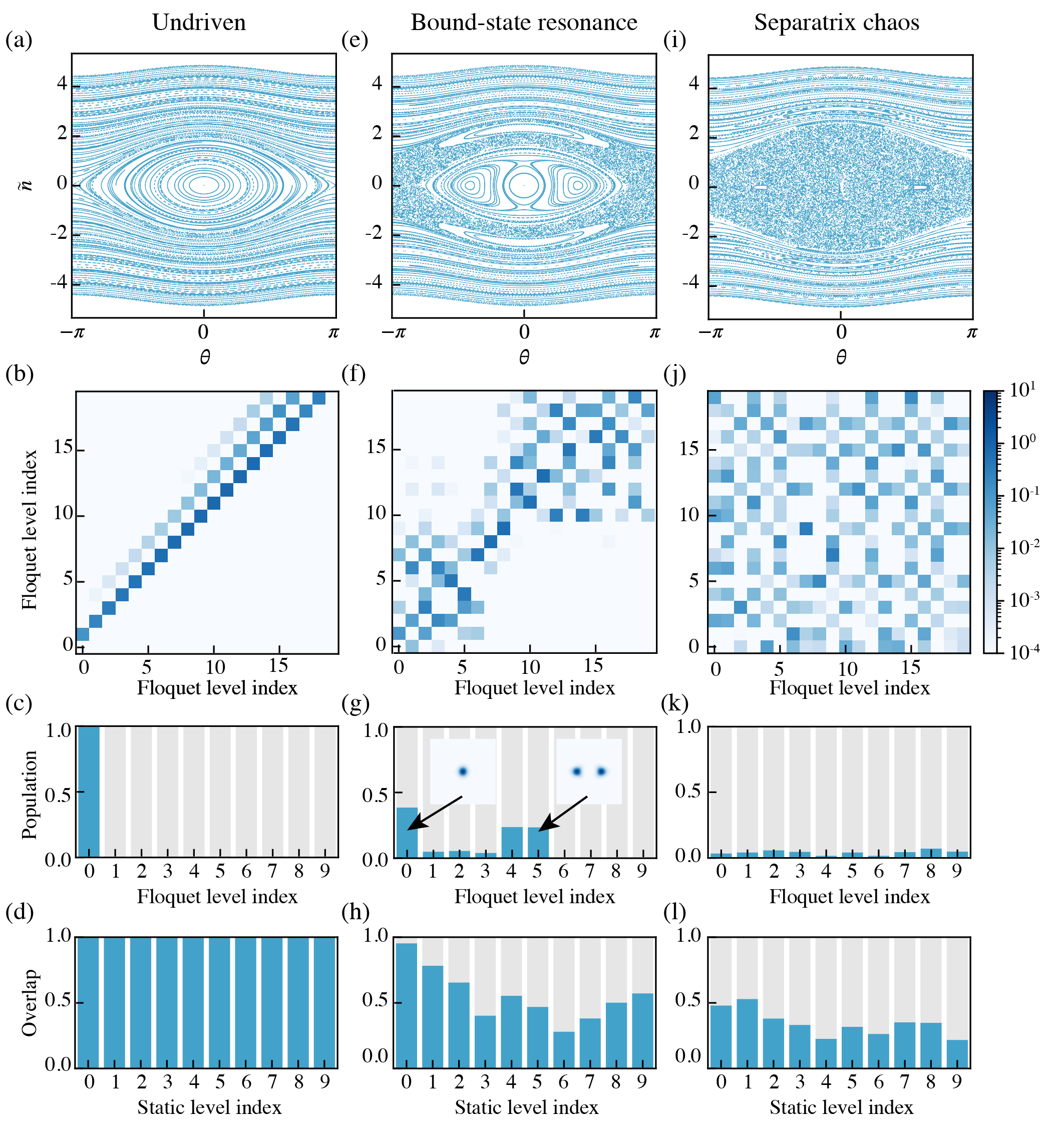}
    \caption{\textbf{Bound-state resonances and separatrix chaos under low-frequency flux drive at zero dc flux bias.}
    (a) Poincar\'e section of the undriven system, sampled stroboscopically at the drive period used for comparison.
    (b) Bath-induced transition rates between Floquet states.
    (c) Steady-state Floquet populations obtained from the Floquet--Markov equation.
    (d) Maximum overlap between each static eigenstate and the Floquet eigenstates.
    (e)--(h) and (i)--(l) show the corresponding quantities for two representative drive conditions, $\phi_\text{ac}=0.2\pi$, $\omega_\text{d}/2\pi=6.3$~GHz, and $\phi_\text{ac}=0.4\pi$, $\omega_\text{d}/2\pi=5.0$~GHz, respectively.
    The inset in (g) shows the Husimi functions of the relevant Floquet states.
    Parameters: $E_\text{C}/2\pi=0.13$~GHz, $E_\text{J}/2\pi=50$~GHz, $n_\text{g}=0$.}
    \label{fig:resonance}
\end{figure*}

For charge-driven transmons, strong-drive leakage has been associated with bound-state resonances and separatrix chaos~\cite{Cohen2023,Dumas2024,Xia2025}.
Studies of inductively coupled readout architectures whose semiclassical dynamics map onto flux modulation have found suppressed multiphoton transitions and reduced separatrix chaos relative to conventional capacitive coupling~\cite{Chapple2025Longitudinal,Mori2026CosPhi}.
Here we map the flux-driven response over drive frequency and amplitude and connect the resulting structures to the Floquet--Markov steady state.
In particular, the narrow impurity features in Fig.~\ref{fig:impurity} arise from localized bound-state resonances, whereas the broad high-impurity region at stronger drive is associated with separatrix chaos.

Retaining all orders of the Jacobi--Anger expansion, we rewrite the system Hamiltonian in Eq.~\eqref{eq:jacobi_flux} as $\hat H_\text{flux}=\hat H_\text{static}+\hat H_\text{pert}(t)$, with
\begin{equation}
    \hat H_\text{static}
    =
    4E_\text{C}\hat n^2
    -
    E_\text{J}J_0(\phi_\text{ac})\cos\hat\theta,
    \label{eq:flux_static}
\end{equation}
and
\begin{equation}
    \hat H_\text{pert}(t)
    =
    -2E_\text{J}
    \sum_{n=1}^{\infty}
    J_{2n}(\phi_\text{ac})
    \cos(2n\omega_\text{d}t)
    \cos\hat\theta.
    \label{eq:flux_pert}
\end{equation}
For the drive amplitudes relevant to the low-frequency resonance features, the lower Bessel orders dominate, so the time-dependent terms provide a useful perturbative picture for identifying the resonance conditions.
The full Hamiltonian is retained in the numerical calculations below.

The motion under $\hat H_\text{static}$ corresponds to that of a transmon with a renormalized Josephson energy $E_\text{J}J_0(\phi_\text{ac})$.
For small phase oscillations, the classical dynamics can be approximated by $\theta(t)\simeq\theta_0\sin[\omega_0(\theta_0)t]$.
Due to the negative anharmonicity, the natural frequency $\omega_0(\theta_0)$ decreases with increasing oscillation amplitude.
To identify the resonances generated by the time-dependent terms, we evaluate the perturbation along this unperturbed classical trajectory.
Applying the Jacobi--Anger expansion again gives
\begin{align}
    H_\text{pert}^\text{cl}(t)
    &\simeq
    -2E_\text{J}
    \sum_{n=1}^{\infty}
    J_{2n}(\phi_\text{ac})
    \cos(2n\omega_\text{d}t)
    \notag\\
    &\quad\times
    \left[
    J_0(\theta_0)
    +
    2\sum_{m=1}^{\infty}
    J_{2m}(\theta_0)
    \cos\left[2m\omega_0(\theta_0)t\right]
    \right]
    \notag\\
    &=
    -2E_\text{J}
    \sum_{n,m=1}^{\infty}
    J_{2n}(\phi_\text{ac})
    J_{2m}(\theta_0)
    \notag\\
    &\quad\times
    \cos\left\{
    2\left[n\omega_\text{d}
    -m\omega_0(\theta_0)\right]t
    \right\}
    \notag\\
    &\quad+\text{fast-rotating terms}.
    \label{eq:bound_res_pert}
\end{align}
The drive becomes resonant when $n\omega_\text{d}=m\omega_0(\theta_0)$, with a strength scaling as $J_{2n}(\phi_\text{ac})J_{2m}(\theta_0)$.

For the low-lying bound-state resonances considered here, the dominant contribution is $n=m=1$, giving $\omega_\text{d}=\omega_0(\theta_0)$.
For sufficiently small oscillation amplitude, the natural frequency can be approximated by
\begin{equation}
    \omega_0(\theta_0\ll1,\phi_\text{ac})
    \simeq
    \sqrt{
    8E_\text{C}E_\text{J}J_0(\phi_\text{ac})
    }.
    \label{eq:res_1}
\end{equation}
Here the dependence on $\phi_\text{ac}$ arises from the drive-induced renormalization of the static Josephson potential.
In Fig.~\ref{fig:impurity}, we plot this resonance condition as a function of drive amplitude as a red dashed line.
It closely follows the main impurity feature around $7$~GHz.
The $n=2,m=1$ resonance, $\omega_\text{d}=\omega_0/2$, and the $n=1,m=2$ resonance, $\omega_\text{d}=2\omega_0$, are also visible at lower and higher drive frequencies, respectively, but produce substantially weaker impurity features.

The classical phase-space structure near resonance, visualized through Poincar\'e sections, provides an intuitive picture of the distinction between localized resonances and separatrix chaos.
A Poincar\'e section is obtained by sampling each classical trajectory stroboscopically at $t=t_0+\ell T$, where $T$ is the fundamental period, and repeating this procedure for many initial conditions across phase space.
Regular trajectories form smooth invariant curves or resonance islands, whereas chaotic trajectories fill extended regions irregularly.
Figure~\ref{fig:resonance}(a) shows the Poincar\'e section of the undriven system.
For visualization, we use the dimensionless charge coordinate $\tilde n=n\sqrt{8E_\text{C}/E_\text{J}}$.
The section shows two types of motion: bounded oscillations inside the Josephson well and above-barrier running trajectories.
These two classes of motion are separated by the primary separatrix.
For the undriven Hamiltonian, the separatrix intersects $\theta=0$ at $n=\pm\sqrt{E_\text{J}/(2E_\text{C})}$, or equivalently $\tilde n=\pm2$ in the rescaled coordinate.

When the drive is close to a resonance, for example $\omega_\text{d}/2\pi=6.3$~GHz and $\phi_\text{ac}=0.2\pi$, the bound motion is significantly altered, as shown in Fig.~\ref{fig:resonance}(e).
The central orbit remains regular, although it is distorted by the drive.
In addition, two resonance islands, corresponding to resonant tori in the full dynamics, appear with fixed points along the horizontal axis.
Since the resonance condition $\omega_\text{d}=\omega_0(\theta_0^*)$ selects a specific oscillation amplitude $\theta_0^*$, the fixed points are located at the corresponding phase-space radius.
Due to the negative anharmonicity, increasing the drive frequency shifts the resonant amplitude, and hence the fixed points, toward the origin.

A chaotic layer is also visible around the separatrix.
At this drive amplitude, however, it remains largely separated from the central low-energy resonance islands and therefore does not dominate the low-energy dynamics.
Two additional resonance islands centered along the vertical direction are also visible.
These are related to the above-barrier resonances analyzed in the high-frequency regime below.

The classical phase-space structure is also reflected in the bath-induced transitions between Floquet states.
Since the SQUID couples to the environment through the charge operator $\hat n$, the corresponding Floquet matrix elements determine the bath-induced transition rates~\cite{Grifoni1998DrivenQuantumTunneling,Verney2019,Cohen2023,You2025}.
At zero bath temperature, the transition rate from Floquet state $j$ to state $i$ is
\begin{equation}
    \Gamma_{ij}
    =
    \sum_k
    \left|n_{ij}^{(k)}\right|^2
    \Theta(\Delta_{ijk})
    J(|\Delta_{ijk}|).
    \label{eq:floquet_rate}
\end{equation}
Here, $\varepsilon_i$ denotes the quasienergy of the Floquet mode $|i(t)\rangle$, while
$\Delta_{ijk}=\varepsilon_j-\varepsilon_i-k\omega_\text{d}$ is the corresponding transition frequency.
The Fourier components of the charge-operator matrix elements are
\begin{equation}
    n_{ij}^{(k)}
    =
    \frac{1}{T}
    \int_0^T
    \mathrm{d}t\,
    e^{-ik\omega_\text{d}t}
    \langle i(t)|\hat n|j(t)\rangle,
    \label{eq:floquet_fourier}
\end{equation}
where $k$ labels the Floquet sideband.
Since the Floquet quasienergy is defined only modulo the drive frequency, we sort the Floquet states by their average energy in the presence of the drive,
\begin{equation}
    E_{\text{avg},i}
    =
    \frac{1}{T}
    \int_0^T
    \mathrm{d}t\,
    \langle i(t)|
    \hat H_\text{flux}
    |i(t)\rangle.
    \label{eq:floquet_avg_eng}
\end{equation}
We here note that $T$ denotes the fundamental period of the Hamiltonian.
At zero dc flux bias, only even temporal harmonics are present, giving $T=\pi/\omega_\text{d}$.
This period is used for both the Floquet calculations and the stroboscopic sampling of the Poincar\'e sections.
When the fundamental temporal harmonic is instead $\omega_\text{d}$, we use $T=2\pi/\omega_\text{d}$.

In the absence of the drive, the zero-temperature bath allows only transitions toward lower-energy states, as shown in Fig.~\ref{fig:resonance}(b).
In the presence of the drive, the transition network changes qualitatively, as shown in Fig.~\ref{fig:resonance}(f).
The allowed transitions can be roughly categorized into two groups: transitions within the lowest $\sim10$ levels and transitions among higher-energy states.
The first group reflects the bound-state resonance described above and contains strongly modified transitions among low-lying Floquet states, including transitions between nonadjacent states.
The second group forms a dense network among higher-energy states, reflecting the chaotic motion observed in the Poincar\'e section.
For this drive condition, the two groups remain largely separated, so the chaotic layer couples only weakly to the central low-energy sector.

Since transitions among the low-energy Floquet states are strongly modified by the drive, the steady-state distribution of the driven-dissipative system differs from the undriven zero-temperature case, where only the ground state is occupied [Fig.~\ref{fig:resonance}(c)].
Instead, as shown in Fig.~\ref{fig:resonance}(g), the lowest-average-energy Floquet state is only partially populated, with substantial populations in several higher Floquet states.
In the inset, we show the Husimi functions of these populated Floquet states.
Compared with the Poincar\'e section in Fig.~\ref{fig:resonance}(e), these states correspond to the central bound-state region and the two horizontal resonance islands.
The dissipative steady state therefore occupies both the central bound-state region and Floquet states localized on the resonance islands, producing the enhanced impurity seen in Fig.~\ref{fig:impurity}.

The overlap metric provides a complementary view of the same regime.
For each static eigenstate, we calculate its largest overlap with the set of Floquet eigenstates, as shown in Fig.~\ref{fig:resonance}(h).
Although the overlap is strongly suppressed for higher levels, the static ground state retains a large maximum overlap, greater than $0.95$.
This is consistent with the persistence of the central regular orbit in the driven Poincar\'e section [Fig.~\ref{fig:resonance}(a) and (e)].
Thus, the static ground state can retain a strong overlap with a Floquet eigenstate even though the dissipative steady state has substantial population in other Floquet states.
Such redistribution can degrade parametric operations even when the ground-state overlap remains large~\cite{Zhang2019Bilinear, Lu2023HighFidelityParametricBeamsplitting,Kim2025}.
The impurity therefore reveals dissipative degradation that is not captured by the ground-state overlap alone.

The first example therefore represents localized leakage through a bound-state resonance, where the drive populates low-lying resonant Floquet states while the separatrix-chaotic layer remains largely disconnected from the central sector.
At stronger drive, the chaotic layer expands and can instead hybridize broadly with the bound-state region.
As a representative example, we consider $\omega_\text{d}/2\pi=5$~GHz and $\phi_\text{ac}=0.4\pi$.
The corresponding Poincar\'e section is shown in Fig.~\ref{fig:resonance}(i).
The regular tori that previously confined the central bound-state motion have disappeared over a broad region, which is instead occupied by a connected chaotic sea.
Classical trajectories in this region wander irregularly through the connected chaotic component rather than remaining confined to invariant tori.
This chaotic region develops from the separatrix layer already visible in Fig.~\ref{fig:resonance}(e).
As the drive amplitude increases, the separatrix-chaotic layer broadens until it strongly overlaps with the central bound-state region.

A quantum analogue is visible in the transition-rate matrix shown in Fig.~\ref{fig:resonance}(j), where a dense network of transitions connects a broad range of Floquet states.
This leads to a steady-state distribution spread over many Floquet states, each carrying only a fraction of the total population.
The populations of the ten lowest Floquet states, ordered by their average energies, are shown in Fig.~\ref{fig:resonance}(k).
This broad redistribution contrasts with the localized population transfer produced by the bound-state resonance in Fig.~\ref{fig:resonance}(g) and identifies the broad low-frequency high-impurity region with separatrix chaos.
The overlap of static eigenstates with Floquet states, shown in Fig.~\ref{fig:resonance}(l), supports the same picture.
The maximum overlaps are strongly reduced across much of the spectrum, consistent with the broad Floquet hybridization associated with the chaotic phase-space region.

Having identified localized bound-state resonances and separatrix chaos as the two mechanisms governing the low-frequency response, we now turn to the high-frequency regime, where both are suppressed and a distinct strong-drive transition emerges.

\subsection{Unbound-resonance transition at high frequency}
\label{subsec:unbound}

\begin{figure*}[t]
    \centering
    \includegraphics[width=\textwidth]{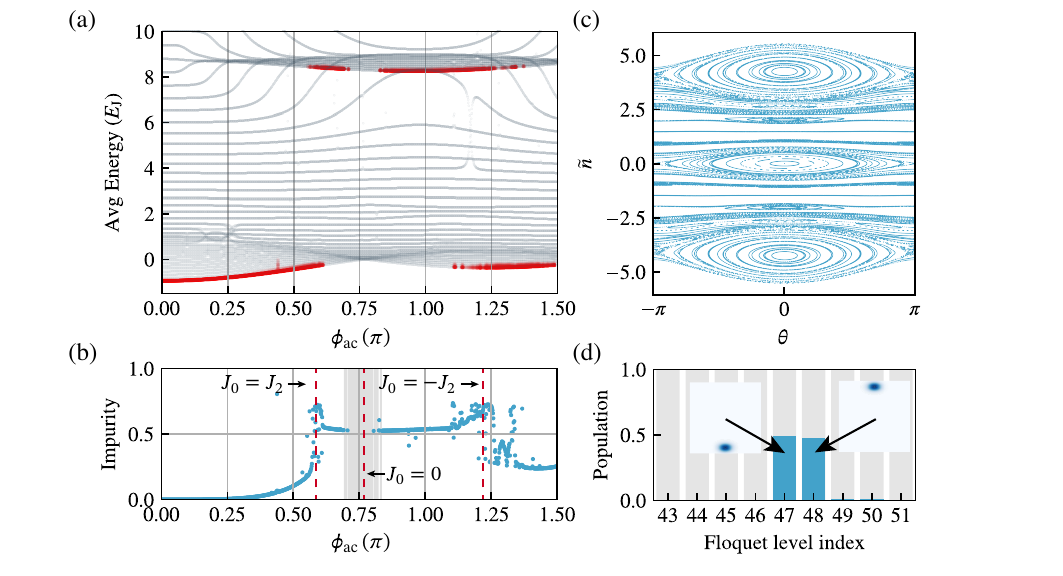}
    \caption{\textbf{Unbound-resonance transition under high-frequency flux drive.}
    (a) Average Floquet energies as a function of drive amplitude for $\omega_{\rm d}/2\pi=15~\mathrm{GHz}$.
    Gray curves show the average energies of the Floquet states, while the opacity of the red curves indicates the steady-state population of each Floquet state obtained from Floquet--Markov simulations.
    (b) Corresponding steady-state impurity.
    The shaded region denotes parameters for which the Floquet--Markov equation admits multiple stationary states.
    (c) Poincar\'e section at the same drive frequency and
    $\phi_{\rm ac}=0.7\pi$.
    The central bound-state island around $\tilde n=0$ is confined by the renormalized Josephson potential, while the two symmetry-related outer resonant tori arise from above-barrier running trajectories resonant with the leading $2\omega_{\rm d}$ drive component.
    (d) Corresponding steady-state Floquet populations.
    The insets show the Husimi functions of the two dominant Floquet states, localized on the outer resonant tori in (c).
    Parameters: $E_\text{C}/2\pi=0.13$~GHz,
    $E_\text{J}/2\pi=50$~GHz, $n_\text{g}=0$.}
    \label{fig:bessel}
\end{figure*}

At high drive frequency, the two low-frequency mechanisms identified above, localized bound-state resonances and separatrix chaos, are both strongly suppressed.
First, the resonance condition $n\omega_\text{d}=m\omega_0(\theta_0)$ requires increasingly large $m$ when the drive frequency exceeds the characteristic bound-state frequency.
The corresponding coupling is strongly suppressed since $J_{2m}(\theta_0)$ rapidly decreases with Bessel order for the relevant bound-state amplitudes.
Second, at sufficiently high frequency the width of the separatrix-chaotic layer is exponentially suppressed~\cite{Zaslavsky1991WeakChaos,Bubner1991QuantumSeparatrix,Cohen2023}.
Instead, Fig.~\ref{fig:impurity} shows a sharp onset near $\phi_\text{ac}=0.6\pi$, beyond which the impurity reaches values near $1/2$.
The location of this onset depends only weakly on drive frequency.
We show below that it marks a transfer of the dissipative steady state from the central bound-state region to outer resonances formed by above-barrier running trajectories.
We refer to this behavior as the \textit{unbound-resonance transition}.
Here, ``unbound'' denotes above-barrier running motion, not literal escape.

We first plot the average energy of the Floquet states using Eq.~\eqref{eq:floquet_avg_eng} as a function of drive amplitude, taking $\omega_\text{d}/2\pi=15$~GHz as a representative high-frequency drive.
Figure~\ref{fig:bessel}(a) reveals two distinct features.
First, the Floquet states associated with the central bound-state region become increasingly compressed in average energy as $\phi_\text{ac}$ approaches the first zero of $J_0(\phi_\text{ac})$, and spread apart again beyond this point.
The corresponding collapse and inversion of the effective Josephson potential has recently been observed experimentally in a strongly driven superconducting circuit~\cite{Deve2026Collapse}.
Second, a group of above-barrier Floquet states converges around an average energy of order $8E_\text{J}$, whereas states away from this energy remain comparatively weakly affected.

To understand these features, we examine the Poincar\'e section at $\phi_\text{ac}=0.7\pi$ in Fig.~\ref{fig:bessel}(c).
Unlike in the low-frequency regime, the section is predominantly regular and contains three distinct regions: a central bound-state island and two outer resonant islands at positive and negative charge.
The central region is the continuation of the bound-state region of the undriven system and is governed, to leading order, by the effective static Hamiltonian $\hat H_\text{static}$ in Eq.~\eqref{eq:flux_static}.
Its reduced area reflects the renormalization of the Josephson energy by $J_0(\phi_\text{ac})$, which weakens the central confinement.
The characteristic small-amplitude frequency is approximately $\sqrt{8E_\text{C}E_\text{J}J_0(\phi_\text{ac})}$ and therefore decreases as the first zero of $J_0$ is approached.
Within this leading-order picture, the Josephson confinement of the central region collapses at the first zero, $\phi_\text{ac}=0.77\pi$.
This accounts for both the compression of the low-energy branches in Fig.~\ref{fig:bessel}(a) and the strongly reduced area of the central island in Fig.~\ref{fig:bessel}(c).
Beyond the first zero, $J_0$ changes sign and the central bound-state region reappears around $\theta=\pi$ rather than $\theta=0$.

The two outer resonant tori have a different origin.
They arise from resonances of above-barrier running trajectories with the leading time-dependent perturbation.
To see this, we consider Hamilton's equations in the above-barrier region.
Since the kinetic energy dominates over the cosine potential, the phase approximately evolves as $\theta(t)\simeq 8E_\text{C}nt$ for a trajectory with charge coordinate $n$.
Substituting this trajectory into Eq.~\eqref{eq:flux_pert} gives
\begin{align}
    H_\text{pert}^{\text{cl}}(t)
    &\simeq
    -2E_\text{J}
    \sum_{m=1}^{\infty}
    J_{2m}(\phi_\text{ac})
    \cos(2m\omega_\text{d}t)
    \cos(8E_\text{C}nt)
    \notag\\
    &=
    -E_\text{J}
    \sum_{m=1}^{\infty}
    J_{2m}(\phi_\text{ac})
    \Big[
    \cos(2m\omega_\text{d}t-8E_\text{C}nt)
    \notag\\
    &\hspace{2.7cm}
    +\cos(2m\omega_\text{d}t+8E_\text{C}nt)
    \Big].
    \label{eq:unbound_res_pert}
\end{align}
The leading $m=1$ term becomes resonant when
$\omega_\text{d}=\pm4E_\text{C}n$.
The resulting outer resonant tori are therefore centered at
$n_\pm=\pm\omega_\text{d}/(4E_\text{C})$, as observed in
Fig.~\ref{fig:bessel}(c).
The kinetic energy at the resonance center is approximately
$4E_\text{C}n_\pm^2=\omega_\text{d}^2/(4E_\text{C})$, explaining why
the affected Floquet states cluster around a well-defined high
average energy in Fig.~\ref{fig:bessel}(a).
The resonant cosine term has amplitude $E_\text{J}J_2(\phi_\text{ac})$.
In a frame co-rotating with the outer resonance, this term becomes static and gives a local resonance barrier of $2E_\text{J}J_2(\phi_\text{ac})$.
As the drive amplitude increases, $J_2(\phi_\text{ac})$ initially grows, so the local outer-resonance confinement strengthens and supports more localized states.
This accounts for the convergence of the high-energy Floquet branches toward the outer-resonance energy.

The outer resonances belong to the above-barrier sector only when their centers lie outside the separatrix of the central effective potential.
Before the first zero of $J_0$, the separatrix intersects $\theta=0$ at
$n_\text{sep}=\sqrt{E_\text{J}J_0(\phi_\text{ac})/(2E_\text{C})}$.
The condition $|n_*|>n_\text{sep}$ therefore gives
\begin{equation}
    \omega_\text{d}
    >
    \sqrt{
    8E_\text{C}E_\text{J}J_0(\phi_\text{ac})
    }.
    \label{eq:high_freq_condition}
\end{equation}
This identifies the high-frequency regime relevant to the unbound-resonance transition, in which the drive frequency exceeds the characteristic frequency of the renormalized central bound-state region.
The boundary is therefore amplitude dependent rather than a fixed frequency.
Outer running-state resonances can also appear at nominally lower drive frequencies once the central confinement has been sufficiently reduced.
In that regime, however, separatrix chaos generally becomes important before the steady-state transfer discussed below, as illustrated by Fig.~\ref{fig:resonance}(e).

Once the outer resonances are established, the central and outer regions are characterized by two competing local confinement scales.
The two structures are described most naturally in different resonant frames, with the $J_0$ component providing the static confinement of the central region and the $J_2$ component becoming static in a frame co-rotating with an outer resonance.
To leading order, the corresponding local Hamiltonians have the same pendulum form and kinetic curvature, with potential amplitudes $E_\text{J}J_0(\phi_\text{ac})$ and $E_\text{J}J_2(\phi_\text{ac})$, respectively.
Their confinement strengths become equal when
\begin{equation}
    J_0(\phi_\text{ac}^*)
    =
    J_2(\phi_\text{ac}^*), \qquad 
    \phi_\text{ac}^*=0.59\pi.
    \label{eq:unbound_threshold_zero_dc}
\end{equation}
Equation~\eqref{eq:unbound_threshold_zero_dc} therefore defines a crossover of local confinement strengths.
Since the central and outer resonances are described in different rotating frames, this condition should not be interpreted as an energetic crossing between them.
Moreover, the Floquet--Markov steady-state populations are determined by the transition network between Floquet states, so equality of the local confinement strengths alone does not determine the steady-state redistribution.
Nevertheless, the steady state transfers from the central region to the outer resonances near this crossover.
As shown below, the corresponding coarse-grained center-to-outer and outer-to-center rates also become comparable in the same drive amplitude, providing a kinetic interpretation of the redistribution.
This correspondence identifies Eq.~\eqref{eq:unbound_threshold_zero_dc} as the analytical criterion for the unbound-resonance threshold.
Since the criterion is independent of $\omega_\text{d}$, it explains the weak frequency dependence of the high-frequency threshold in Fig.~\ref{fig:impurity}.
Importantly, the unbound-resonance threshold occurs at $\phi_\text{ac}^*=0.59\pi$, before the first zero of $J_0(\phi_\text{ac})$ near $0.77\pi$, and is therefore distinct from the subsequent collapse of the central effective Josephson potential~\cite{Deve2026Collapse}.

Below $\phi_\text{ac}^*$, the dissipative steady state is concentrated in the central bound-state region.
Above $\phi_\text{ac}^*$, it transfers predominantly to the two symmetry-related outer resonant regions.
We confirm this picture from the steady-state population as a function of drive amplitude.
The results are shown in Fig.~\ref{fig:bessel}(a) as red curves, with opacity indicating the population.
At small amplitude, the population is concentrated in the lowest state associated with the central bound-state region.
Near $\phi_\text{ac}^*$, it transfers sharply to a symmetry-related pair of Floquet states localized on the outer resonances, with average energies of order $8E_\text{J}$.
Figure~\ref{fig:bessel}(d) shows the Husimi functions of these two populated states, confirming their localization on the well-separated outer resonant tori in Fig.~\ref{fig:bessel}(c).
This phase-space separation is also important for the Floquet--Markov treatment.
Although the two outer states are degenerate, the matrix elements of the local system--bath coupling operator $\hat n$ between them are negligible since the states are localized on well-separated regions of phase space.
Their coherences therefore do not appreciably couple to the population dynamics, so the secular approximation remains valid.

The same steady-state redistribution gives rise to the sharp increase in impurity.
Figure~\ref{fig:bessel}(b) shows the impurity at $\omega_\text{d}/2\pi=15$~GHz, corresponding to the populations in Fig.~\ref{fig:bessel}(a).
As the drive amplitude increases, the impurity initially increases gradually due to drive-induced excitation within the central bound-state region.
Near the threshold $\phi_\text{ac}^*$, the impurity rises sharply, reaching a maximum close to $2/3$.
This value is consistent with approximately equal populations in three dominant Floquet states localized in the central and two outer resonant regions, since the impurity of an equal three-state mixture is $1-3(1/3)^2=2/3$.
Beyond $\phi_\text{ac}^*$, the impurity decreases toward $1/2$, consistent with equal population of the two symmetry-related outer-resonance states.
The outer resonant tori underlying this redistribution are features of the driven Hamiltonian and do not depend on the particular dissipative model used to determine their steady-state occupation.
Additional noise channels, such as flux noise, may broaden the sharp redistribution in drive amplitude seen in Fig.~\ref{fig:bessel}(b), without changing the underlying outer resonant structures.

Near the first zero of $J_0$, around $\phi_\text{ac}=0.77\pi$, the Floquet--Markov generator develops two stationary states [shaded region in Fig.~\ref{fig:bessel}(b)], each localized predominantly on one of the two outer resonant regions.
As discussed later in Sec.~\ref{sec:timescale}, equilibration between the two outer regions is governed by the outer-antisymmetric mode.
Near the first zero of $J_0$, its transition rate approaches zero as the effective outer-to-center and outer-to-outer channels are strongly suppressed.
The two outer regions consequently become effectively disconnected on the timescale resolved by the Floquet--Markov dynamics, producing the multiple stationary states.
See Appendix~\ref{app:bistability} for further discussion.

Beyond the first Bessel zero, the central bound-state region reappears, but its center shifts from $\theta=0$ to $\theta=\pi$ in the Poincar\'e section (not shown).
This shift follows from the sign change of $J_0(\phi_\text{ac})$, which exchanges the minima and maxima of the effective cosine potential.
At still larger amplitude, around $\phi_\text{ac}=1.2\pi$, $|J_0(\phi_\text{ac})|$ again exceeds $J_2(\phi_\text{ac})$.
The central bound-state region, now centered near $\theta=\pi$, becomes dominant again and the impurity decreases.
The remaining impurity of approximately $0.2$ reflects population distributed among excited states within this central region.

The analysis in this section has assumed zero dc flux bias, for which the symmetry of the flux-driven Hamiltonian constrains the strong-drive structure described above. Finite dc flux bias changes the relative weights of the even- and odd-order Bessel contributions and therefore modifies both the low-frequency separatrix-chaos threshold and the high-frequency unbound-resonance threshold. We examine this dc-flux-bias dependence next.

\section{DC-flux-bias dependence of the strong-drive thresholds}
\label{sec:dc}

\begin{figure*}[t]
    \centering
    \includegraphics[width=\textwidth]{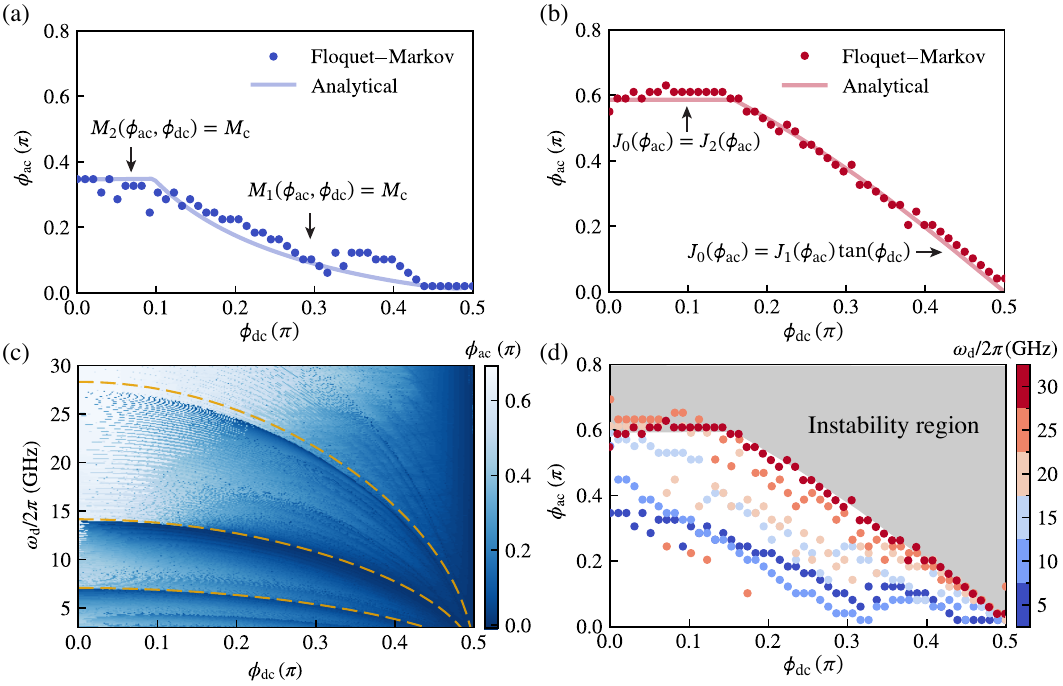}
\caption{\textbf{DC-flux-bias dependence of the strong-drive thresholds.}
    (a) Critical ac amplitude $\phi_{\rm ac}$ as a function of dc flux bias $\phi_{\rm dc}$ in the low-frequency regime.
    The dots are extracted from Floquet--Markov simulations using a fixed impurity threshold of $1/2$ at $\omega_\text{d}/2\pi=5$~GHz, while the solid line is the analytical estimate from the separatrix-chaos criterion in Eq.~\eqref{eq:chaos_threshold_dc}.
    (b) Critical ac amplitude $\phi_{\rm ac}$ in the high-frequency regime, obtained from Floquet--Markov simulations at $\omega_\text{d}/2\pi=30$~GHz and compared with the analytical unbound-resonance threshold in Eq.~\eqref{eq:unbound_threshold_dc}.
    (c) Critical ac amplitude $\phi_{\rm ac}$ at fixed impurity threshold as a function of dc flux bias and drive frequency.
    The color scale gives the extracted critical ac amplitude.
    The dashed curves show the bound-state resonance conditions
    $\omega_\text{d}=\omega_0$, $2\omega_0$, and $4\omega_0$
    of the effective static potential.
    (d) Critical ac amplitude as a function of dc flux bias for selected drive frequencies.
    The color indicates $\omega_\text{d}/2\pi$.
    The upper-right region indicates the instability regime.
    Parameters: $E_{\rm C}/2\pi=0.13$~GHz, $E_{\rm J}/2\pi=50$~GHz, $n_\text{g}=0$.}
    \label{fig:dc_ac}
\end{figure*}

We now examine how a finite dc flux bias modifies the strong-drive thresholds established at zero bias.
This situation is relevant in practice, for example when operating a nonlinear element away from zero flux to activate otherwise forbidden mixing processes or to access Kerr-free operating points~\cite{Chapman2023HighOnOffRatio}.
As shown in Sec.~\ref{subsec:jacobi_anger}, finite dc flux bias activates the odd-order Bessel sector in addition to the even-order sector present at zero bias.
Consistent with this harmonic structure, finite flux bias has been shown to activate MIST resonances that are symmetry-forbidden at zero bias in an inductively coupled readout architecture~\cite{Mori2026CosPhi}.
Here we instead determine how dc flux bias controls the broader low-frequency separatrix-chaos threshold and the distinct high-frequency unbound-resonance threshold.

Writing $J_n\equiv J_n(\phi_{\rm ac})$ for brevity and retaining the leading static, even, and odd contributions gives
\begin{align}
\hat H_{\rm flux}
\simeq{}&
4E_{\rm C}\hat n^2
-
E_{\rm J}J_0\cos\phi_{\rm dc}\cos\hat\theta
\notag\\
&-
2E_{\rm J}J_2\cos\phi_{\rm dc}
\cos(2\omega_{\rm d}t)\cos\hat\theta
\notag\\
&+
2E_{\rm J}J_1\sin\phi_{\rm dc}
\sin(\omega_{\rm d}t)\cos\hat\theta.
\label{eq:finite_dc_leading}
\end{align}
Finite dc flux bias also weakens the effective static potential through the factor $\cos\phi_{\rm dc}$, with effective Josephson energy
$E_{\rm J}J_0(\phi_{\rm ac})\cos\phi_{\rm dc}$
and effective plasma frequency
$\omega_0(\phi_{\rm ac},\phi_{\rm dc})
=\sqrt{8E_{\rm C}E_{\rm J}J_0(\phi_{\rm ac})\cos\phi_{\rm dc}}$.
Thus, over the parameter range considered here, increasing either $\phi_{\rm ac}$ or $\phi_{\rm dc}$ reduces the bound-region area of the effective static potential.
At the same time, finite dc flux bias activates the perturbation proportional to $J_1(\phi_{\rm ac})\sin\phi_{\rm dc}$.

The two leading time-dependent channels play different roles.
The even channel oscillates at $2\omega_{\rm d}$ with strength proportional to $J_2(\phi_{\rm ac})\cos\phi_{\rm dc}$, whereas the odd channel oscillates at $\omega_{\rm d}$ with strength proportional to $J_1(\phi_{\rm ac})\sin\phi_{\rm dc}$.
Since $J_1(\phi_{\rm ac})$ grows linearly at small amplitude while $J_2(\phi_{\rm ac})$ grows quadratically, even a modest dc flux bias can make the odd channel important.
These changes modify the low- and high-frequency strong-drive mechanisms in different ways.
We therefore consider the separatrix-chaos and unbound-resonance thresholds separately.

\subsection{Low-frequency separatrix-chaos threshold}
\label{subsec:low_freq_dc}

At low frequency, finite dc flux bias modifies both the localized bound-state resonances and the broader separatrix-chaos boundary identified in Sec.~\ref{subsec:bound}.
Since the bound-state resonances remain narrow and strongly frequency dependent, we focus here on the broad separatrix-chaos threshold that limits the strong-drive range over a wider frequency interval.
The localized resonance features remain visible in the full threshold map discussed below.

The static part of Eq.~\eqref{eq:finite_dc_leading} has a separatrix separating bound oscillations from above-barrier running motion.
A periodic perturbation produces a stochastic layer around this separatrix, whose width can be estimated from the Melnikov amplitude~\cite{Zaslavsky1991WeakChaos,Bubner1991QuantumSeparatrix}.
For the two leading channels, the dimensionless separatrix splittings are
\begin{equation}
    M_1
    =
    4\pi
    \tan\phi_{\rm dc}
    \frac{J_1(\phi_{\rm ac})}{J_0(\phi_{\rm ac})}
    \Omega^2
    \operatorname{csch}\left(\frac{\pi\Omega}{2}\right),
    \label{eq:M1_dc}
\end{equation}
and
\begin{equation}
    M_2
    =
    16\pi
    \frac{J_2(\phi_{\rm ac})}{J_0(\phi_{\rm ac})}
    \Omega^2
    \operatorname{csch}\left(\pi\Omega\right),
    \label{eq:M2_dc}
\end{equation}
where $\Omega=\omega_{\rm d}/\omega_0(\phi_{\rm ac},\phi_{\rm dc})$.
Here $M_1$ is generated by the $J_1$ channel at frequency $\omega_{\rm d}$, while $M_2$ is generated by the $J_2$ channel at frequency $2\omega_{\rm d}$.
They characterize the separatrix-energy splitting normalized by the effective Josephson energy.

We use $M=\max\{|M_1|,|M_2|\}$ as a measure of the stochastic-layer growth, thereby selecting the dominant channel.
The phase-space-area estimate in Appendix~\ref{app:chaos_area} shows that the stochastic layer becomes a nonperturbative fraction of the bound region once $M$ is no longer small. This motivates the threshold criterion
\begin{equation}
    \max\{|M_1|,|M_2|\}=M_{\mathrm c}.
    \label{eq:chaos_threshold_dc}
\end{equation}
The numerical constant $M_{\mathrm c}$ is not fixed by the leading-log Melnikov treatment, and we determine $M_{\mathrm c}=1$ from comparison with the Floquet--Markov threshold. Changing $M_{\mathrm c}$ shifts the absolute threshold but does not alter its dc-flux-bias dependence or the crossover between the $J_1$ and $J_2$ channels.

Figure~\ref{fig:dc_ac}(a) compares this estimate with the Floquet--Markov threshold at $\omega_{\rm d}/2\pi=5$~GHz.
For each dc flux bias, we define the numerical threshold as the drive amplitude at which the impurity first reaches $1/2$.
For $\phi_{\rm dc}\lesssim0.1\pi$, $M_2$ is the larger contribution and the threshold depends only weakly on dc flux bias.
Beyond this crossover, $M_1$ becomes dominant due to its $\tan\phi_{\rm dc}$ dependence, and the threshold decreases with increasing dc flux bias.
With $M_\text{c}=1$, the Melnikov-based estimate from Eq.~\eqref{eq:chaos_threshold_dc} reproduces the overall dc-flux-bias dependence of the numerical threshold over a broad range, supporting the interpretation that the broad low-frequency threshold is controlled by the growth of separatrix chaos.

The Melnikov estimate is expected to be quantitatively controlled when the frequency of the relevant drive channel is comparable to or larger than the effective plasma frequency, namely $\omega_{\rm d}/\omega_0\gtrsim1$ for the $J_1$ channel and $2\omega_{\rm d}/\omega_0\gtrsim1$ for the $J_2$ channel.
In parts of the parameter range this condition is only marginally satisfied, so the analytical curve should be viewed as a separatrix-chaos estimate rather than a strict asymptotic threshold.

Localized bound-state resonances remain superimposed on this broader boundary and can modify the extracted threshold when the system approaches a resonance.
This additional structure is visible in the full threshold map in Fig.~\ref{fig:dc_ac}(c), where narrow resonance features appear on top of the smooth separatrix-chaos boundary.
At small ac amplitude, their dc-flux-bias dependence approximately follows integer multiples of the plasma frequency
$\omega_0(\phi_{\rm dc})=\sqrt{8E_{\rm C}E_{\rm J}\cos\phi_{\rm dc}}$.
The corresponding $\omega_0$, $2\omega_0$, and $4\omega_0$ curves are shown as dashed lines.

Figure~\ref{fig:dc_ac}(d) overlays the critical ac amplitude as a function of dc flux bias for several drive frequencies.
At low and intermediate frequencies, the threshold follows the same overall trend predicted by Eq.~\eqref{eq:chaos_threshold_dc}, with larger dc flux bias lowering the critical ac amplitude since it both enhances the $J_1$ perturbation and reduces the effective well depth.
At sufficiently high drive frequency, however, the Melnikov factors in Eqs.~\eqref{eq:M1_dc} and \eqref{eq:M2_dc} are exponentially suppressed.
In this limit, separatrix chaos no longer sets the first broad strong-drive threshold, and the unbound-resonance mechanism identified in Sec.~\ref{subsec:unbound} becomes relevant.

\subsection{High-frequency unbound-resonance threshold}
\label{subsec:high_freq_dc}

In this high-frequency regime, the strong-drive threshold is governed by competition between the central bound-state region and drive-induced unbound resonances.
Generalizing the high-frequency discussion at zero dc flux bias, finite dc flux bias produces two pairs of outer resonant tori.
The $J_2(\phi_{\rm ac})$ perturbation generates tori centered at
$n_{\pm,2}=\pm\omega_{\rm d}/(4E_{\rm C})$, while the $J_1(\phi_{\rm ac})$ perturbation generates an additional pair centered at
$n_{\pm,1}=\pm\omega_{\rm d}/(8E_{\rm C})$ since the $J_1$ term oscillates at $\omega_{\rm d}$ rather than $2\omega_{\rm d}$.
Together with the central bound-state region, these give five relevant phase-space structures in the high-frequency regime considered here.
At small drive amplitude, the central region provides the strongest confinement and the steady state remains localized there.
Increasing the amplitude weakens the central confinement while strengthening the outer resonances.

Following the local-confinement picture introduced in Sec.~\ref{subsec:unbound}, we characterize each phase-space structure by the barrier height of its corresponding local potential.
Before the first zero of $J_0$, the corresponding local confinement scales are
$U_0=2E_{\rm J}J_0(\phi_{\rm ac})\cos\phi_{\rm dc}$ for the central region,
$U_2=2E_{\rm J}J_2(\phi_{\rm ac})\cos\phi_{\rm dc}$ for each $J_2$-generated outer resonance, and
$U_1=2E_{\rm J}J_1(\phi_{\rm ac})\sin\phi_{\rm dc}$ for each $J_1$-generated outer resonance.
These quantities characterize the local confinement of each phase-space structure in the frame where it is stationary, rather than directly comparable absolute energies.
Thus, dc flux bias suppresses both the central and $J_2$ confinement through the same factor $\cos\phi_{\rm dc}$, while simultaneously increasing the $J_1$ confinement through $\sin\phi_{\rm dc}$.

As in the zero-dc-flux-bias case, these resonances belong to the unbound sector only when their centers lie outside the separatrix of the effective central potential.
Before the first zero of $J_0$, the separatrix intersects $\theta=0$ at
\begin{equation}
    n_{\rm sep}
    =
    \sqrt{
    \frac{
    E_{\rm J}J_0(\phi_{\rm ac})\cos\phi_{\rm dc}
    }{
    2E_{\rm C}
    }}.
    \label{eq:finite_dc_separatrix}
\end{equation}
For the $J_2$-generated resonances, whose centers are
$n_{\pm,2}$, requiring
$|n_{\pm,2}|>n_{\rm sep}$ gives
\begin{equation}
    \omega_{\rm d}
    >
    \sqrt{
    8E_{\rm C}E_{\rm J}
    J_0(\phi_{\rm ac})\cos\phi_{\rm dc}
    }
    =
    \omega_0.
    \label{eq:j2_unbound_condition_dc}
\end{equation}
The $J_1$-generated resonances lie closer to the central region, with centers at
$n_{\pm,1}$.
Requiring these resonances to lie outside the separatrix gives the more stringent condition
\begin{equation}
    \omega_{\rm d}
    >
    \sqrt{
    32E_{\rm C}E_{\rm J}
    J_0(\phi_{\rm ac})\cos\phi_{\rm dc}
    }
    =
    2\omega_0.
    \label{eq:j1_unbound_condition_dc}
\end{equation}
For both the $J_1$- and $J_2$-generated resonance pairs to lie in the above-barrier sector, the more stringent $J_1$ condition must therefore be satisfied,
$\omega_{\rm d}>2\omega_0$.
Which of these conditions is relevant depends on which outer-resonance family controls the local-confinement crossover.

For small dc flux bias, the $J_1$ contribution is suppressed by $\sin\phi_{\rm dc}$, and the relevant confinement crossover is between the central region and the $J_2$-generated outer resonances.
The condition is $J_0(\phi_{\rm ac})=J_2(\phi_{\rm ac})$.
The common factor $\cos\phi_{\rm dc}$ cancels, so the critical amplitude remains approximately independent of dc flux bias in this regime, with $\phi_{\rm ac}^*=0.59\pi$.
For larger dc flux bias, the $J_1$-generated outer resonances provide the larger outer confinement scale, and the corresponding crossover satisfies
$J_0(\phi_{\rm ac})\cos\phi_{\rm dc}
=
J_1(\phi_{\rm ac})\sin\phi_{\rm dc}$.
Combining the two limits gives
\begin{equation}
    J_0(\phi_{\rm ac})
    =
    \max\left\{
    J_2(\phi_{\rm ac}),
    J_1(\phi_{\rm ac})\tan\phi_{\rm dc}
    \right\}.
    \label{eq:unbound_threshold_dc}
\end{equation}

Figure~\ref{fig:dc_ac}(b) shows that the above criterion accurately tracks the Floquet--Markov steady-state transition with $\omega_\text{d}/2\pi=30$~GHz.
We use a higher drive frequency than in the zero-bias example of Sec.~\ref{subsec:unbound} since the $J_1$-generated resonances lie closer to the central region and enter the above-barrier sector only when $\omega_{\rm d}>2\omega_0$.
The critical amplitude initially remains near the solution of $J_0(\phi_{\rm ac})=J_2(\phi_{\rm ac})$, producing a weakly dc-flux-bias-dependent plateau.
At larger dc flux bias, the $J_1$ channel takes over and the threshold follows the condition
$J_0(\phi_{\rm ac})=J_1(\phi_{\rm ac})\tan\phi_{\rm dc}$.
The analytical result agrees well with the numerical threshold, confirming that the high-frequency unbound-resonance transition tracks the crossover between the local confinement scales of the central and drive-induced outer resonances.
As in the zero-bias case, after the transition the Floquet--Markov steady state is dominated by two states localized on the relevant outer resonant tori.

Having established the dc-flux-bias dependence of both thresholds numerically and analytically, we next test the predicted low-frequency dependence experimentally in a flux-driven SQUID.

\subsection{Experimental verification of the low-frequency dc-flux-bias dependence}
\label{subsec:exp_dc}

\begin{figure}[htbp]
    \centering
    \includegraphics[width=\columnwidth]{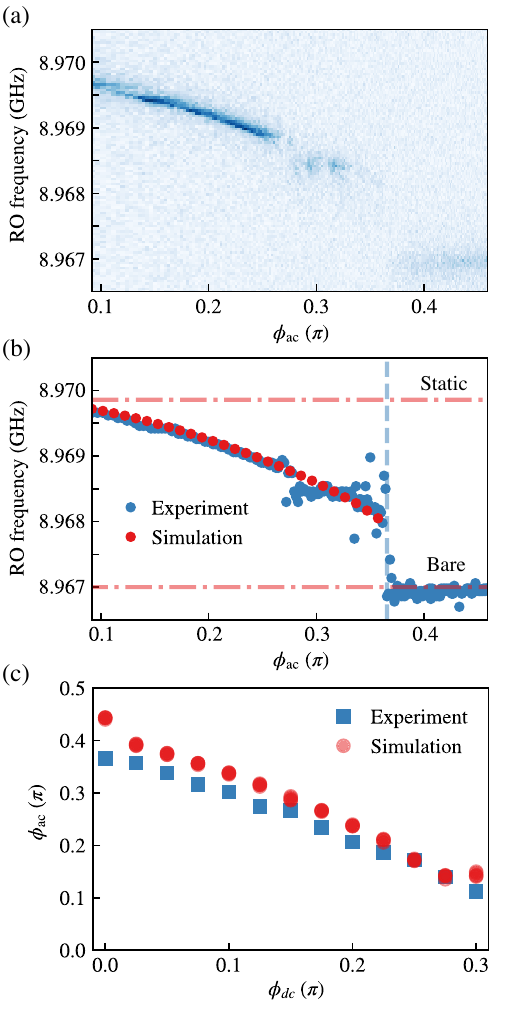}
    \caption{\textbf{Experimental dc-flux-bias dependence of the low-frequency strong-drive threshold.}
(a) Readout resonator spectrum versus ac flux-drive amplitude at zero dc flux bias.
(b) Extracted resonance frequency (blue dots), experimentally identified threshold (blue dashed line), and simulated ac-Stark-shifted resonance frequency (red dots). Red dash-dot lines mark the undriven and bare-resonator frequencies.
(c) Critical ac amplitude versus dc flux bias. Blue squares show experiment, and red dots show Floquet--Markov results for $n_\text{g}=0,0.1,\ldots,0.4$, demonstrating weak offset-charge dependence.
Parameters: $E_\text{C}/2\pi=0.09$~GHz, $E_\text{J}/2\pi=66$~GHz, $\omega_\text{r}/2\pi=8.967$~GHz, $g/2\pi=39$~MHz.}
\label{fig:dc_ac_exp}
\end{figure}

The device consists of a SQUID with $E_\text{C}/2\pi=0.09$~GHz and $E_\text{J}/2\pi=66$~GHz capacitively coupled to a readout resonator with coupling strength $g/2\pi=39$~MHz.
The SQUID is flux driven through a buffer mode at $\omega_\text{d}/2\pi=2.03$~GHz.
Additional details of the experiment and device are given in Appendix~\ref{app:exp}.

Figure~\ref{fig:dc_ac_exp}(a) shows a representative readout spectrum at zero dc flux bias as the ac flux-drive amplitude is increased.
At small drive amplitude, the readout frequency decreases gradually, consistent with the negative ac Stark shift induced by the SQUID anharmonicity.
Above a critical drive amplitude, the readout frequency shifts abruptly toward its bare frequency and becomes relatively insensitive to further increases in drive amplitude.
We use this abrupt change in the resonator response to define the experimental strong-drive threshold.

Similar behavior has been reported in strongly driven transmon--resonator systems~\cite{Pietikainen2017BlochSiegert,Cohen2023}.
Population of many highly excited states can suppress the total dispersive frequency pull on the resonator, leaving it close to its bare frequency~\cite{Cohen2023}.
Population distributed over many SQUID states is also consistent with a strongly mixed steady state and therefore large impurity.
The abrupt loss of dispersive frequency pull thus provides an experimental proxy for the impurity onset used to define the numerical threshold above.

To extract the threshold from the raw spectrum, we identify the resonance frequency corresponding to the maximum resonator response at each drive power, shown as blue dots in Fig.~\ref{fig:dc_ac_exp}(b).
The blue dashed line marks the experimentally identified critical amplitude.
The red dots show the simulated Stark-shifted resonator frequency, which we use to calibrate the conversion from applied drive power to the ac flux amplitude $\phi_\text{ac}$.
The red dash-dot lines indicate the resonator frequency without drive and the bare resonator frequency in the absence of coupling to the SQUID.

We repeat this procedure for different dc flux biases and plot the extracted critical ac amplitude as blue squares in Fig.~\ref{fig:dc_ac_exp}(c).
At larger dc flux bias, the measured threshold decreases, consistent with the theoretical mechanism described above, as increasing $\phi_\text{dc}$ both weakens the effective static potential and activates the odd-Bessel perturbation channel proportional to $J_1(\phi_\text{ac})\sin\phi_\text{dc}$.

For quantitative comparison, we calculate the Floquet--Markov steady-state impurity of the driven SQUID as a function of ac amplitude for each dc flux bias.
The simulated critical amplitudes, extracted from the sharp impurity onset, are shown as red dots in Fig.~\ref{fig:dc_ac_exp}(c).
At each dc flux bias, we overlay results for five offset charges, $n_\text{g}=0,0.1,\ldots,0.4$.
The resulting spread is small compared with the systematic shift induced by the dc flux bias, indicating that the offset-charge dependence of the critical amplitude is negligible in this parameter regime. 
The Floquet--Markov calculation reproduces the measured dc-flux-bias dependence of the critical amplitude, supporting the connection between the observed threshold and the low-frequency strong-drive behavior described above.

The remaining discrepancy is largest near zero dc flux.
At zero bias, the fundamental-frequency channel (drive term at $\omega_\text{d}$) vanishes in the ideal symmetric model, making the threshold particularly sensitive to residual junction asymmetry or common-mode drive.
Thus, even in a device designed with nominally symmetric junctions and a predominantly differential flux-drive architecture, a small junction-energy mismatch or common-mode drive component can activate this channel and lower the threshold relative to the ideal prediction (see Appendix~\ref{app:asymmetry}).
At finite dc flux, the corresponding $J_1$ channel is already present in the ideal symmetric model, reducing the relative importance of these corrections and possibly explaining the improved agreement.

The experimental comparison above concerns the low-frequency strong-drive regime.
We now return to the high-frequency unbound-resonance threshold.
Rather than the onset of chaos, this threshold marks a transfer from the central bound-state region to localized outer resonances.
This distinction raises an important practical question: can the intended parametric process persist beyond the unbound-resonance threshold?
We address this question next.

\section{Parametric operation across the unbound-resonance threshold}
\label{sec:parametric}

As a representative process, we consider excitation exchange between a flux-driven Josephson coupler and a cavity mode.
Flux modulation is used to activate sideband transitions and parametrically controlled exchange processes, including qubit--resonator sidebands and beamsplitter interactions between bosonic modes~\cite{PhysRevLett.119.150502, PhysRevB.87.220505,Lu2023HighFidelityParametricBeamsplitting,Petrescu2023StrongDrive}.
Below the threshold, the dissipative steady state is localized in the central bound-state region, whereas above the threshold it is transferred to the unbound-resonance region formed by above-barrier running trajectories.
We show that parametric exchange can persist in this outer-resonance regime, but the achievable rate is lower than the maximum attained on the bound-state side of the threshold for the processes studied here.

\subsection{Parametric exchange in the central bound-state region}

We consider a flux-driven SQUID at zero dc flux bias capacitively coupled to a cavity mode $\hat a$,
\begin{equation}
  \hat H
  =
  4\EC \hat n^2
  -
  \EJ \cos\hat\theta
  \cos[\phi_{\rm ac}\sin(\wdrv t)]
  +
  g\hat n\hat n_\text{a}
  +
  \omega_\text{a}\hat a^\dagger\hat a,
  \label{eq:Htotal}
\end{equation}
where $g$ is the coupling strength, $\hat n_\text{a}$ is the cavity charge operator, and $\omega_\text{a}$ is the cavity frequency.
Using the Jacobi--Anger expansion and retaining the leading terms,
\begin{equation}
  \cos[\phi_{\rm ac}\sin(\wdrv t)]
  \simeq
  J_0(\phi_{\rm ac})
  +
  2J_2(\phi_{\rm ac})\cos(2\wdrv t).
  \label{eq:cosJ0J2}
\end{equation}
This is the same decomposition used in Sec.~\ref{subsec:unbound} to describe the central bound-state region and the outer resonant tori associated with above-barrier motion.

We first consider the regime below the unbound-resonance threshold, where the dissipative steady state is localized in the central bound-state region.
For amplitudes below the first zero of $J_0$, this region is centered at $\theta=0$.
The derivation below describes the parametric process in this steady-state regime, but remains applicable above the threshold as long as the central well and its associated localized Floquet state remain well defined.
In that case, the coupler can be initialized in the central localized state and support coherent parametric operation, even though dissipation ultimately transfers it to the unbound-resonance steady state.
We return to this possibility in the numerical comparison below.

The leading static component defines the effective Hamiltonian $\hat{H}_\text{static}$ as in Eq.~\eqref{eq:flux_static}.
Expanding near $\theta=0$ gives
\begin{equation}
  \hat{H}_\text{static}
  \simeq
  4\EC\hat n^2
  +
  \frac{1}{2}\EJ J_0(\phi_{\rm ac})\hat\theta^2,
\end{equation}
with local harmonic frequency
\begin{equation}
  \bar\omega_{\mathrm q}^{\rm c}
  =
  \omega_{\mathrm q}^{(0)}
  \sqrt{J_0(\phi_{\rm ac})},
  \qquad
  \omega_{\mathrm q}^{(0)}
  =
  \sqrt{8\EC\EJ}.
  \label{eq:omega_center}
\end{equation}
The remaining leading time-dependent term is
\begin{equation}
  \hat V(t)
  =
  -2\EJ J_2(\phi_{\rm ac})
  \cos(2\wdrv t)\cos\hat\theta.
\end{equation}
Near the minimum, this term produces a modulation of the instantaneous local oscillator frequency.
With the zero-point fluctuation
\begin{equation}
  \theta_{\rm zpf}^\text{c}
  =
  \left[
  \frac{2\EC}
  {\EJ J_0(\phi_{\rm ac})}
  \right]^{1/4},
\end{equation}
we obtain
\begin{equation}
  \omega_{\mathrm q}(t)
  \simeq
  \bar\omega_{\mathrm q}^{\rm c}
  +
  \delta\omega_{\mathrm q}^{\rm c}\cos(2\wdrv t),
\end{equation}
where
\begin{equation}
  \delta\omega_{\mathrm q}^{\rm c}
  =
  2\EJ J_2(\phi_{\rm ac})(\theta_{\rm zpf}^\text{c})^2
  =
  \omega_{\mathrm q}^{(0)}
  \frac{J_2(\phi_{\rm ac})}
  {\sqrt{J_0(\phi_{\rm ac})}},
  \label{eq:freq_mod}
\end{equation}
and the superscript `c' labels quantities associated with the central region.

Following Ref.~\onlinecite{PhysRevB.87.220505}, we move to the interaction picture generated by the time-dependent oscillator frequency using
\begin{equation}
  \hat U(t)
  =
  \exp\left[
  -i\Phi(t)\hat c^\dagger\hat c
  -i\omega_\text{a} t\hat a^\dagger\hat a
  \right],
  \label{eq:Utransmon}
\end{equation}
where $\hat c$ is the local coupler annihilation operator and
\begin{equation}
  \Phi(t)
  =
  \int_0^t\omega_{\mathrm q}(t')\,dt'
  =
  \bar\omega_{\mathrm q}^{\rm c}t
  +
  \frac{\delta\omega_{\mathrm q}^{\rm c}}
  {2\wdrv}
  \sin(2\wdrv t).
\end{equation}
The resulting phase factor can be expanded as
\begin{equation}
  e^{i\Phi(t)}
  =
  e^{i\bar\omega_{\mathrm q}^{\rm c}t}
  \sum_{m=-\infty}^{\infty}
  J_m(\beta_{\rm c})
  e^{i2m\wdrv t},
\end{equation}
with modulation index
\begin{equation}
  \beta_{\rm c}
  =
  \frac{\delta\omega_{\mathrm q}^{\rm c}}{2\wdrv}
  =
  \frac{\omega_{\mathrm q}^{(0)}}{2\wdrv}
  \frac{J_2(\phi_{\rm ac})}
  {\sqrt{J_0(\phi_{\rm ac})}}.
  \label{eq:beta}
\end{equation}

After applying the rotating-wave approximation, the charge--charge interaction becomes
\begin{equation}
  g_\text{a}
  \hat c^\dagger\hat a\,
  e^{i(\bar\omega_{\mathrm q}^{\rm c}-\omega_\text{a})t}
  \sum_m
  J_m(\beta_{\rm c})e^{i2m\wdrv t}
  +
  {\rm h.c.},
\end{equation}
where $g_\text{a}=g\,n_{\rm zpf}^\text{c}n_{\text{a},\rm zpf}$ is the effective coupling strength,
$n_{\text{a},\rm zpf}$ is the charge zero-point fluctuation of the cavity mode, and
\begin{equation}
  n_{\rm zpf}^\text{c}
  =
  \frac{1}{2}
  \left[
  \frac{\EJ J_0(\phi_{\rm ac})}
  {2\EC}
  \right]^{1/4}
  \label{eq:nzpf_std}
\end{equation}
is the corresponding charge zero-point fluctuation of the coupler mode.
Substituting this expression for $n_{\rm zpf}^\text{c}$, the effective coupling can be written as
\begin{equation}
  g_\text{a}
  =
  g_\text{a}^{(0)}
  [J_0(\phi_{\rm ac})]^{1/4},
  \qquad
  g_\text{a}^{(0)}
  \equiv
  \frac{g}{2}
  \left(\frac{\EJ}{2\EC}\right)^{1/4}
  n_{\text{a},\rm zpf}.
\end{equation}
On the $m$-th sideband,
\begin{equation}
  \bar\omega_{\mathrm q}^{\rm c}
  -
  \omega_\text{a}
  +
  2m\wdrv
  =
  0,
\end{equation}
the effective coupling is therefore
\begin{equation}
  g_{\rm eff}^{\text{c},m}
  =
  g_\text{a}^{(0)}
  [J_0(\phi_{\rm ac})]^{1/4}
  J_m(\beta_{\rm c}).
  \label{eq:g_center_general}
\end{equation}
For a resonant beamsplitter interaction
$g_{\rm eff}(\hat c^\dagger\hat a+\hat c\hat a^\dagger)$,
the population oscillates at the exchange oscillation rate
\begin{equation}
  \Omega_{\rm ex}
  =
  2|g_{\rm eff}|.
  \label{eq:exchange_rate_definition}
\end{equation}

Equation~\eqref{eq:g_center_general} has the same sideband structure as the familiar result for a sinusoidally frequency-modulated qubit~\cite{PhysRevB.87.220505}, but its dependence on the applied drive amplitude is different.
For a modulation
$\omega_{\mathrm q}(t)=\omega_{\mathrm q}^{(0)}+A\cos\omega t$,
the resulting sideband rates are controlled by $J_m(A/\omega)$.
Here, the frequency-modulation amplitude itself depends nonlinearly on the applied ac flux amplitude through
$J_2(\phi_{\rm ac})/\sqrt{J_0(\phi_{\rm ac})}$.
At small amplitude, $J_2(\phi_{\rm ac})\propto\phi_{\rm ac}^2$, so the sideband argument is quadratic rather than linear in the applied flux amplitude.
In addition, the coupler charge zero-point fluctuation, and hence the coupler--cavity coupling, acquires a factor $J_0^{1/4}(\phi_{\rm ac})$.
Both effects determine the drive-amplitude dependence of the parametric rate in the present flux-driven circuit.

\subsection{Parametric exchange in the unbound-resonance region}
\label{subsec:outer_parametric}

Above the unbound-resonance threshold, the dissipative steady state transfers predominantly to the two outer resonant regions centered at
$n_\pm=\pm\wdrv/(4\EC)$.
To analyze coherent parametric operation, we consider a Floquet state localized in one of these regions and choose the positive-$n$ branch for definiteness.
In a frame co-rotating with this resonance, the $J_2$ component becomes the leading static confinement, whereas the $J_0$ component becomes time dependent.
At this level, the roles played by $J_0$ and $J_2$ in the central-region analysis are interchanged.
This already suggests a reversal of the drive-amplitude dependence since $J_2$ grows while $J_0$ decreases over the range surrounding the unbound-resonance threshold.

Using Eq.~\eqref{eq:cosJ0J2}, the coupler potential can be written as
\begin{equation}
\begin{aligned}
  &-\EJ J_0(\phi_{\rm ac})\cos\hat\theta \\
  &\quad
  -\EJ J_2(\phi_{\rm ac})
  \big[
    \cos(\hat\theta-2\wdrv t)
    +
    \cos(\hat\theta+2\wdrv t)
  \big].
\end{aligned}
\label{eq:cos_split}
\end{equation}
For the positive-$n$ outer resonance, we apply
\begin{equation}
  \hat U_{\mathrm F}(t)
  =
  \exp[-i2\wdrv t\,\hat n].
  \label{eq:UF}
\end{equation}
The phase operator transforms as
\begin{equation}
  \hat U_{\mathrm F}^\dagger(t)
  \hat\theta
  \hat U_{\mathrm F}(t)
  =
  \hat\theta+2\wdrv t,
\end{equation}
so that
\begin{align}
  \cos(\hat\theta-2\wdrv t)
  &\rightarrow
  \cos\hat\theta,
  \\
  \cos\hat\theta
  &\rightarrow
  \cos(\hat\theta+2\wdrv t),
  \\
  \cos(\hat\theta+2\wdrv t)
  &\rightarrow
  \cos(\hat\theta+4\wdrv t).
\end{align}
The kinetic term becomes
\begin{equation}
  4\EC\hat n^2
  \rightarrow
  4\EC
  \left(
  \hat n-\frac{\wdrv}{4\EC}
  \right)^2
  -
  \frac{\wdrv^2}{4\EC}.
  \label{eq:kinetic_shifted}
\end{equation}
The kinetic minimum is therefore displaced to the charge coordinate
$n_+=\wdrv/(4\EC)$,
consistent with the location of the outer resonant torus found in Sec.~\ref{subsec:unbound}.
Ignoring the constant in Eq.~\eqref{eq:kinetic_shifted}, the rotating-frame Hamiltonian is
\begin{align}
  \hat{H}_+
  ={}&
  4\EC(\hat n-n_+)^2
  -
  \EJ J_2(\phi_{\rm ac})\cos\hat\theta
  \notag\\
  &-
  \EJ J_0(\phi_{\rm ac})
  \cos(\hat\theta+2\wdrv t)
  \notag\\
  &-
  \EJ J_2(\phi_{\rm ac})
  \cos(\hat\theta+4\wdrv t).
  \label{eq:Hcos_flipped}
\end{align}
The first line defines the local static confinement associated with the outer resonance. In the drive-amplitude range considered here, $J_2(\phi_{\rm ac})>0$, and its local plasma frequency is
\begin{equation}
  \bar\omega_{\mathrm q}^{\rm o}
  =
  \sqrt{
  8\EC\EJ J_2(\phi_{\rm ac})
  },
  \label{eq:omega_outer}
\end{equation}
where the superscript `o' labels quantities associated with the outer region.
The corresponding zero-point fluctuations are
\begin{equation}
  \theta_{\rm zpf}^{\rm o}
  =
  \left[
  \frac{2\EC}
  {\EJ J_2(\phi_{\rm ac})}
  \right]^{1/4},
  \qquad
  n_{\rm zpf}^{\rm o}
  =
  \frac{1}{2}
  \left[
  \frac{\EJ J_2(\phi_{\rm ac})}
  {2\EC}
  \right]^{1/4}.
  \label{eq:zpf_flipped}
\end{equation}

The two time-dependent terms in Eq.~\eqref{eq:Hcos_flipped} do not factorize into a purely spatial term multiplied by a purely temporal modulation.
Using
\begin{align}
  \cos(\hat\theta+2\wdrv t)
  &=
  \cos\hat\theta\cos(2\wdrv t)
  -
  \sin\hat\theta\sin(2\wdrv t),
  \\
  \cos(\hat\theta+4\wdrv t)
  &=
  \cos\hat\theta\cos(4\wdrv t)
  -
  \sin\hat\theta\sin(4\wdrv t),
\end{align}
the terms proportional to $\sin\hat\theta$ generate linear drives on the local outer-resonance oscillator.
These linear terms displace the classical coupler orbit and, through the charge--charge interaction, induce a small coherent displacement of the cavity.
They can be removed by successive time-dependent displacements of the coupler and cavity, as detailed in Appendix~\ref{app:parametric_displacement}.
For drive components sufficiently detuned from both local mode frequencies, these displacements remain perturbative and do not modify the leading parametric coupling derived below.

After removing the linear terms, the quadratic part produces the frequency modulation
\begin{equation}
  \delta\omega_{\mathrm q}^{\rm o}(t)
  =
  \EJ{(\theta_{\rm zpf}^{\rm o})}^2
  \left[
  J_0(\phi_{\rm ac})\cos(2\wdrv t)
  +
  J_2(\phi_{\rm ac})\cos(4\wdrv t)
  \right].
\end{equation}
The two modulation amplitudes are therefore
\begin{align}
  \delta\omega_{\mathrm q}^{\text{o},2\wdrv}
  &=
  \sqrt{2\EC\EJ}\,
  \frac{J_0(\phi_{\rm ac})}
  {\sqrt{J_2(\phi_{\rm ac})}},
  \\
  \delta\omega_{\mathrm q}^{\text{o},4\wdrv}
  &=
  \sqrt{2\EC\EJ J_2(\phi_{\rm ac})}.
  \label{eq:delta_w_flipped}
\end{align}
The associated modulation indices are
\begin{align}
  \beta_1
  &=
  \frac{\delta\omega_{\mathrm q}^{\text{o},2\wdrv}}{2\wdrv}
  =
  \frac{\omega_{\mathrm q}^{(0)}}{4\wdrv}
  \frac{J_0(\phi_{\rm ac})}
  {\sqrt{J_2(\phi_{\rm ac})}},
  \\
  \beta_2
  &=
  \frac{\delta\omega_{\mathrm q}^{\text{o},4\wdrv}}{4\wdrv}
  =
  \frac{\omega_{\mathrm q}^{(0)}}{8\wdrv}
  \sqrt{J_2(\phi_{\rm ac})}.
  \label{eq:beta_flipped}
\end{align}
Introducing the phase factor
\begin{equation}
  \Phi_{\rm o}(t)
  =
  \bar\omega_{\mathrm q}^{\rm o}t
  +
  \beta_1\sin(2\wdrv t)
  +
  \beta_2\sin(4\wdrv t),
\end{equation}
and applying the Jacobi--Anger expansion to both components gives
\begin{equation}
  e^{i\Phi_{\rm o}(t)}
  =
  e^{i\bar\omega_{\mathrm q}^{\rm o}t}
  \sum_{m_1,m_2}
  J_{m_1}(\beta_1)
  J_{m_2}(\beta_2)
  e^{2i(m_1+2m_2)\wdrv t}.
  \label{eq:double_JA}
\end{equation}
Multiple pairs $(m_1,m_2)$ therefore contribute coherently to a physical sideband with index $\ell=m_1+2m_2$.
Defining
\begin{equation}
  \mathcal B_\ell(\beta_1,\beta_2)
  =
  \sum_{m_2=-\infty}^{\infty}
  J_{\ell-2m_2}(\beta_1)
  J_{m_2}(\beta_2),
  \label{eq:Bcomb}
\end{equation}
the corresponding component is
\begin{equation}
  e^{i\bar\omega_{\mathrm q}^{\rm o}t}
  e^{i2\ell\wdrv t}
  \mathcal B_\ell(\beta_1,\beta_2).
\end{equation}
The local coupler--cavity coupling carries the drive-dependent zero-point fluctuation
\begin{equation}
  g_\text{a}^{\rm o}
  =
  g\, n_{\rm zpf}^\text{o}n_{\text{a},\rm zpf}
  =
  g_\text{a}^{(0)}
  [J_2(\phi_{\rm ac})]^{1/4}.
\end{equation}
On the $\ell$-th sideband,
\begin{equation}
  \bar\omega_{\mathrm q}^{\rm o}
  -
  \omega_\text{a}
  +
  2\ell\wdrv
  =
  0,
\end{equation}
the effective coupling is
\begin{equation}
  g_{\rm eff}^{\text{o}, \ell}
  =
  g_\text{a}^{(0)}
  [J_2(\phi_{\rm ac})]^{1/4}
  \mathcal B_\ell(\beta_1,\beta_2).
  \label{eq:g_outer_general}
\end{equation}

\subsection{Exchange-rate comparison across the unbound-resonance threshold}

We now compare the exchange rates supported by the central bound-state region and the outer resonant tori.
For the central bound-state region, the derivation above assumed $J_0(\phi_{\rm ac})>0$.
Beyond the first zero of $J_0$, the central bound-state region reappears around $\theta=\pi$.
Its curvature and zero-point fluctuations are then controlled by $|J_0(\phi_{\rm ac})|$.
The sign change reverses the phase of the quadratic modulation but does not affect the magnitude of the exchange rate.
Away from the immediate vicinity of the zero, we can therefore write
\begin{equation}
  \bar\omega_{\mathrm q}^{\rm c}
  =
  \omega_{\mathrm q}^{(0)}
  \sqrt{|J_0(\phi_{\rm ac})|}.
  \label{eq:omega_center_abs}
\end{equation}
For the $m=1$ central-region sideband,
$2\wdrv=\omega_\text{a}-\bar\omega_{\mathrm q}^{\rm c}$,
the exchange oscillation rate is
\begin{equation}
\begin{aligned}
  \Omega_{\rm ex}^{{\rm c},1}
  =
  2g_\text{a}^{(0)}
  |J_0(\phi_{\rm ac})|^{1/4}
  \left|
  J_1\left[
  \frac{\omega_{\mathrm q}^{(0)}}{2\wdrv}
  \frac{J_2(\phi_{\rm ac})}
  {\sqrt{|J_0(\phi_{\rm ac})|}}
  \right]
  \right|.
\end{aligned}
\label{eq:g_std}
\end{equation}
For the outer resonant tori, the general result derived above gives
\begin{equation}
  \Omega_{\rm ex}^{{\rm o},\ell}
  =
  2g_\text{a}^{(0)}
  [J_2(\phi_{\rm ac})]^{1/4}
  |\mathcal B_\ell|,
\end{equation}
where $\mathcal B_\ell$ is defined in Eq.~\eqref{eq:Bcomb}.

The different drive-amplitude dependence of the central and outer-resonance processes reflects the interchange of the roles of $J_0$ and $J_2$.
For operation in the central bound-state region, the confinement is set by $J_0$, while the leading parametric modulation is set by $J_2$.
For the outer resonant tori, the confinement is instead set by $J_2$.
The leading $2\wdrv$ modulation relevant to the $\ell=1$ process is controlled primarily by $J_0$.
This interchange produces the opposite overall drive-amplitude trend of the $\ell=1$ exchange rate near the unbound-resonance threshold.
The $\ell=2$ process is instead driven by the $4\wdrv$ modulation associated with $J_2$.
Its modulation index is proportional to $\sqrt{J_2}$ but contains the larger denominator $8\wdrv$, which suppresses the corresponding exchange rate.

\subsection{Numerical verification of parametric exchange rates}
\label{subsec:numeric_parametric}

\begin{figure}[t]
    \centering
    \includegraphics[width=\columnwidth]{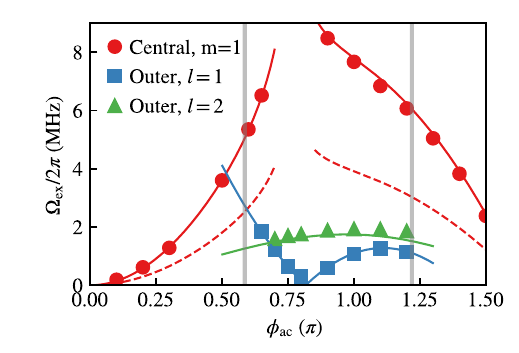}
   \caption{\textbf{Parametric exchange rates across the unbound-resonance threshold.}
    Exchange oscillation rate $\Omega_{\rm ex}$ between the flux-driven coupler and the cavity as a function of drive amplitude $\phi_{\rm ac}$.
    Markers denote rates extracted from coherent time-domain simulations, and solid lines denote analytical predictions at $\omega_{\rm d}/2\pi=15$~GHz.
    Red circles show exchange for a coupler localized in the central bound-state region, with the analytical prediction given by Eq.~\eqref{eq:g_std}.
    Blue squares and green triangles show the $\ell=1$ and $\ell=2$ processes, respectively, for a coupler localized on an outer resonant torus, with analytical predictions given by Eqs.~\eqref{eq:new_1} and \eqref{eq:new_2}.
    The red dashed curve shows the analytical central-region prediction from Eq.~\eqref{eq:g_std} for $\omega_{\rm d}/2\pi=30$~GHz.
    The two gray vertical lines mark the boundaries of the unbound-resonance regime.
    Parameters: $E_{\rm C}/2\pi=0.13$~GHz, $E_{\rm J}/2\pi=50$~GHz, $n_\text{g}=0$, $gn_{\rm a,\rm zpf}/2\pi=0.1$~GHz.}
    \label{fig:parametric}
\end{figure}

To verify these predictions, we simulate coherent excitation exchange between the flux-driven coupler and a cavity.
We use $E_{\rm J}/2\pi=50$~GHz and $E_{\rm C}/2\pi=0.13$~GHz, as in the preceding high-frequency examples, set $\phi_{\rm dc}=0$, and take $\omega_{\rm d}/2\pi=15$~GHz.
For each drive amplitude and each parametric channel, the cavity frequency is chosen to satisfy the corresponding sideband resonance condition.
The resulting curves therefore compare the on-resonance exchange rates supported by the different localized Floquet states.
The cavity frequency is calibrated at each drive amplitude and for each channel, so the curves do not represent the response of a single cavity with fixed $\omega_{\rm a}$ as the drive amplitude is varied.
We initialize the coupler in a Floquet state localized either in the central bound-state region or on one of the outer resonant tori and evolve the coupled system coherently.
This protocol also allows us to evaluate a central localized Floquet state above the unbound-resonance threshold whenever that state remains well defined, as discussed above.
The population dynamics are fitted to an oscillatory function, and the fitted angular frequency is reported as the exchange oscillation rate $\Omega_{\rm ex}$ defined in Eq.~\eqref{eq:exchange_rate_definition}.

For quantitative comparison near the first zero of $J_0$, we additionally retain the $J_4(\phi_{\rm ac})$ term of the Jacobi--Anger expansion.
Although omitted from the leading-order derivation for clarity, in the frame rotating with the $J_2$ resonance this term contributes an additional component at $2\wdrv$.
The corresponding corrected modulation index is
\begin{equation}
  \beta_1^{\rm corr}
  =
  \frac{\omega_{\mathrm q}^{(0)}}{4\wdrv}
  \frac{
  J_0(\phi_{\rm ac})+J_4(\phi_{\rm ac})
  }{
  \sqrt{J_2(\phi_{\rm ac})}
  },
  \label{eq:beta1_corr}
\end{equation}
while the second modulation index remains unchanged.
For the parameters used here, numerical evaluation of the full Bessel sums shows that the $\ell=1$ process is dominated by the $(m_1,m_2)=(1,0)$ contribution to Eq.~\eqref{eq:Bcomb}.
On resonance,
$2\wdrv=\omega_\text{a}-\bar\omega_{\mathrm q}^{\rm o}$,
the exchange oscillation rate is therefore
\begin{equation}
\begin{aligned}
  \Omega_{\rm ex}^{{\rm o},1}
  =
  2g_\text{a}^{(0)}
  [J_2(\phi_{\rm ac})]^{1/4}
  \left|
  J_1(\beta_1^{\rm corr})
  J_0(\beta_2)
  \right|.
\end{aligned}
\label{eq:new_1}
\end{equation}
Similarly, the $\ell=2$ process is dominated by the $(m_1,m_2)=(0,1)$ contribution.
For
$4\wdrv=\omega_\text{a}-\bar\omega_{\mathrm q}^{\rm o}$,
we obtain
\begin{equation}
\begin{aligned}
  \Omega_{\rm ex}^{{\rm o},2}
  =
  2g_\text{a}^{(0)}
  [J_2(\phi_{\rm ac})]^{1/4}
  \left|
  J_0(\beta_1^{\rm corr})
  J_1(\beta_2)
  \right|.
\end{aligned}
\label{eq:new_2}
\end{equation}

Figure~\ref{fig:parametric} compares these rates with the analytical predictions.
At small drive amplitude, the exchange process associated with the central bound-state region increases rapidly with $\phi_{\rm ac}$.
The red circles agree well with Eq.~\eqref{eq:g_std}.
We omit the narrow amplitude interval around the first zero of $J_0$, where the central harmonic confinement collapses and the local approximation used to derive Eq.~\eqref{eq:g_std} is not valid.
Beyond the zero, the central bound-state region reappears around $\theta=\pi$, and the continuation based on $|J_0|$ again agrees with the coherent simulation.
Once localized Floquet states associated with the outer resonant tori are available, the $\ell=1$ and $\ell=2$ parametric processes can also be activated.
The $\ell=1$ exchange rate, shown by blue squares, decreases toward the vicinity of the first zero of $J_0$ and then increases on the other side, exhibiting the opposite overall drive-amplitude dependence from the central-region process over this range.
Equation~\eqref{eq:new_1} agrees quantitatively with the numerical data.
The $\ell=2$ process, shown by green triangles, is slower and is also well reproduced by Eq.~\eqref{eq:new_2}.

The analytical expressions also show that increasing drive frequency reduces the parametric rate through the modulation indices, which scale inversely with $\omega_{\rm d}$.
The red dashed curve in Fig.~\ref{fig:parametric} shows the central-region prediction for $\omega_{\rm d}/2\pi=30$~GHz.
Away from the immediate vicinity of the $J_0$ zero, the achievable exchange rate is reduced compared with the 15-GHz case.
Thus, increasing the drive frequency weakens the parametric sideband coupling for this process.

The analysis above assumes zero dc flux bias. At finite dc flux bias,
the odd harmonics activate $J_1$-generated resonant regions and
additional sidebands with fundamental spacing $\wdrv$. The
corresponding leading exchange rates are derived in
Appendix~\ref{app:parametric_finite_dc}.

These results show that parametric operation can persist above the unbound-resonance threshold in two distinct settings.
In the long-time dissipative steady state, population is transferred to localized Floquet states associated with the outer resonant tori, where coherent parametric exchange remains possible.
For the processes studied here, however, the achievable exchange rate in this outer-resonance sector remains below the maximum rate reached on the central bound-state side of the threshold.
The unbound-resonance threshold therefore places an effective upper bound on the parametric operation speed without abruptly suppressing coherent exchange.
This speed limit does not by itself determine the fidelity of a practical operation.
Flux noise~\cite{Didier2019AcSweetSpot,Hong2020AcSweetSpot} and other decoherence mechanisms~\cite{Mollenhauer2025Network,Copetudo2026CPHASE} can impose additional fidelity constraints at drive amplitudes below this limit.

The coherent simulations also show that a localized Floquet state in the central bound-state region can support parametric operation above the threshold whenever that state remains well defined.
However, such operation is transient in an open system since the central state is no longer the long-time dissipative steady state.
Its practical relevance therefore depends on how rapidly population transfers from the central region to the outer resonant tori.
We examine this transfer timescale, together with the relaxation back to the bound-state manifold after the drive is removed, in the next section.

\section{Transfer and relaxation dynamics across the unbound-resonance transition}
\label{sec:timescale}

\subsection{Transfer into the unbound-resonance regime}
\label{subsec:transfertime}

\begin{figure}[t]
    \centering
    \includegraphics[width=\columnwidth]{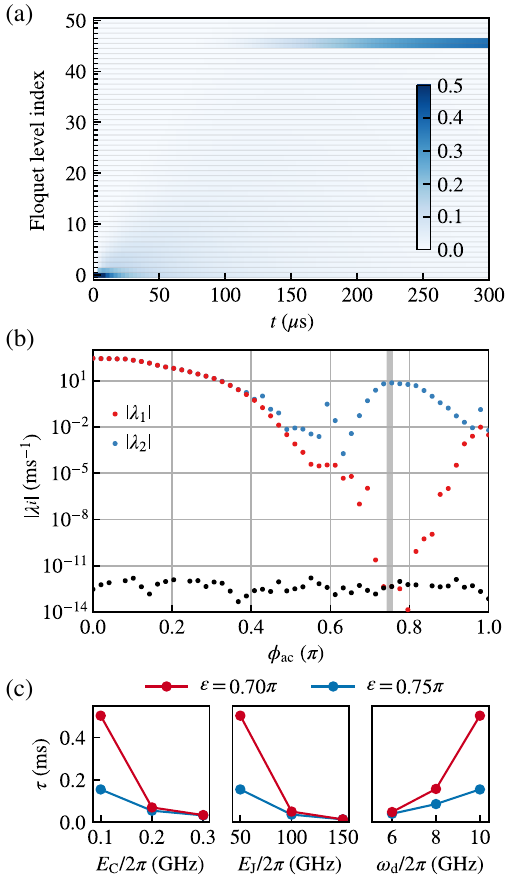}
    \caption{\textbf{Dynamics and timescale of transfer into the unbound-resonance regime.}
    (a) Floquet-mode populations after initializing the system in the lowest local Floquet state of the central bound-state region at $\phi_{\rm ac}=0.75\pi$ and $\omega_{\rm d}/2\pi=12$~GHz.
    (b) Magnitudes of the two nonzero eigenvalues of the Floquet--Markov generator closest to zero as a function of drive amplitude.
    The shaded vertical line marks $\phi_{\rm ac}=0.75\pi$ used in (a).
    (c) Transfer time $\tau$ extracted from a cascade-model fit to the population growth of the dominant outer-resonance Floquet mode as a function of $E_{\rm C}$, $E_{\rm J}$, and $\omega_{\rm d}$, for $\phi_{\rm ac}=0.70\pi$ and $0.75\pi$.
    Unless varied, the parameters in (c) are $E_{\rm C}/2\pi=0.1$~GHz, $E_{\rm J}/2\pi=50$~GHz, $n_\text{g}=0$, $\omega_{\rm d}/2\pi=10$~GHz, and $T_1=5\,\mu$s.}
    \label{fig:transfer}
\end{figure}

We first quantify the transfer from the central bound-state region to the outer resonant tori above the unbound-resonance threshold. This timescale has two practical implications. For operation using a prepared central localized state, it sets the available time before dissipative transfer to the outer resonant tori and therefore constrains the achievable fidelity and number of operations. Conversely, if operation in the unbound-resonance regime is desired, it determines the preparation time required for the outer-torus steady state to be established.

For the representative dynamics in Figs.~\ref{fig:transfer}(a) and \ref{fig:transfer}(b), we take $\omega_{\rm d}/2\pi=12$~GHz, $E_{\rm C}/2\pi=0.1$~GHz, and $E_{\rm J}/2\pi=50$~GHz. The parameter sweeps in Fig.~\ref{fig:transfer}(c) use $\omega_{\rm d}/2\pi=10$~GHz as the default drive frequency unless $\omega_{\rm d}$ itself is varied.
At $\phi_{\rm ac}=0.75\pi$, where the dissipative steady state is supported by the outer resonant tori, we identify the Floquet state localized near the bottom of each of the three effective potential wells.
We then initialize the system in the central bound-state region and evolve the system under the Floquet--Markov dynamics.
The system--bath coupling is chosen such that the undriven coupler has $T_1=5~\mu{\rm s}$.
Figure~\ref{fig:transfer}(a) shows the resulting time evolution, with the Floquet-mode indices sorted by their average energies.
The population initially occupies the low-energy Floquet mode associated with the central bound-state region.
Over a timescale of order $100~\mu{\rm s}$, population transfers to Floquet modes near index 50, which correspond to the states associated with the two outer resonant tori.
The color scale is limited to $1/2$ since the long-time population is approximately equally distributed between the two outer tori.

The characteristic timescales can also be identified directly from the eigenvalues of the Floquet--Markov generator, without explicitly propagating the population dynamics.
Since the Floquet--Markov populations obey a rate equation, each nonzero eigenvalue of the generator defines a characteristic timescale.
We therefore diagonalize the generator as a function of drive amplitude.
The generator always has a zero eigenvalue, $\lambda_0=0$, corresponding to a stationary state, while the two nonzero eigenvalues closest to zero, $\lambda_1$ and $\lambda_2$, describe the slowest population dynamics.
Their magnitudes are shown in Fig.~\ref{fig:transfer}(b), with characteristic timescales $\tau_i=1/|\lambda_i|$.
At smaller drive amplitudes, $|\lambda_1|$ and $|\lambda_2|$ are nearly degenerate and decrease together with increasing drive amplitude.
Near $\phi_{\rm ac}^*=0.59\pi$, where the central and outer confinement scales become comparable, the two eigenvalues separate, with $|\lambda_1|$ continuing to decrease while $|\lambda_2|$ turns upward.
Their separation becomes increasingly pronounced toward the first zero of $J_0$, where $|\lambda_1|$ approaches zero while $|\lambda_2|$ reaches a maximum.
The vanishing of $|\lambda_1|$ near the first zero of $J_0$ is consistent with the emergence of an additional stationary state discussed in Fig.~\ref{fig:bessel}, and is further discussed in Appendix~\ref{app:bistability}.

The physical meaning of these slow modes can be understood from the corresponding right eigenvectors $\nu_i$ of the Floquet--Markov generator.
Once the outer resonances are sufficiently developed, we coarse grain the slow dynamics into the central bound-state region, $\mathrm{c}$, and the two symmetry-related outer resonant regions, $\mathrm{t}$ and $\mathrm{b}$.
This three-region description is not intended for the weak-drive regime, where the slowest modes can instead be associated with dynamics within the central region.
We introduce effective rates $\Gamma_{\mathrm{c}\rightarrow\mathrm{o}}$ from the central region to each outer region, $\Gamma_{\mathrm{o}\rightarrow\mathrm{c}}$ for the reverse process, and $\Gamma_{\mathrm{o}\leftrightarrow\mathrm{o}}$ for redistribution between the two outer regions.
These rates describe net transfer between metastable phase-space regions and coarse grain the many Floquet-state transitions involved in the microscopic dynamics.
For the population vector $\mathbf{p}=(p_{\mathrm{c}},p_{\mathrm{t}},p_{\mathrm{b}})^T$, the corresponding effective population dynamics obey $\mathrm{d}\mathbf{p}/\mathrm{d}t=\mathcal{R}_{\mathrm{eff}}\mathbf{p}$, with
\begin{equation}
    \mathcal{R}_{\mathrm{eff}}
    =
    \begin{pmatrix}
        -2\Gamma_{\mathrm{c}\rightarrow\mathrm{o}}
        &
        \Gamma_{\mathrm{o}\rightarrow\mathrm{c}}
        &
        \Gamma_{\mathrm{o}\rightarrow\mathrm{c}}
        \\
        \Gamma_{\mathrm{c}\rightarrow\mathrm{o}}
        &
        -\Gamma_{\mathrm{o}\rightarrow\mathrm{c}}-\Gamma_{\mathrm{o}\leftrightarrow\mathrm{o}}
        &
        \Gamma_{\mathrm{o}\leftrightarrow\mathrm{o}}
        \\
        \Gamma_{\mathrm{c}\rightarrow\mathrm{o}}
        &
        \Gamma_{\mathrm{o}\leftrightarrow\mathrm{o}}
        &
        -\Gamma_{\mathrm{o}\rightarrow\mathrm{c}}-\Gamma_{\mathrm{o}\leftrightarrow\mathrm{o}}
    \end{pmatrix}.
\end{equation}
Diagonalizing this effective generator gives an outer-antisymmetric mode
\begin{equation}
    \nu_{\mathrm{a}}
    \simeq
    \begin{pmatrix}
        0 & 1/2 & -1/2
    \end{pmatrix}^{T},
    \qquad
    \gamma_{\mathrm{a}}
    =
    \Gamma_{\mathrm{o}\rightarrow\mathrm{c}}
    +
    2\Gamma_{\mathrm{o}\leftrightarrow\mathrm{o}},
\end{equation}
and a center--outer symmetric mode
\begin{equation}
    \nu_{\mathrm{s}}
    \simeq
    \begin{pmatrix}
        -1 & 1/2 & 1/2
    \end{pmatrix}^{T},
    \qquad
    \gamma_{\mathrm{s}}
    =
    2\Gamma_{\mathrm{c}\rightarrow\mathrm{o}}
    +
    \Gamma_{\mathrm{o}\rightarrow\mathrm{c}}.
\end{equation}
The eigenvector structure of the full Floquet--Markov generator allows us to associate these effective decay rates with the two slow eigenvalues in Fig.~\ref{fig:transfer}(b).
The antisymmetric mode describes relaxation of a population imbalance between the two outer regions, whereas the symmetric mode describes population transfer between the central and outer regions.
Near the first zero of $J_0$, the stationary eigenvector is concentrated predominantly in the two outer regions,
\begin{equation}
    \nu_0
    \simeq
    \begin{pmatrix}
        0 & 1/2 & 1/2
    \end{pmatrix}^{T}.
\end{equation}
At $\phi_{\rm ac}=0.75\pi$, the center--outer symmetric mode has $\gamma_{\mathrm{s}}\approx10~\mathrm{kHz}$, corresponding to a characteristic timescale of order $100~\mu{\rm s}$, consistent with the transfer observed in Fig.~\ref{fig:transfer}(a).

The drive-amplitude dependence of the effective rates explains the evolution of the two slow eigenvalues.
Below the confinement crossover, $\Gamma_{\mathrm{o}\rightarrow\mathrm{c}}$ provides the dominant contribution to both relaxation modes, while $\Gamma_{\mathrm{c}\rightarrow\mathrm{o}}$ is small.
$\Gamma_{\mathrm{o}\leftrightarrow\mathrm{o}}$ provides an additional contribution to the antisymmetric mode and becomes progressively less important as the outer resonances are separated further in phase space at higher drive frequency.
Consequently, $\gamma_{\mathrm{a}}$ and $\gamma_{\mathrm{s}}$ are nearly degenerate.
As the drive amplitude increases, the central confinement scale $2E_{\rm J}J_0(\phi_{\rm ac})$ decreases while the outer-resonance confinement scale $2E_{\rm J}J_2(\phi_{\rm ac})$ increases.
Consistent with this opposing evolution, the effective center-to-outer rate $\Gamma_{\mathrm{c}\rightarrow\mathrm{o}}$ increases, whereas the outer-to-center rate $\Gamma_{\mathrm{o}\rightarrow\mathrm{c}}$ decreases.
Although the periodically driven system is out of equilibrium and these rates are not fixed by an equilibrium detailed-balance relation, we find that they become comparable near the same drive amplitude $\phi_{\rm ac}^*$ at which the two confinement scales cross.
Beyond this point, the growing $\Gamma_{\mathrm{c}\rightarrow\mathrm{o}}$ drives $\gamma_{\mathrm{s}}$ upward, whereas the decreasing $\Gamma_{\mathrm{o}\rightarrow\mathrm{c}}$ drives $\gamma_{\mathrm{a}}$ downward.
Near the first zero of $J_0$, the central confinement vanishes and $\gamma_{\mathrm{s}}$ reaches its maximum, consistent with rapid depletion of population from the central region, while $\gamma_{\mathrm{a}}$ approaches zero, giving rise to bistability in the steady state.

Having identified the slow center--outer relaxation mode, we next vary the circuit parameters $E_{\rm C}$ and $E_{\rm J}$, as well as the drive frequency, to examine what controls the center-to-outer transfer timescale.
Figure~\ref{fig:transfer}(c) shows the transfer time extracted by fitting the growth of the outer-resonance population to a cascade model for these parameter sweeps.
We consider both $\phi_{\rm ac}=0.75\pi$ and a slightly smaller amplitude $\phi_{\rm ac}=0.70\pi$.
At $\phi_{\rm ac}=0.75\pi$, increasing $E_{\rm C}$ by a factor of three reduces the transfer time monotonically from hundreds to tens of microseconds.
The same trend is observed at the smaller amplitude $\phi_{\rm ac}=0.70\pi$.
The difference between the two drive amplitudes also decreases rapidly with increasing $E_{\rm C}$, and their transfer times approach one another at the largest $E_{\rm C}$ considered.
A similar monotonic reduction of the transfer time occurs when $E_{\rm J}$ is increased or the drive frequency is reduced.
Thus, although the high-frequency threshold itself depends weakly on these parameters, the timescale for reaching the outer-resonance steady state can vary substantially.

These trends can be understood qualitatively from a simple phenomenological phase-space picture.
Larger phase-space separations among the central and outer resonant regions are associated with smaller effective inter-region rates and hence slower population redistribution.
This picture is not a microscopic derivation of the coarse-grained rates, but provides a qualitative interpretation of the parameter dependence observed numerically.
The outer resonant tori are centered at $n_\pm=\pm\omega_{\rm d}/(4E_{\rm C})$.
Reducing $\omega_{\rm d}$ or increasing $E_{\rm C}$ therefore moves them toward the central region, consistent with the observed reduction of the transfer time.
Increasing $E_{\rm J}$ instead expands the central separatrix in the charge direction, whose characteristic extent scales as $\sqrt{E_{\rm J}/E_{\rm C}}$.
The boundary of the central region therefore moves closer to the outer resonant tori as $E_{\rm J}$ increases, again correlating with a shorter transfer time.
Increasing $E_{\rm C}$ has the opposite effect on the extent of the central region, but this dependence scales only as $E_{\rm C}^{-1/2}$, whereas the outer-torus position scales as $E_{\rm C}^{-1}$.
Over the parameter range studied, the inward motion of the outer resonant tori therefore provides the dominant trend.
Thus, the phase-space separation provides a qualitative indicator of the inter-region relaxation rates, and hence of the transfer timescale.
Having characterized the approach to the outer-resonance steady state under sustained strong drive, we next consider the reverse process of relaxation back to the undriven low-energy manifold after the drive is removed.

\subsection{Relaxation back to the bound-state regime}

\begin{figure}[t]
    \centering
    \includegraphics[width=\columnwidth]{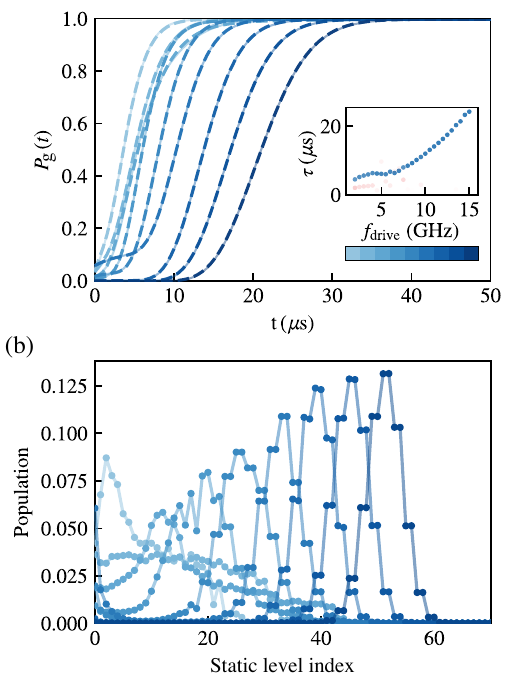}
    \caption{\textbf{Relaxation from strongly driven states to the undriven ground state.}
    (a) Recovery of the undriven ground-state population after the drive is switched off for different initial drive frequencies.
    Solid curves are simulations and dashed curves are fits to the mixed fast-plus-cascade model in Eq.~\eqref{eq:mixed_fast_gamma_fit}.
    The inset shows the extracted fast timescale $\tau_{\rm fast}$ and cascade timescale $t_{\rm cas}$, with marker transparency indicating the fitted weight of each component.
    (b) Driven steady-state populations immediately before switch-off, projected onto the eigenbasis of the undriven Hamiltonian.
    Parameters: $E_{\rm C}/2\pi=0.13$~GHz, $E_{\rm J}/2\pi=50$~GHz, $n_\text{g}=0$, $\phi_{\rm ac}=0.63\pi$, $T_1=5\,\mu$s.}
    \label{fig:relax}
\end{figure}

Returning the coupler to its undriven ground state is important before subsequent operations.
Unlike the transfer dynamics considered above, this recovery process is relevant across the full strong-drive regime, from low-frequency separatrix chaos to the high-frequency unbound-resonance regime.
We therefore examine how the recovery dynamics depend on drive frequency across the crossover between these regimes.

For drive frequencies from $2$ to $15$~GHz, we first prepare the initial state as the steady-state density matrix obtained from the Floquet--Markov equation at $\phi_{\rm ac}=0.63\pi$.
At this strong drive amplitude, the frequency sweep extends from the low-frequency chaos-dominated regime to the high-frequency unbound-resonance regime.
We then switch off the drive and evolve the system under the undriven dissipative dynamics, with the system--bath coupling chosen such that the undriven coupler has $T_1=5~\mu{\rm s}$.
The population of the undriven ground state, $P_{\rm g}(t)$, is shown in Fig.~\ref{fig:relax}(a).
For all drive frequencies, the recovery exhibits an S-shaped profile, indicating delayed population transfer into the ground state.
The recovery slows substantially as the drive frequency is increased.

To quantify this relaxation, we fit the time-dependent ground-state population using a mixed fast-plus-cascade model,
\begin{equation}
\begin{aligned}
    P_\text{g}(t)
    &=
    P_\text{g}(0)
    +
    A_{\mathrm{fast}}
    \left(1-e^{-t/\tau_{\mathrm{fast}}}\right) \\
    &\quad
    +
    \left[1-P_{\rm g}(0)-A_{\mathrm{fast}}\right]
    \frac{\gamma(k,t/\tau)}{\Gamma(k)}.
\end{aligned}
\label{eq:mixed_fast_gamma_fit}
\end{equation}
Here, $P_\text{g}(0)$ is fixed by the initial ground-state population.
The second term describes a fast relaxation channel with amplitude $A_{\mathrm{fast}}$ and characteristic time $\tau_{\mathrm{fast}}$.
The last term describes a delayed cascade contribution through the normalized lower incomplete gamma function, where $k$ is an effective cascade order and $\tau$ is an effective step time.
The fast channel is associated with direct relaxation into the ground state from initially populated states with appreciable transition matrix elements to the ground state.
The cascade channel describes population initially stored in higher-lying states that relax only indirectly through a sequence of transitions between nearby states.
Since the microscopic transition rates between eigenstates are generally nonuniform, $k$ and $\tau$ should be interpreted phenomenologically rather than as a literal number of identical decay steps.
We therefore use
$t_{\rm cas}=k\tau$
as the characteristic timescale of the cascade component.

The dashed curves in Fig.~\ref{fig:relax}(a) show the fitted results, which closely reproduce the simulated dynamics.
The inset shows the extracted fast timescale $\tau_{\rm fast}$ and cascade timescale $t_{\rm cas}$ as red and blue markers, respectively.
The marker transparency indicates the fitted weight of each component.
The fast component has a small but nonzero weight at low drive frequency and becomes negligible in the high-frequency regime.
The dominant cascade timescale $t_{\rm cas}$ remains near $5~\mu{\rm s}$, comparable to the undriven $T_1$, throughout the low-frequency regime, but increases to tens of microseconds as the drive frequency enters the high-frequency regime.

The origin of this frequency dependence is visible by expressing the driven steady state immediately before switch-off in the eigenbasis of the undriven Hamiltonian.
Figure~\ref{fig:relax}(b) shows the resulting populations of the undriven eigenstates.
At low drive frequency, the population is broadly distributed over many undriven eigenstates, consistent with the broad hybridization of the chaos-dominated regime discussed in Sec.~\ref{subsec:bound}.
In contrast, at high drive frequency, the population becomes narrowly distributed over several pairs of degenerate higher-lying states, and the center of the distribution shifts upward with increasing drive frequency.
This behavior follows naturally from the localization of the high-frequency steady state on the outer resonant tori.
As shown by Eq.~\eqref{eq:kinetic_shifted}, these tori are centered at
$n_\pm=\pm\omega_{\rm d}/(4E_{\rm C})$,
so their distance from the central region increases linearly with drive frequency.
Consequently, after the drive is removed, the population is concentrated progressively farther from the ground state in the undriven eigenbasis and is associated with a longer multistep cascade.
This provides a qualitative explanation for the increasing recovery time in Fig.~\ref{fig:relax}(a).

\section{Strong-drive thresholds under charge drive}
\label{sec:charge_drive}

We now compare the charge-driven response with the flux-driven results above.
Related studies of inductively coupled readout architectures have found suppressed multiphoton transitions and reduced chaos relative to conventional capacitive coupling~\cite{Chapple2025Longitudinal,Mori2026CosPhi}.
Here we extend this comparison across drive frequency and determine the distinct low- and high-frequency strong-drive limits of the two drive channels.
At low frequency, charge drive exhibits bound-state resonances and separatrix chaos, while at high frequency its dissipative steady state undergoes an analogous unbound-resonance transition.

\begin{figure}[t]
    \centering
    \includegraphics[width=\columnwidth]{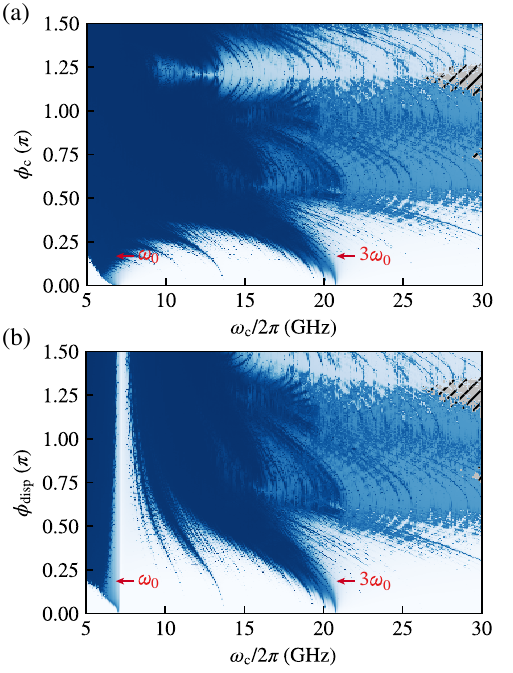}
    \caption{\textbf{Steady-state impurity under charge drive.}
    (a) Impurity as a function of charge-drive frequency $\omega_\text{c}$ and effective amplitude $\phi_\text{c}=\epsilon_\text{c}/\omega_\text{c}$.
    (b) Same results plotted using the linearized phase displacement $\phi_\text{disp}$ as the vertical coordinate.
    Parameters: $E_\text{C}/2\pi=0.13$~GHz, $E_\text{J}/2\pi=50$~GHz, $n_\text{g}=0$.}
    \label{fig:charge_impurity}
\end{figure}

Figure~\ref{fig:charge_impurity}(a) shows the steady-state impurity obtained from Floquet--Markov simulations of the charge-driven Hamiltonian in Eq.~\eqref{eq:charge_ham}.
The horizontal axis is the drive frequency $\omega_\text{c}$, and the vertical axis is
$\phi_\text{c}=\epsilon_\text{c}/\omega_\text{c}$, the argument of the Bessel functions in Eq.~\eqref{eq:charge_jacobi}.
This representation therefore allows a direct comparison with the flux-driven case using the Bessel-sector decomposition introduced in Sec.~\ref{subsec:jacobi_anger}.

As discussed in Sec.~\ref{subsec:jacobi_anger}, $\phi_\text{c}$ is not generally equal to the linearized phase displacement $\phi_\text{disp}$, although the two approach one another in the high-frequency limit.
Since parametric interaction strengths are naturally characterized in terms of this displacement, Fig.~\ref{fig:charge_impurity}(b) replots the same results using $\phi_\text{disp}$ as the vertical coordinate.
Both representations show the same qualitative structure, with narrow high-impurity features at low frequency and a broad high-impurity region at high frequency.
Their main difference occurs near the SQUID frequency $\omega_0$ ($\sim$7~GHz), where the expression for $\phi_\text{disp}$ becomes resonant.
The apparent gap in Fig.~\ref{fig:charge_impurity}(b) therefore results from the singular mapping between $\phi_\text{c}$ and $\phi_\text{disp}$ near resonance rather than from an additional physical regime.
We use $\phi_\text{c}$ below since it directly controls the Bessel amplitudes responsible for the strong-drive structure.

\subsection{Low-frequency resonance structure}

At low frequency, the narrow high-impurity stripes arise from resonances within the bound-state sector, analogous to those identified for flux drive in Sec.~\ref{subsec:bound}.
Compared with flux drive at zero dc flux bias in Fig.~\ref{fig:impurity}, charge drive exhibits prominent features near $\omega_\text{c}=\omega_0$ and $\omega_\text{c}=3\omega_0$.
The additional resonance structure follows from the simultaneous presence of the even and odd sectors in Eq.~\eqref{eq:charge_jacobi}.

For a bound trajectory approximated by
$\theta(t)\simeq\theta_0\sin[\omega_0(\theta_0)t]$, the slowly rotating contributions are
\begin{equation}
\begin{aligned}
H_{\mathrm{pert.}}^{\mathrm{charge}}
&\simeq
-2E_\text{J}
\sum_{n,m=1}^{\infty}
J_{2n}(\phi_\text{c})J_{2m}(\theta_0)
\\
&\quad\times
\cos\left[
2n\omega_\text{c}t
-
2m\omega_0(\theta_0)t
\right]
\\
&\quad+
2E_\text{J}
\sum_{n=1}^{\infty}
\sum_{m=0}^{\infty}
J_{2n-1}(\phi_\text{c})J_{2m+1}(\theta_0)
\\
&\quad\times
\cos\left[
(2n-1)\omega_\text{c}t
-
(2m+1)\omega_0(\theta_0)t
\right].
\end{aligned}
\end{equation}
The even sector therefore gives resonances satisfying
$n\omega_\text{c}=m\omega_0$, whereas the odd sector gives
$(2n-1)\omega_\text{c}=(2m+1)\omega_0$.
The resonance near $\omega_\text{c}=\omega_0$ is allowed by both sectors.
In particular, the even sector contains the contribution
$J_2(\phi_\text{c})J_2(\theta_0)$ also present for zero-bias flux drive, while the odd sector contains the lower-order contribution
$J_1(\phi_\text{c})J_1(\theta_0)$.
The odd sector additionally produces the prominent resonance near
$\omega_\text{c}=3\omega_0$ through
$J_1(\phi_\text{c})J_3(\theta_0)$.

Overall, the low-frequency charge-driven response is qualitatively consistent with the picture established for flux drive.
Here we focus on the bound-state resonance structure, which differs since the odd- and even-Bessel sectors of charge drive provide additional resonance channels.
In particular, these sectors account for the prominent features near $\omega_\text{c}=\omega_0$ and $3\omega_0$.

\subsection{Unbound-resonance transition at high frequency}

\begin{figure*}[t]
    \centering
    \includegraphics[width=\textwidth]{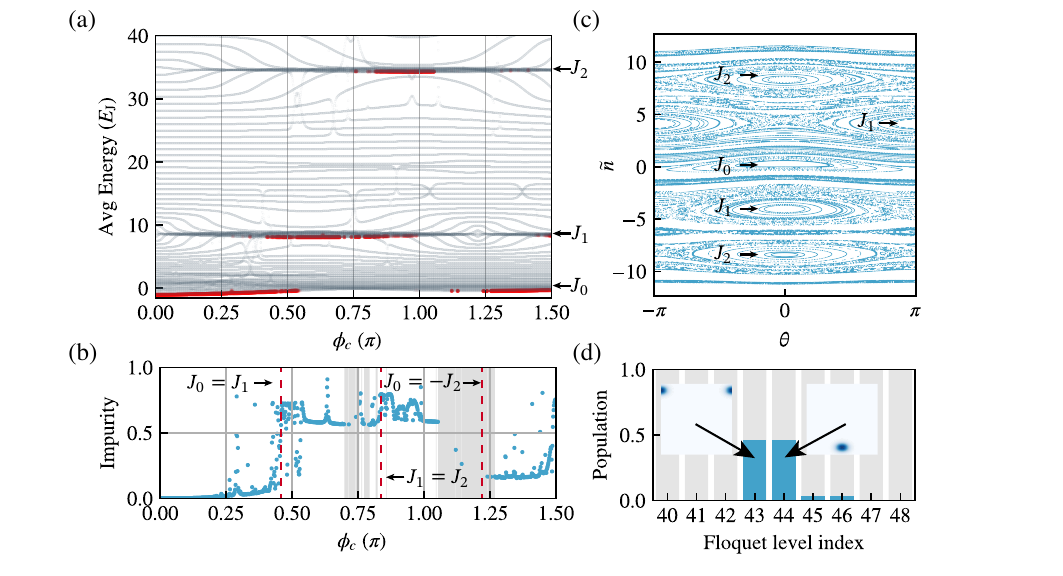}
    \caption{\textbf{High-frequency unbound resonances under charge drive.}
    (a) Average Floquet energies as a function of charge-drive amplitude $\phi_\text{c}$, with red opacity indicating the Floquet--Markov steady-state population.
    The arrows identify the branches associated with the central $J_0$ region and the $J_1$- and $J_2$-generated outer resonances.
    (b) Corresponding steady-state impurity.
    The dashed vertical lines mark the Bessel-function crossings discussed in the text, and the shaded regions indicate parameter intervals with multiple steady states.
    (c) Poincar\'e section at $\omega_\text{c}/2\pi=30$~GHz and $\phi_\text{c}=0.7\pi$, showing the central bound-state region and the $J_1$- and $J_2$-generated pairs of outer resonant tori.
    (d) Floquet--Markov steady-state populations at the parameters of (c).
    The insets show the Husimi functions of the two predominantly populated states localized on the $J_1$-generated outer resonant tori.
    Parameters: $E_\text{C}/2\pi=0.13$~GHz, $E_\text{J}/2\pi=50$~GHz, $n_\text{g}=0$.}
    \label{fig:bessel_charge}
\end{figure*}

At high frequency, the low-frequency bound-state resonances and separatrix-chaos effects are suppressed, and, as in the flux-driven case, the dissipative steady state transfers from the central bound-state region to above-barrier outer resonances.
The resulting high-impurity region begins at a smaller amplitude than for zero-bias flux drive.
In addition, charge drive produces two pairs of outer resonant tori rather than one, leading to a richer sequence of steady-state redistributions at larger amplitude.

To resolve this structure, we take $\omega_\text{c}/2\pi=30$~GHz in Fig.~\ref{fig:bessel_charge}.
As discussed in Sec.~\ref{subsec:high_freq_dc}, the $J_1$-generated resonances lie closer to the central region than the $J_2$-generated resonances and therefore require a higher drive frequency to enter the above-barrier sector.
The choice of $30$~GHz places both leading pairs of resonances in the above-barrier sector while further suppressing the low-frequency resonance and chaos mechanisms.

The origin of the five structures visible in the Poincar\'e section in Fig.~\ref{fig:bessel_charge}(c) follows directly from the odd and even sectors of the charge-driven Hamiltonian.
For an above-barrier running trajectory, the Josephson potential gives only a small correction to the motion and $\theta(t)\simeq 8E_\text{C}nt$.
The leading odd contribution in Eq.~\eqref{eq:charge_jacobi} is
\begin{equation}
    \hat{H}_{J_1}
    =
    2E_\text{J}J_1(\phi_\text{c})
    \sin(\omega_\text{c}t)\sin\hat{\theta}.
\end{equation}
Substituting the running trajectory gives
\begin{equation}
\begin{aligned}
    H_{J_1}
    \simeq
    E_\text{J}J_1(\phi_\text{c})
    \Big[
    &\cos\left[(\omega_\text{c}-8E_\text{C}n)t\right]
    \\
    -&
    \cos\left[(\omega_\text{c}+8E_\text{C}n)t\right]
    \Big].
\end{aligned}
\end{equation}
The corresponding pair of resonances is centered at $n_{\pm,1}=\pm\omega_\text{c}/(8E_\text{C})$.
The opposite signs of the two resonant terms shift the stable fixed points of the positive- and negative-$n$ resonances by $\pi$ relative to one another along the phase coordinate.
This relative shift is visible in the Poincar\'e section in Fig.~\ref{fig:bessel_charge}(c).

The leading even contribution is
\begin{equation}
    \hat{H}_{J_2}
    =
    -2E_\text{J}J_2(\phi_\text{c})
    \cos(2\omega_\text{c}t)\cos\hat{\theta}.
\end{equation}
For the same running trajectory,
\begin{equation}
\begin{aligned}
    H_{J_2}
    \simeq
    -E_\text{J}J_2(\phi_\text{c})
    \Big[
    &\cos\left[(2\omega_\text{c}-8E_\text{C}n)t\right]
    \\
    +&
    \cos\left[(2\omega_\text{c}+8E_\text{C}n)t\right]
    \Big],
\end{aligned}
\end{equation}
giving a second pair of outer resonances centered at $n_{\pm,2}=\pm\omega_\text{c}/(4E_\text{C})$.
These resonances therefore lie twice as far from the center in charge as the $J_1$-generated pair.
Together with the central bound-state region, the two pairs account for the five phase-space structures labeled in Fig.~\ref{fig:bessel_charge}(c).

Before the first zero of $J_0$, the leading local confinement scales associated with these three types of phase-space structures are
\begin{equation}
    U_0=2E_\text{J}J_0(\phi_\text{c}),\quad
    U_1=2E_\text{J}J_1(\phi_\text{c}),\quad
    U_2=2E_\text{J}J_2(\phi_\text{c}),
\end{equation}
for the central region, each $J_1$-generated outer resonance, and each $J_2$-generated outer resonance, respectively.
Charge drive therefore realizes the same competition between central and outer local confinement as flux drive, but with both outer-resonance channels intrinsically present.

At small drive amplitude, the central $J_0$ region provides the strongest local confinement and supports the dissipative steady state.
As the drive amplitude increases, the $J_1$ confinement grows more rapidly than the $J_2$ confinement and first becomes equal to the central confinement at the $J_0=J_1$ crossover marked in Fig.~\ref{fig:bessel_charge}(b),
\begin{equation}
    J_0(\phi_\text{c}^*)
    =
    J_1(\phi_\text{c}^*),
    \qquad
    \phi_\text{c}^*=0.46\pi.
    \label{eq:unbound_threshold_charge}
\end{equation}
The Floquet--Markov steady-state redistribution coincides with this local-confinement crossover, identifying Eq.~\eqref{eq:unbound_threshold_charge} as the analytical criterion for the charge-drive unbound-resonance threshold.
Near the transition, the steady state is approximately distributed among the central region and the two $J_1$-generated outer resonances, giving an impurity approaching $2/3$, consistent with an approximately equal three-state mixture.
For zero-bias flux drive, the corresponding criterion is instead $J_0(\phi_\text{ac}^*)=J_2(\phi_\text{ac}^*)$, with $\phi_\text{ac}^*=0.59\pi$.
The additional $J_1$ channel under charge drive therefore shifts the unbound-resonance threshold to a smaller drive amplitude.

Above the threshold, the steady state becomes dominated by the $J_1$-generated pair of outer resonant tori, and the impurity remains close to $1/2$.
Figure~\ref{fig:bessel_charge}(d) shows the steady-state populations at $\phi_\text{c}=0.7\pi$.
Two Floquet states each carry population close to $1/2$.
Their Husimi functions are localized at the positions
$n_{\pm,1}=\pm\omega_\text{c}/(8E_\text{C})$ and exhibit the relative $\pi$ phase shift expected from the $J_1$ resonance, confirming their identification with the $J_1$-generated pair in Fig.~\ref{fig:bessel_charge}(c).

Increasing the drive amplitude further brings the second pair of outer resonances into competition with the first.
Near $\phi_\text{c}=0.84\pi$, the $J_1=J_2$ crossing marked in Fig.~\ref{fig:bessel_charge}(b) makes the two pairs of outer resonances comparable in local confinement strength.
The steady-state population then spreads over the four outer-resonance states, and the impurity rises toward $3/4$, consistent with an approximately equal four-state mixture.

Beyond the first zero of $J_0$, the central confinement revives with opposite sign and is centered around $\theta=\pi$, as in the flux-driven case.
At still larger amplitude, the magnitude of this inverted central confinement grows and the steady state eventually returns predominantly to the central region.
The condition $J_0=-J_2$, marked in Fig.~\ref{fig:bessel_charge}(b), provides a useful reference for where the magnitude of the inverted central local confinement becomes comparable to that of the $J_2$-generated outer resonances.
In this amplitude range, however, the Floquet--Markov dynamics supports multiple steady states, so this crossing does not define a unique sharp transfer threshold.
After the return to the central region, the impurity in Fig.~\ref{fig:bessel_charge}(b) remains finite, at approximately $0.2$.
As in the flux-driven case, this residual impurity arises from excitation within the central well rather than from occupation of the outer resonances.
The return to the central region is also visible directly in the populated Floquet branches in Fig.~\ref{fig:bessel_charge}(a).

The charge-driven system therefore exhibits the same two-regime strong-drive picture as flux drive.
At low frequency, bound-state resonances and separatrix chaos remain present, with a modified resonance structure arising from the simultaneous odd- and even-Bessel sectors.
At high frequency, the first broad instability is again an unbound-resonance transition from the central bound-state region to above-barrier outer resonances, but with the threshold controlled by $J_0=J_1$ rather than $J_0=J_2$. The $J_2$-generated resonances instead modify the post-threshold redistribution at larger drive amplitude.

\section{Strong-drive thresholds under multitone flux drive}\label{sec:multitone}

Having shown that the strong-drive picture extends from flux to charge drive, we now extend the analysis to multitone driving for parametric operations.
A beamsplitter interaction between two cavities can be generated by bichromatic coupler drives whose frequency difference matches the cavity detuning~\cite{Lu2023HighFidelityParametricBeamsplitting}.
Experiments with simultaneous parametric operations have reported additional decoherence at strong pump amplitudes~\cite{Zhou2023Router}.
A bichromatic flux drive is the simplest such setting and introduces mixing channels that depend on both drive amplitudes.
We therefore use it to examine how single-tone strong-drive thresholds are modified by multiple temporal channels.

\subsection{Resonance structure under multitone drive}

For a symmetric SQUID subject to two flux-drive tones, the Hamiltonian is
\begin{equation}
\begin{aligned}
\hat{H}_{\mathrm{flux}}(t)
=&\,4E_\text{C}\hat{n}^2
-E_\text{J}\cos\hat{\theta}\,
\cos\bigl[
\phi_\text{dc}
+\phi_1\sin(\omega_1t+\varphi_1)
\\
&\qquad\qquad\qquad
+\phi_2\sin(\omega_2t+\varphi_2)
\bigr],
\end{aligned}
\end{equation}
where $\phi_{1,2}$ and $\omega_{1,2}$ denote the amplitudes and frequencies of the two ac drives, and $\varphi_{1,2}$ their phases.
For compactness, in this section we write $\phi_{1,2}\equiv\phi_{\text{ac},1,2}$.

Applying the Jacobi--Anger expansion to both tones gives
\begin{equation}
\begin{aligned}
\hat{H}_{\mathrm{flux}}(t)
&=
4E_\text{C}\hat{n}^2
-E_\text{J}\cos\hat{\theta}
\sum_{p,q=-\infty}^{\infty}
J_p(\phi_1)J_q(\phi_2)
\\
&\quad\times
\cos\!\left[
(p\omega_1+q\omega_2)t
+p\varphi_1+q\varphi_2
+\phi_\text{dc}
\right].
\end{aligned}
\end{equation}
Folding the indices to $p,q\geq0$ and sorting by their parities,
\begin{equation}\label{eq:jacobi_full}
\begin{aligned}
\hat{H}_{\mathrm{flux}}(t)
=&\,4E_\text{C}\hat{n}^2
\\
&-\frac{E_\text{J}}{2}\cos\hat{\theta}\cos(\phi_\text{dc})
\sum_{\substack{p,q\geq0\\p,q\ \mathrm{even}}}
\mathcal N_p\mathcal N_qJ_p(\phi_1)J_q(\phi_2)
\\
&\qquad\times
\bigl[
\cos(\Omega_+t+\Phi_+)
+\cos(\Omega_-t+\Phi_-)
\bigr]
\\
&-\frac{E_\text{J}}{2}\cos\hat{\theta}\cos(\phi_\text{dc})
\sum_{\substack{p,q\geq1\\p,q\ \mathrm{odd}}}
\mathcal N_p\mathcal N_qJ_p(\phi_1)J_q(\phi_2)
\\
&\qquad\times
\bigl[
\cos(\Omega_+t+\Phi_+)
-\cos(\Omega_-t+\Phi_-)
\bigr]
\\
&+\frac{E_\text{J}}{2}\cos\hat{\theta}\sin(\phi_\text{dc})
\sum_{\substack{p\geq0\ \mathrm{even}\\q\geq1\ \mathrm{odd}}}
\mathcal N_p\mathcal N_qJ_p(\phi_1)J_q(\phi_2)
\\
&\qquad\times
\bigl[
\sin(\Omega_+t+\Phi_+)
-\sin(\Omega_-t+\Phi_-)
\bigr]
\\
&+\frac{E_\text{J}}{2}\cos\hat{\theta}\sin(\phi_\text{dc})
\sum_{\substack{p\geq1\ \mathrm{odd}\\q\geq0\ \mathrm{even}}}
\mathcal N_p\mathcal N_qJ_p(\phi_1)J_q(\phi_2)
\\
&\qquad\times
\bigl[
\sin(\Omega_+t+\Phi_+)
+\sin(\Omega_-t+\Phi_-)
\bigr],
\end{aligned}
\end{equation}
where $\Omega_\pm=p\omega_1\pm q\omega_2$, $\Phi_\pm=p\varphi_1\pm q\varphi_2$, and $\mathcal N_k=2-\delta_{k0}$ is the Neumann factor.
Equation~\eqref{eq:jacobi_full} shows that, unlike monochromatic drive, the temporal components occur at the combination frequencies $|p\omega_1\pm q\omega_2|$ rather than at harmonics of a single drive frequency.
These components can generate resonances with either bound motion or above-barrier running trajectories.

\paragraph{Bound-state resonances.}

For compactness, an individual temporal component in Eq.~\eqref{eq:jacobi_full} can be written as
\begin{equation}
\hat{H}_{pq,\pm}^{\mathrm{pert}}
=
A_{pq,\pm}
\cos(\Omega_\pm t+\Psi_{pq,\pm})\cos\hat{\theta},
\end{equation}
where $A_{pq,\pm}$ includes the corresponding Bessel amplitudes, dc-flux-bias dependence, and overall sign.
For cosine components $\Psi_{pq,\pm}=\Phi_\pm$, while a sine component can be written in the same form with $\Psi_{pq,\pm}=\Phi_\pm-\pi/2$.

For a bound trajectory approximated by
$\theta(t)\simeq\theta_0\sin[\omega_0(\theta_0)t]$, the slowly rotating contribution is
\begin{equation}
\begin{aligned}
H_{pq,\pm}^{\mathrm{pert}}
\simeq
A_{pq,\pm}
\sum_{m=1}^{\infty}
J_{2m}(\theta_0)
\cos\bigl[
&(\Omega_\pm-2m\omega_0(\theta_0))t
\\
&+\Psi_{pq,\pm}
\bigr].
\end{aligned}
\end{equation}
A bound-state resonance therefore occurs when
\begin{equation}
2m\omega_0(\theta_0)
=
|p\omega_1\pm q\omega_2|,
\end{equation}
with strength proportional to
$|A_{pq,\pm}J_{2m}(\theta_0)|$.
Thus, the bound-state resonance mechanism extends directly to multitone drive through the replacement of the single-tone harmonics by combination frequencies.
When one of these frequencies is sufficiently low, bound-state resonances analogous to those discussed in Sec.~\ref{subsec:bound} become relevant.

\paragraph{Unbound-state resonances.}

The same temporal components can resonate with above-barrier running trajectories, $\theta(t)\simeq 8E_\text{C}nt+\theta_{\rm i}$.
Substituting this motion into a nonzero temporal component gives
\begin{equation}
\begin{aligned}
H_{pq,\pm}^{\mathrm{pert}}
\simeq
\frac{A_{pq,\pm}}{2}
\Big\{
&\cos\bigl[
(\Omega_\pm-8E_\text{C}n)t
+\Psi_{pq,\pm}-\theta_{\rm i}
\bigr]
\\
+&
\cos\bigl[
(\Omega_\pm+8E_\text{C}n)t
+\Psi_{pq,\pm}+\theta_{\rm i}
\bigr]
\Big\}.
\end{aligned}
\end{equation}
Each distinct temporal component therefore generates a pair of running-state resonances centered at
\begin{equation}
n_\pm=\pm\frac{\Omega_\pm}{8E_\text{C}}.
\end{equation}
Their locations along the phase coordinate are set by $\Psi_{pq,\pm}$ and by the sign of $A_{pq,\pm}$.
For a cosine component, the two phase centers are $\theta_\pm=\pm\Phi_\pm$, up to a common $\pi$ shift when the sign of the effective potential is reversed.
For a sine component, the corresponding centers are shifted by $\pi/2$.
The confinement depth of each resonant pair is $|A_{pq,\pm}|$.
These relations provide a compact description of the outer resonant tori generated by all four parity sectors in Eq.~\eqref{eq:jacobi_full}.

\subsection{Bichromatic strong-drive threshold}

\begin{figure*}[t]
    \centering
    \includegraphics[width=\textwidth]{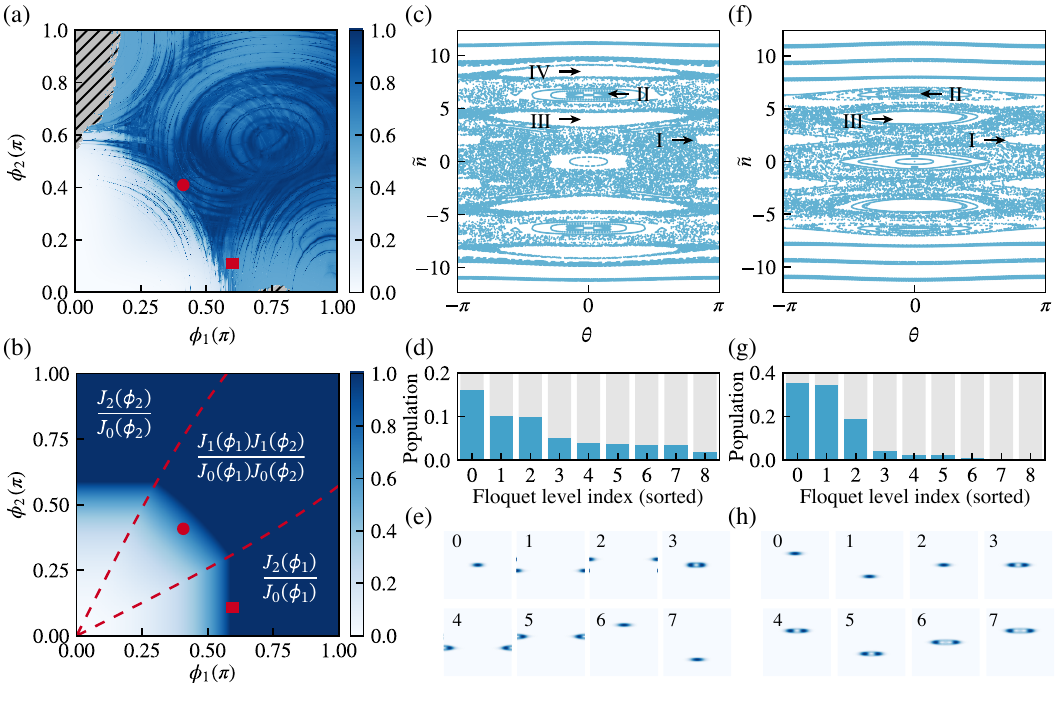}
    \caption{\textbf{Strong-drive threshold under bichromatic flux drive.}
    (a) Floquet--Markov steady-state impurity as a function of the two drive amplitudes for $\omega_1/2\pi=15$~GHz and $\omega_2/2\pi=30$~GHz at zero dc flux bias.
    The hatched region denotes parameter values for which the Floquet--Markov generator has multiple stationary states.
    (b) Dominant outer-resonance channel predicted from the leading local confinement scales.
The color indicates the ratio of the largest outer-resonance local confinement scale to that of the central bound-state region, and the dashed lines separate regions in which different outer-resonance channels dominate.
    (c) and (f) Poincar\'e sections at the circle and square markers in (a) and (b), respectively.
    The leading outer resonances are labeled as channels I--IV, corresponding to $\omega_1-\omega_2$, $\omega_1+\omega_2$, $2\omega_1$, and $2\omega_2$, respectively.
    (d) and (g) Corresponding Floquet--Markov steady-state populations, with Floquet states sorted by population for display.
    (e) and (h) Husimi functions of the populated Floquet states.
    Parameters: $E_\text{C}/2\pi=0.13$~GHz, $E_\text{J}/2\pi=50$~GHz, $n_\text{g}=0$.}
    \label{fig:multitone}
\end{figure*}

To illustrate how the competing resonance channels determine the strong-drive threshold, we choose
$\omega_1/2\pi=15$~GHz and $\omega_2/2\pi=30$~GHz, with $\phi_\text{dc}=0$ and $\varphi_{1,2}=0$.
The two frequencies are commensurate, $\omega_2=2\omega_1$, so the Hamiltonian remains periodic with common period $2\pi/\omega_1$ and can be treated using the same Floquet--Markov formalism as the single-tone case.

The steady-state impurity as a function of $\phi_1$ and $\phi_2$ is shown in Fig.~\ref{fig:multitone}(a).
The impurity remains low when both amplitudes are small and increases across a well-defined boundary in the $(\phi_1,\phi_2)$ plane, revealing a tradeoff between the two drive amplitudes.
As shown below, different portions of this boundary are associated with different dominant outer-resonance channels.
For the selected frequencies and amplitude range, the onset of the broad high-impurity region is governed by transfer from the central bound-state region to unbound resonances.

As already encountered for finite dc flux bias and charge drive, several outer resonances can coexist.
Multitone driving further enriches this structure since the combination frequencies $|p\omega_1\pm q\omega_2|$ generate additional resonance channels whose relative importance depends on both drive amplitudes.
Keeping the leading Bessel contributions over the range shown in Fig.~\ref{fig:multitone}, four outer-resonance channels are most relevant.

Channel I corresponds to the difference-frequency component
$\Omega_-=\omega_1-\omega_2$ of the $p=q=1$ odd-odd sector.
It produces a pair of outer resonances centered at
\begin{equation}
    \theta_\pm=\pi,
    \qquad
    n_\pm=\pm\frac{|\omega_1-\omega_2|}{8E_\text{C}},
\end{equation}
with leading local confinement scale
\begin{equation}
    U_\text{I}
    =
    2E_\text{J}
    J_1(\phi_1)J_1(\phi_2).
\end{equation}
Channel II corresponds to the sum-frequency component
$\Omega_+=\omega_1+\omega_2$ of the same $p=q=1$ sector.
It produces a pair centered at
\begin{equation}
    \theta_\pm=0,
    \qquad
    n_\pm=\pm\frac{\omega_1+\omega_2}{8E_\text{C}},
\end{equation}
with the same leading local confinement scale,
\begin{equation}
    U_\text{II}
    =
    2E_\text{J}
    J_1(\phi_1)J_1(\phi_2).
\end{equation}

Channels III and IV arise from the single-tone second-harmonic components $2\omega_1$ and $2\omega_2$, corresponding to $(p,q)=(2,0)$ and $(0,2)$, respectively.
Their leading local confinement scales are
\begin{equation}
\begin{aligned}
    U_\text{III}
    &=
    2E_\text{J}
    J_2(\phi_1)J_0(\phi_2),
    \\
    U_\text{IV}
    &=
    2E_\text{J}
    J_0(\phi_1)J_2(\phi_2),
\end{aligned}
\end{equation}
and the corresponding pairs are centered at $\theta_\pm=0$ and
\begin{equation}
    n_\pm=\pm\frac{\omega_1}{4E_\text{C}},
    \qquad
    n_\pm=\pm\frac{\omega_2}{4E_\text{C}},
\end{equation}
respectively.
The central bound-state region has leading local confinement scale
\begin{equation}
    U_0
    =
    2E_\text{J}
    J_0(\phi_1)J_0(\phi_2).
\end{equation}

These quantities characterize the leading local barriers of the corresponding resonant phase-space structures, each expressed in the rotating frame in which that structure is stationary.
At larger drive amplitudes, several outer resonances broaden simultaneously and can overlap, producing chaotic regions through the standard resonance-overlap mechanism~\cite{Chirikov1979}.
Here we focus on the onset of the high-impurity region, where the dominant outer resonances remain isolated and the comparison of local confinement scales provides a useful analytical description of the steady-state transfer.

Figure~\ref{fig:multitone}(b) identifies the dominant leading outer-resonance channel and shows the ratio
\begin{equation}
    R(\phi_1,\phi_2)
    =
    \frac{
    \max(U_{\rm I,II},U_{\rm III},U_{\rm IV})
    }{
    U_0
    }.
\end{equation}
Channels I and II have the same leading local confinement scale and therefore enter this comparison together.

\paragraph{Comparable drive amplitudes.}

Near the diagonal $\phi_1\approx\phi_2$, between the two dashed lines in Fig.~\ref{fig:multitone}(b), channels I and II dominate over channels III and IV.
The leading local-confinement crossover then occurs when
\begin{equation}\label{eq:multitone_threshold_mixed}
    J_1(\phi_1^*)J_1(\phi_2^*)
    =
    J_0(\phi_1^*)J_0(\phi_2^*),
\end{equation}
which matches the onset of the high-impurity region in Fig.~\ref{fig:multitone}(a).
To illustrate the case where the two drive amplitudes are comparable, we consider a representative point near the predicted transition at $(\phi_1,\phi_2)=(0.41\pi,0.41\pi)$, marked by the red circle in Figs.~\ref{fig:multitone}(a) and (b).
The corresponding Poincar\'e section is shown in Fig.~\ref{fig:multitone}(c), where the leading outer resonances are labeled as channels I--IV, corresponding to $\omega_1-\omega_2$, $\omega_1+\omega_2$, $2\omega_1$, and $2\omega_2$, respectively.
The steady-state populations and corresponding Husimi functions are shown in Figs.~\ref{fig:multitone}(d) and \ref{fig:multitone}(e).
The Floquet states are sorted by steady-state population for display.
In this transition region, appreciable population remains in states localized in the central region, labeled 0 and 3, while substantial population has already transferred to channel I, labeled 1, 2, 4, and 5, and channel II, labeled 6 and 7.
Population associated with channels III and IV remains negligible.
Over the transition region studied here, the contour in Eq.~\eqref{eq:multitone_threshold_mixed} gives an approximately linear tradeoff between the two drive amplitudes.
Increasing one tone therefore reduces the amplitude of the other that can be applied before reaching the unbound-resonance threshold.

\paragraph{Asymmetric drive amplitudes.}

When one drive amplitude is much larger than the other, the dominant channel approaches the corresponding monochromatic limit.
For $\phi_1\gg\phi_2$, channel III dominates and the local-confinement criterion becomes
\begin{equation}
    J_2(\phi_1^*)=J_0(\phi_1^*),
\end{equation}
recovering the zero-dc-bias single-tone threshold.
The analogous relation holds with $1\leftrightarrow2$ when the second tone dominates.
For example, the Poincar\'e section at
$(\phi_1,\phi_2)=(0.6\pi,0.1\pi)$, marked by the red square in Figs.~\ref{fig:multitone}(a) and (b), is shown in Fig.~\ref{fig:multitone}(f).
Since $\phi_2$ is small, channel IV is strongly suppressed and is no longer visible, while channel III dominates the outer-resonance structure.
The corresponding steady-state populations and Husimi functions are shown in Figs.~\ref{fig:multitone}(g) and \ref{fig:multitone}(h).
In this case, the population is concentrated primarily in the central region and channel III.

The bichromatic example therefore extends the single-tone threshold criterion to two drive amplitudes.
When one tone dominates, the threshold approaches the corresponding monochromatic condition, $J_2(\phi_i^*)=J_0(\phi_i^*)$.
When the two amplitudes are comparable, the mixed $J_1(\phi_1)J_1(\phi_2)$ channel becomes dominant, and the transition approximately follows Eq.~\eqref{eq:multitone_threshold_mixed}.
Over the transition region studied here, this produces an approximately linear tradeoff between the two drive amplitudes.
The two amplitudes therefore cannot be chosen independently, since increasing one reduces the available range of the other before the steady state transfers from the central bound-state region to the outer resonances.

\section{Strong-drive thresholds and junction critical current}
\label{sec:current}

\begin{figure}[t]
    \centering
    \includegraphics[width=\columnwidth]{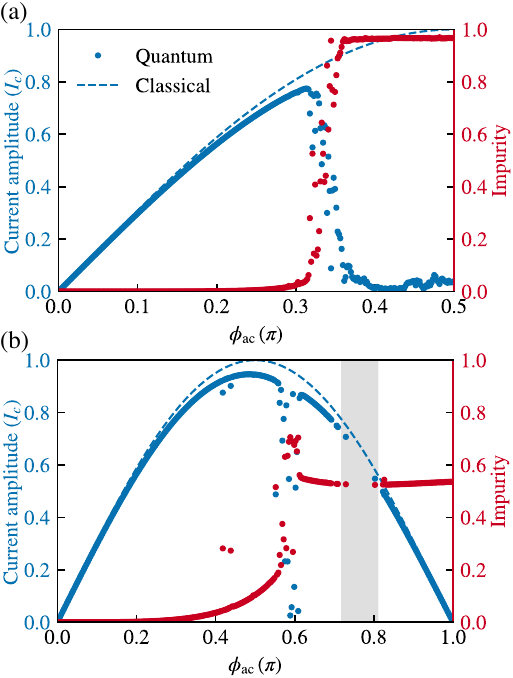}
    \caption{\textbf{Steady-state junction current across the strong-drive thresholds.}
    (a) Steady-state junction current at the maximum applied flux (blue, left axis) and impurity (red, right axis) as functions of the flux-drive amplitude for $\omega_\text{d}/2\pi=5$~GHz.
    The blue points are obtained from the Floquet--Markov steady state, while the dashed line shows the classical result for the central trajectory.
    (b) Corresponding results for $\omega_\text{d}/2\pi=15$~GHz.
    The shaded region indicates the parameter interval with multiple Floquet--Markov steady states.
    Parameters: $E_\text{C}/2\pi=0.13$~GHz, $E_\text{J}/2\pi=50$~GHz, $n_\text{g}=0$.}
    \label{fig:current}
\end{figure}

A familiar circuit-level estimate of the strong-drive range is to monitor the phase excursion or branch current of each Josephson junction~\cite{Frattini2018SNAIL}.
Around a stable operating minimum, small phase excursions correspond to junction currents well below $I_\text{c}$.
This defines the regime in which low-order expansions of the Josephson potential are controlled~\cite{Eichler2014DynamicRange,Kochetov2015HigherOrder}.
For a junction phase $\delta$, the current-phase relation $I(\delta)=I_\text{c}\sin\delta$ gives the differential Josephson inductance
\begin{equation}
    L_\text{J}(\delta)
    =
    \frac{\Phi_0}{2\pi}
    \left(
    \frac{\partial I(\delta)}{\partial\delta}
    \right)^{-1}
    =
    \frac{\Phi_0}{2\pi I_\text{c}\cos\delta},
\end{equation}
which diverges at $\delta=\pi/2$, effectively opening the junction in the small-signal sense.
We compare this heuristic with the averaged junction current obtained from Floquet--Markov steady-state calculations for a flux-driven SQUID at zero dc flux bias.
The low- and high-frequency regimes provide qualitatively different tests of this current-based picture.

\subsection{Low-frequency junction-current response}

The current through an individual junction of the SQUID is described by
\begin{equation}
    \frac{\hat{I}(t)}{I_\text{c}}
    =
    \sin\left[
    \hat{\theta}
    +\phi_\text{dc}
    +\phi_\text{ac}\sin(\omega_\text{d}t)
    \right],
\end{equation}
where
$\hat{\delta}=\hat{\theta}+\phi_\text{dc}+\phi_\text{ac}\sin(\omega_\text{d}t)$
is the phase across that junction.
We calculate the steady-state expectation value as
$\langle\hat{I}(t)\rangle
=\Tr[\rho_\text{s}(t)\hat{I}(t)]$.
We evaluate the current at $t=T/4$, when the flux modulation reaches its maximum.
Maximizing the current over the full drive period gives the same qualitative threshold behavior, with quantitative differences only at larger drive amplitudes.

The resulting current for $\omega_\text{d}/2\pi=5$~GHz is shown in Fig.~\ref{fig:current}(a).
The dashed line shows the corresponding classical result for the central trajectory initialized at $(\theta,n)=(0,0)$.
This trajectory remains at $\theta=0$, giving
$I(T/4)/I_\text{c}=\sin\phi_\text{ac}$.
At weak drive, the quantum steady-state current closely follows this sinusoidal dependence, with a small reduction in magnitude.
This behavior can be understood directly from the symmetry of the steady state.
At zero dc flux bias, reflection symmetry gives
$\langle\sin\hat{\theta}\rangle=0$, so
\begin{equation}
    \frac{\langle\hat{I}(t)\rangle}{I_\text{c}}
    =
    \langle\cos\hat{\theta}\rangle
    \sin\left[
    \phi_\text{ac}\sin(\omega_\text{d}t)
    \right].
\end{equation}
At weak drive, the steady state remains localized near the center of the Josephson well, and $\langle\cos\hat{\theta}\rangle$ remains close to unity, with the small reduction arising from the finite phase extent of the localized state.
The junction current is therefore dominated by the explicit flux modulation in the junction phase rather than by the internal phase response of the junction.
At $t=T/4$, this gives an approximately $\sin\phi_\text{ac}$ dependence, consistent with the classical reference.

As the drive enters the separatrix-chaos regime discussed in previous sections, the steady state becomes strongly mixed and populates many higher-lying states.
These higher-lying states extend over a broader range of phase and therefore reduce $\langle\cos\hat{\theta}\rangle$.
As a result, the steady-state junction current is strongly suppressed.
This produces the sharp drop in Fig.~\ref{fig:current}(a), even though the current remains well below $I_\text{c}$.
The classical central trajectory continues to follow the sinusoidal reference and therefore no longer represents the dissipative steady state.

Thus, at low frequency the strong-drive instability appears while the steady-state junction current is still substantially below $I_\text{c}$.
In this regime, the critical-current heuristic does not reproduce the separatrix-chaos threshold.

\subsection{High-frequency junction-current response}

We next consider the current response at $\omega_\text{d}/2\pi=15$~GHz, shown in Fig.~\ref{fig:current}(b).
Below the unbound-resonance threshold, the steady state remains localized in the central bound-state region.
As in the low-frequency case, the current approximately follows $\sin\phi_\text{ac}$ and approaches its maximum near $\phi_\text{ac}=0.5\pi$, where the quantum result is close to $I_\text{c}$.
The unbound-resonance threshold occurs at the larger amplitude $\phi_\text{ac}^*=0.59\pi$.

Above the threshold, the dissipative steady state is dominated by the pair of outer resonant tori, whose running trajectories are approximately
$\theta_\pm(t)\simeq\pm2\omega_\text{d}t$.
Approximating the symmetry-related outer states as equally populated gives
\begin{equation}
\begin{aligned}
    \frac{\langle\hat{I}(t)\rangle}{I_\text{c}}
    &=
    \left\langle
    \sin\left[
    \hat{\theta}
    +\phi_\text{ac}\sin(\omega_\text{d}t)
    \right]
    \right\rangle
    \\
    &=
    \sum_{\pm}
    \frac{1}{2}
    \langle\sin\theta\rangle_\pm
    \cos\left[
    \phi_\text{ac}\sin(\omega_\text{d}t)
    \right]
    \\
    &\quad+
    \sum_{\pm}
    \frac{1}{2}
    \langle\cos\theta\rangle_\pm
    \sin\left[
    \phi_\text{ac}\sin(\omega_\text{d}t)
    \right]
    \\
    &=
    \cos(2\omega_\text{d}t)
    \sin\left[
    \phi_\text{ac}\sin(\omega_\text{d}t)
    \right],
\end{aligned}
\end{equation}
where the subscript $\pm$ denotes the expectation value taken with respect to the state localized on the corresponding outer resonant torus.
At $t=T/4$, the magnitude of this expression is again $\sin\phi_\text{ac}$.
Thus, although the steady state has transferred from the central region to the outer resonances, the averaged current recovers approximately the same dependence on drive amplitude.
Near the transition, the steady state contains appreciable population in both the central and outer regions.
At $t=T/4$, their current contributions have opposite signs, producing the pronounced dip in the averaged current around the transition in Fig.~\ref{fig:current}(b).

At high frequency, the averaged junction current approaches $I_\text{c}$ near the unbound-resonance transition, making the critical-current heuristic appear qualitatively consistent with the observed strong-drive scale.
The two conditions are nevertheless different.
The current reaches its maximum near $\phi_\text{ac}=\pi/2$, whereas the unbound-resonance threshold is set by $J_0(\phi_\text{ac}^*)=J_2(\phi_\text{ac}^*)$ at $\phi_\text{ac}^*=0.59\pi$.
Together with the low-frequency result, this comparison shows that proximity to the junction critical current can provide an intuitive scale at high frequency but does not constitute a general criterion for the strong-drive threshold.

\section{Strong-drive response under inductive confinement}
\label{sec:linc}

\begin{figure*}[t]
    \centering
    \includegraphics[width=\textwidth]{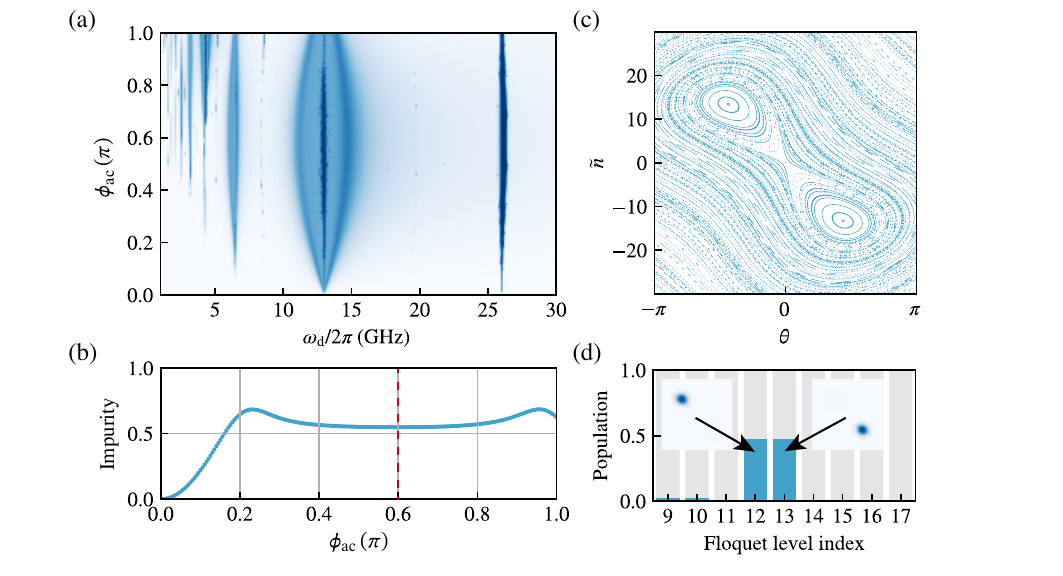}
    \caption{\textbf{Strong-drive response of a LINC.}
    (a) Floquet--Markov steady-state impurity as a function of flux-drive amplitude and frequency.
    (b) Line cut of (a) at $\omega_\text{d}/2\pi=14$~GHz.
    (c) Poincar\'e section for $\omega_\text{d}/2\pi=14$~GHz and $\phi_\text{ac}=0.6\pi$, marked by the red dashed line in (b).
    (d) Corresponding steady-state Floquet populations for the drive parameters in (c).
    The insets show the Husimi functions of the two dominantly populated Floquet states, which are localized on the two resonance tori shown in (c).
    Parameters: $E_\text{L}/2\pi=52.8$~GHz, $E_\text{J}/2\pi=15.84$~GHz, $E_\text{C}/2\pi=0.10$~GHz~\cite{Maiti2025LINC}.}
    \label{fig:linc}
\end{figure*}

The preceding results show that strong-drive instabilities are closely tied to the phase-space structure created by the circuit potential and the applied drive.
This motivates asking whether these instabilities can be engineered at the circuit level, rather than only by changing the drive conditions~\cite{PhysRevApplied.18.064044,Zobrist2026}.
We therefore extend the analysis to an inductively shunted circuit and examine how the resulting linear confinement modifies the phase-space structures responsible for strong-drive instabilities.

Adding a linear inductive shunt was previously proposed as a strategy for suppressing structural instabilities in strongly driven Josephson circuits~\cite{Verney2019}.
Inductive shunting alone, however, does not preclude strong-drive transitions.
Related transitions have been studied in fluxonium circuits and other inductively shunted circuits, including theoretical analyses, direct experimental observations, and processes involving internal modes of the superinductor array~\cite{Hassani2023IST,Nesterov2024FluxoniumMIST,Singh2025FluxoniumArrayModes,Bista2026FluxoniumLeakage,Zobrist2026,Chapple2026FluxoniumMIST,Zwanenburg2026FluxoniumMIST}.

Here we consider the Linear INductive Coupler (LINC), an inductively shunted SQUID that retains the flux-driven architecture studied above while adding linear confinement~\cite{Maiti2025LINC}.
It therefore provides a natural device-level extension of our main analysis and allows us to isolate how an inductive shunt modifies the strong-drive phase-space structure without changing the basic SQUID geometry.
The LINC is designed with a shunting inductor with inductive energy $E_\text{L}>2E_\text{J}$ and operated at $\phi_\text{dc}=\pi/2$, where the undriven Hamiltonian is exactly harmonic and the potential remains single-well throughout the flux cycle.
The differential-drive symmetry of the LINC also enforces a parity-protection selection rule that suppresses a broad class of parasitic nonlinear processes.
We focus here on how the added inductive confinement changes the strong-drive response relative to the unshunted flux-driven SQUID.

The LINC Hamiltonian is
\begin{equation}
    \begin{aligned}
        \hat H_\text{LINC}(t)
        ={}& 4E_\text{C}\hat n^2
        +\frac{E_\text{L}}{2}\hat\theta^2 \\
        &-2E_\text{J}
        \cos\left[
        \phi_\text{dc}
        +\phi_\text{ac}\sin(\omega_\text{d}t)
        \right]
        \cos\hat\theta.
    \end{aligned}
    \label{eq:linc_ham_general}
\end{equation}
At the operating point $\phi_\text{dc}=\pi/2$, this becomes
\begin{equation}
    \hat H_\text{LINC}(t)
    =
    4E_\text{C}\hat n^2
    +\frac{E_\text{L}}{2}\hat\theta^2
    +2E_\text{J}
    \sin\left[
    \phi_\text{ac}\sin(\omega_\text{d}t)
    \right]
    \cos\hat\theta.
    \label{eq:linc_ham_sweet}
\end{equation}
The undriven Hamiltonian therefore describes a harmonic oscillator with frequency
$\omega_0=\sqrt{8E_\text{C}E_\text{L}}$.
This is the central structural difference from the flux-driven SQUID, where the static confinement is provided by the Josephson cosine potential $E_\text{J}$ and is renormalized by the Bessel factor $J_0(\phi_\text{ac})$.

The Jacobi--Anger expansion of Eq.~\eqref{eq:linc_ham_sweet} gives
\begin{equation}
\begin{aligned}
    \hat H_\text{LINC}(t)
    &=
    4E_\text{C}\hat n^2
    +\frac{E_\text{L}}{2}\hat\theta^2
    \\
    &\quad+
    4E_\text{J}
    \sum_{\ell=0}^{\infty}
    J_{2\ell+1}(\phi_\text{ac})
    \sin\left[
    (2\ell+1)\omega_\text{d}t
    \right]
    \cos\hat\theta.
\end{aligned}
\label{eq:linc_bessel}
\end{equation}
To identify the resulting resonances, we substitute the classical harmonic trajectory
$\theta(t)\simeq\theta_0\sin(\omega_0t)$ into the perturbation.
Using
\begin{equation}
\begin{aligned}
    \cos\left[
    \theta_0\sin(\omega_0t)
    \right]
    &=
    J_0(\theta_0)
    \\
    &\quad+
    2\sum_{m=1}^{\infty}
    J_{2m}(\theta_0)
    \cos(2m\omega_0t),
\end{aligned}
\end{equation}
the slowly rotating terms satisfy $(2\ell+1)\omega_\text{d}=2m\omega_0$.
The leading small-amplitude resonance corresponds to $\ell=0$ and $m=1$, has strength proportional to $J_1(\phi_\text{ac})J_2(\theta_0)$, and occurs near $\omega_\text{d}=2\omega_0$.
This should be contrasted with the flux-driven SQUID at zero dc flux bias, where the leading perturbation is proportional to
$J_2(\phi_\text{ac})\cos(2\omega_\text{d}t)$ and the dominant low-frequency resonance appears near $\omega_\text{d}=\omega_0$.

Figure~\ref{fig:linc}(a) shows the Floquet--Markov steady-state impurity as a function of drive frequency and amplitude.
Its overall structure is qualitatively different from that of the flux-driven SQUID.
Instead of a broad impurity region that bends toward the first zero of $J_0$, the LINC displays relatively narrow resonance bands centered at nearly fixed drive frequencies.
The most prominent band occurs near $\omega_\text{d}=2\omega_0$, consistent with the leading resonance identified above.
Its weak dependence on drive amplitude follows from the fact that the intrinsic oscillator frequency is mainly set by the linear shunt inductance and is not renormalized by the ac flux amplitude.
In the SQUID, by contrast, the ac drive renormalizes the Josephson energy through $J_0(\phi_\text{ac})$, reducing the effective plasma frequency as the drive amplitude approaches the first zero of $J_0$.
The SQUID resonance bands therefore bend toward lower drive frequencies.

To examine the dominant LINC resonance, Fig.~\ref{fig:linc}(b) shows a line cut at $\omega_\text{d}/2\pi=14$~GHz, close to $\omega_\text{d}=2\omega_0$.
Starting from zero drive amplitude, the impurity increases and then remains close to $1/2$ over a broad range of amplitudes before decreasing again.
The plateau near $1/2$ indicates that the steady state is not a highly mixed chaotic state involving many Floquet modes.
Rather, it is approximately an equal mixture of two Floquet states.
The decrease in impurity at larger $\phi_\text{ac}$ follows the decrease of $J_1(\phi_\text{ac})$ beyond its maximum, which weakens the leading resonance.

The origin of this two-state mixture is visible in the classical phase space.
Figure~\ref{fig:linc}(c) shows the Poincar\'e section at $\omega_\text{d}/2\pi=14$~GHz and $\phi_\text{ac}=0.6\pi$.
The section contains two symmetry-related resonance tori associated with the leading resonance.
Their origin can be understood from the classical equation of motion.
Keeping only the $J_1$ term in Eq.~\eqref{eq:linc_bessel} gives
\begin{equation}
    \ddot\theta+\omega_0^2\theta
    =
    \Lambda_{\rm L}\sin(\omega_\text{d}t)\sin\theta,
    \quad
    \Lambda_{\rm L}=32E_\text{C}E_\text{J}J_1(\phi_\text{ac}).
    \label{eq:linc_classical_eom}
\end{equation}
Expanding $\sin\theta$ to cubic order gives
\begin{equation}
    \ddot\theta+\omega_0^2\theta
    -\Lambda_{\rm L}\sin(\omega_\text{d}t)\theta
    +\frac{\Lambda_{\rm L}}{6}\sin(\omega_\text{d}t)\theta^3
    =
    0.
    \label{eq:linc_mathieu_duffing}
\end{equation}
This is a nonlinear Mathieu-type equation~\cite{Kovacic2018Mathieu}.
Near $\omega_\text{d}=2\omega_0$, the linearized equation exhibits a parametric resonance.
Retaining the nonlinear dependence on $\theta$ produces the finite-amplitude, symmetry-related resonance tori observed in Fig.~\ref{fig:linc}(c).

The quantum steady state reflects this classical resonance structure.
Figure~\ref{fig:linc}(d) shows the Floquet-state populations for the same drive parameters as in Fig.~\ref{fig:linc}(c).
Two Floquet states dominate the steady state, each with population close to $1/2$.
Their Husimi functions are localized on the two resonance tori observed in the Poincar\'e section.
The impurity plateau near $1/2$ in Fig.~\ref{fig:linc}(b) therefore arises from a mixture of two phase-space-localized states rather than broad population over many Floquet states.

The comparatively clean LINC response has two complementary origins.
First, the parity selection rule identified in Ref.~\onlinecite{Maiti2025LINC} suppresses many parasitic transition channels.
Second, the inductive confinement changes the global phase-space structure.
In the unshunted SQUID, the static cosine potential contains a primary separatrix between bound oscillations and above-barrier running trajectories.
Periodic driving creates a chaotic layer around this separatrix, which can broaden and hybridize with nearby resonances.
At high frequency, the steady state can instead transfer from the central bound-state region to outer resonances formed from running trajectories.
In the LINC at $\phi_\text{dc}=\pi/2$, the undriven potential is harmonic, while the condition $E_\text{L}>2E_\text{J}$ keeps the instantaneous potential single-well throughout the drive cycle. Consequently, there is no primary separatrix between bound and running motion and no above-barrier sector into which the steady state can transfer. Drive-induced resonance islands and their associated separatrices can still form, but over the parameter range studied here they remain isolated and do not develop into the broad chaotic regions observed in the unshunted SQUID. The LINC therefore exhibits neither the low-frequency route to broad hybridization through separatrix chaos nor the nearly frequency-independent high-frequency unbound-resonance threshold. Instead, appreciable impurity occurs primarily within narrow resonance bands at specific drive frequencies, while the steady state remains nearly pure over a broad range of amplitudes away from these resonances. This frequency selectivity allows strong driving over a wider amplitude range by choosing the drive frequency between the resonance bands.
The suppression of broad separatrix chaos and the unbound-resonance transition does not remove other practical limitations under strong driving.
In particular, flux-noise-induced dephasing can remain a constraint on operation fidelity even away from the resonance bands~\cite{Maiti2025LINC}.

\section{Conclusion}
\label{sec:conclusion}

We have developed a unified description of strong-drive limits in flux- and charge-driven Josephson circuits across drive channel, frequency, and dc flux bias.
At low frequency, the response is governed by bound-state resonances and separatrix chaos.
The former produce localized transitions, whereas the latter causes broad Floquet hybridization and breakdown of the intended low-energy dynamics.
At high frequency, both mechanisms are suppressed, but the strong-drive limit does not disappear.
Instead, the dissipative steady state transfers from the central bound-state sector to outer resonances formed by above-barrier running trajectories.
This unbound-resonance transition therefore represents a qualitatively distinct high-frequency limitation for both flux and charge drive rather than a continuation of the low-frequency chaotic regime.
The specific resonance structure and threshold criterion, however, differ between the two drive channels.

For flux drive, we find that the low-frequency chaos threshold depends strongly on dc flux bias, reflecting both the weakening of the effective Josephson confinement and the enhancement of the odd-harmonic drive channels.
We derive analytical criteria for this threshold, validate them using Floquet--Markov simulations, and experimentally verify the predicted dc-flux-bias dependence in a flux-driven SQUID.
At high frequency, the unbound-resonance threshold is instead associated with a crossover between the local confinement scales of the shrinking central region and the drive-induced outer resonances.
Over the regime studied, this threshold is nearly independent of drive frequency and circuit parameters and is controlled primarily by dc flux bias.
Coherent simulations show that parametric exchange can persist beyond the transition, but at a lower achievable rate in the outer-resonance sector, placing an effective upper bound on the exchange rate.
The transfer and relaxation dynamics further determine how rapidly the outer-resonance steady state is established and how long the circuit requires to return to the bound-state manifold after the drive is removed.

Several extensions illustrate how the framework can guide more complex drives and circuit designs.
Under bichromatic flux modulation, mixed resonance channels make the strong-drive threshold depend jointly on the two drive amplitudes, while the junction critical current provides a simple but qualitative indicator of the high-frequency limit.
Inductive confinement offers a more direct route to improved stability by removing the primary separatrix and, over the regime studied, suppressing both broad separatrix chaos and the unbound-resonance transition, although isolated resonances remain.
These results show that the strong-drive behavior can be substantially modified through drive and circuit design, but the thresholds identified here do not exhaust the practical constraints on operation.
Other decoherence mechanisms, including flux-noise-induced dephasing during flux-modulated operation, can impose additional fidelity constraints at drive amplitudes below these thresholds~\cite{Didier2019AcSweetSpot,Hong2020AcSweetSpot,Maiti2025LINC}.
More generally, the high-frequency regime should be understood in terms of relative rather than absolute frequency scales.
Some of the high-frequency examples considered here use drive frequencies above those typical of conventional superconducting-circuit operation, but the distinction between the low- and high-frequency regimes is set by the drive frequency relative to the characteristic circuit frequencies rather than by an absolute frequency scale.
Lower-frequency qubits and couplers can therefore access the same regime at lower absolute drive frequencies, consistent with recent experiments using strongly detuned high-frequency readout and strong high-frequency driving in superconducting circuits~\cite{Kurilovich2025,Dai2026,Dixit2026,Deve2026Collapse}.
An important next step is to experimentally test the predicted high-frequency unbound-resonance threshold and its consequences for parametric operation.

\begin{acknowledgments}
We thank John W. O. Garmon and Robert J. Schoelkopf for illuminating discussions. We also thank Robert J. Schoelkopf for supporting the experimental work at Yale that helped shape this study.
This work was supported by the U.S. Department of Energy, Office of Science, National Quantum Information Science Research Centers, Superconducting Quantum Materials and Systems Center (SQMS), under Contract No. 89243024CSC000002. Fermilab is operated by Fermi Forward Discovery Group, LLC under Contract No. 89243024CSC000002 with the U.S. Department of Energy, Office of Science, Office of High Energy Physics.
Y.L. and X.Y. were supported in part by the DOE Early Career Research Program of Y.L.
\end{acknowledgments}

\section*{Data Availability}

The data and source files supporting the findings of this work are available in the Zenodo repository~\cite{You2026DatasetStrongDrive}.

\appendix

\section{Effects of junction asymmetry and a charge-drive component in a flux-driven SQUID}
\label{app:asymmetry}

In Sec.~\ref{sec:charg_and_flux}, we considered flux and charge
driving separately and assumed identical junctions. In a realistic
flux-driven SQUID, two additional effects may be present. First, the
two junctions may have slightly different Josephson energies. Second, a nominal
flux drive may be accompanied by a weak charge-drive component, for
example through unintended capacitive coupling or through the
electromotive force generated by the time-dependent magnetic
flux~\cite{You2019TimeDependentFlux,Riwar2022TimeDependentFields,Bryon2023TimeDependentFlux,Lu2023HighFidelityParametricBeamsplitting,Lu2025SystematicConstruction}.
We refer to these two effects as junction asymmetry and drive
asymmetry, respectively.

We first treat the charge component phenomenologically as a general
weak drive with an arbitrary relative phase. This description does
not assume a particular microscopic origin and also applies when an
intentional charge tone is applied together with the primary flux
drive. We then determine how the charge component and junction
asymmetry modify the effective static potential and the leading
time-dependent drive channels. Finally, we derive conditions under
which a phase-controlled charge component can compensate for the
junction asymmetry and explain the special phase relation that arises
when the charge component is generated by the electromotive force of
the same flux drive.
Throughout this appendix, we set $n_{\rm g}=0$ and assume that the
charge and flux components have the same frequency $\omega_{\rm d}$.

\subsection{Combined flux and charge drive in an asymmetric SQUID}

We define the total Josephson energy and the signed junction
asymmetry as
\begin{equation}
    E_{\rm J}
    =
    E_{\rm J1}+E_{\rm J2},
    \qquad
    d
    =
    \frac{
        E_{\rm J1}-E_{\rm J2}
    }{
        E_{\rm J1}+E_{\rm J2}
    },
    \label{eq:asymmetry_definition}
\end{equation}
such that
\begin{equation}
    E_{\rm J1}
    =
    \frac{E_{\rm J}}{2}(1+d),
    \qquad
    E_{\rm J2}
    =
    \frac{E_{\rm J}}{2}(1-d).
    \label{eq:asymmetry_junction_energies}
\end{equation}
We use the same flux-drive convention as in
Sec.~\ref{sec:charg_and_flux},
\begin{equation}
    \frac{\Phi_{\rm ext}(t)}{2\phi_0}
    =
    \phi_{\rm dc}
    +
    \phi_{\rm ac}\sin(\omega_{\rm d}t),
    \label{eq:asym_flux_drive}
\end{equation}
and include a charge component with the same drive frequency,
\begin{equation}
    \epsilon(t)
    =
    \epsilon_{\rm c}
    \cos(\omega_{\rm d}t+\varphi).
    \label{eq:asym_charge_drive}
\end{equation}
Here, $\epsilon_{\rm c}$ is the drive amplitude, while $\varphi$
specifies its temporal phase relative to the flux drive. At this
stage, $\varphi$ is treated as an arbitrary parameter.

The lab-frame Hamiltonian is
\begin{align}
    \hat H
    ={}&
    4E_{\rm C}\hat n^2
    +
    \epsilon_{\rm c}
    \cos(\omega_{\rm d}t+\varphi)\hat n
    \notag\\
    &-
    \frac{E_{\rm J}}{2}(1+d)
    \cos\left[
        \hat\theta
        +
        \phi_{\rm dc}
        +
        \phi_{\rm ac}\sin(\omega_{\rm d}t)
    \right]
    \notag\\
    &-
    \frac{E_{\rm J}}{2}(1-d)
    \cos\left[
        \hat\theta
        -
        \phi_{\rm dc}
        -
        \phi_{\rm ac}\sin(\omega_{\rm d}t)
    \right].
    \label{eq:asym_lab_hamiltonian}
\end{align}
As in the charge-drive derivation of
Sec.~\ref{sec:charg_and_flux}, we define
\begin{equation}
    \phi_{\rm c}
    =
    \frac{\epsilon_{\rm c}}{\omega_{\rm d}}
    \label{eq:asym_charge_phase_amplitude}
\end{equation}
and perform the time-dependent unitary transformation
\begin{equation}
    \hat U(t)
    =
    \exp\left[
        -i\hat n\phi_{\rm c}
        \sin(\omega_{\rm d}t+\varphi)
    \right].
    \label{eq:asym_charge_unitary}
\end{equation}
The explicit charge-drive term is removed, while the phase operator
is translated according to
\begin{equation}
    \hat U^\dagger\hat\theta\hat U
    =
    \hat\theta
    +
    \phi_{\rm c}\sin(\omega_{\rm d}t+\varphi).
\end{equation}
Writing $\xi=\omega_{\rm d}t$, we define
\begin{equation}
    \chi_\sigma(\xi)
    =
    \sigma\phi_{\rm ac}\sin\xi
    +
    \phi_{\rm c}\sin(\xi+\varphi),
    \qquad
    \sigma=\pm,
    \label{eq:asym_chi_sigma}
\end{equation}
where $\sigma=+$ and $\sigma=-$ correspond to junctions 1 and 2,
respectively. The transformed Hamiltonian then becomes
\begin{equation}
    \hat H'
    =
    4E_{\rm C}\hat n^2
    -
    \frac{E_{\rm J}}{2}
    \sum_{\sigma=\pm}
    (1+\sigma d)
    \cos\left[
        \hat\theta
        +
        \sigma\phi_{\rm dc}
        +
        \chi_\sigma(\xi)
    \right].
    \label{eq:asym_transformed_hamiltonian}
\end{equation}
The flux modulation enters the two junction phases with opposite
signs, whereas the charge-induced phase displacement enters them
with the same sign. Expanding Eq.~\eqref{eq:asym_chi_sigma} gives
\begin{align}
    \chi_\sigma(\xi)
    ={}&
    \left(
        \sigma\phi_{\rm ac}
        +
        \phi_{\rm c}\cos\varphi
    \right)\sin\xi
    \notag\\
    &+
    \phi_{\rm c}\sin\varphi\cos\xi.
    \label{eq:asym_combined_modulation_expanded}
\end{align}
We write the combined modulation as
\begin{equation}
    \chi_\sigma(\xi)
    =
    r_\sigma\sin(\xi+\delta_\sigma),
    \label{eq:asym_combined_modulation}
\end{equation}
where
\begin{equation}
    r_\sigma
    =
    \sqrt{
        \phi_{\rm ac}^2
        +
        \phi_{\rm c}^2
        +
        2\sigma
        \phi_{\rm ac}\phi_{\rm c}\cos\varphi
    },
    \label{eq:asym_rsigma}
\end{equation}
and the phase $\delta_\sigma$ is defined by
\begin{align}
    r_\sigma\cos\delta_\sigma
    &=
    \sigma\phi_{\rm ac}
    +
    \phi_{\rm c}\cos\varphi,
    \label{eq:asym_delta_cos}
    \\
    r_\sigma\sin\delta_\sigma
    &=
    \phi_{\rm c}\sin\varphi.
    \label{eq:asym_delta_sin}
\end{align}

Equations~\eqref{eq:asym_rsigma}--\eqref{eq:asym_delta_sin} show
that simultaneous charge and flux driving generally modifies both
the amplitudes and phases of the modulations experienced by the two
junctions. Even when the bare junctions are identical, the two
junctions experience different modulation amplitudes,
\begin{equation}
    r_+\neq r_-,
\end{equation}
whenever $\cos\varphi\neq0$. When
$\varphi=\pm\pi/2$, by contrast,
\begin{equation}
    r_+=r_-
    =
    \sqrt{
        \phi_{\rm ac}^2+\phi_{\rm c}^2
    },
\end{equation}
although their modulation phases $\delta_\pm$ remain different.

\subsection{Effective static and time-dependent Hamiltonians}

Using the Jacobi--Anger expansion, the exact zero-frequency
component of Eq.~\eqref{eq:asym_transformed_hamiltonian} is
\begin{align}
    \hat H_{\rm static}
    ={}&
    4E_{\rm C}\hat n^2
    \notag\\
    &-
    \frac{E_{\rm J}}{2}(1+d)
    J_0(r_+)
    \cos(\hat\theta+\phi_{\rm dc})
    \notag\\
    &-
    \frac{E_{\rm J}}{2}(1-d)
    J_0(r_-)
    \cos(\hat\theta-\phi_{\rm dc}).
    \label{eq:asym_static_exact}
\end{align}
The two junctions therefore acquire the signed effective Josephson
energies
\begin{equation}
    E_{\rm J1}^{\rm eff}
    =
    \frac{E_{\rm J}}{2}(1+d)J_0(r_+),
    \qquad
    E_{\rm J2}^{\rm eff}
    =
    \frac{E_{\rm J}}{2}(1-d)J_0(r_-).
    \label{eq:asym_effective_junction_energies}
\end{equation}
It is useful to introduce the sum and difference combinations
\begin{align}
    \mathcal J_\Sigma
    ={}&
    \frac{1+d}{2}J_0(r_+)
    +
    \frac{1-d}{2}J_0(r_-),
    \label{eq:asym_jsigma}
    \\
    \mathcal J_\Delta
    ={}&
    \frac{1+d}{2}J_0(r_+)
    -
    \frac{1-d}{2}J_0(r_-).
    \label{eq:asym_jdelta}
\end{align}
These coefficients multiply the even- and odd-in-$\hat\theta$
components of the effective static potential, respectively. The
effective static Hamiltonian can then be written as
\begin{equation}
    \hat H_{\rm static}
    =
    4E_{\rm C}\hat n^2
    -
    E_{\rm J}
    \left[
        \mathcal J_\Sigma
        \cos\phi_{\rm dc}\cos\hat\theta
        -
        \mathcal J_\Delta
        \sin\phi_{\rm dc}\sin\hat\theta
    \right].
    \label{eq:asym_static_quadratures}
\end{equation}
Away from points at which $\mathcal J_\Sigma=0$, it is useful to
define an effective junction asymmetry,
\begin{equation}
    d_{\rm eff}
    =
    \frac{\mathcal J_\Delta}{\mathcal J_\Sigma}.
    \label{eq:asym_deff}
\end{equation}
In terms of this quantity,
\begin{equation}
    \hat H_{\rm static}
    =
    4E_{\rm C}\hat n^2
    -
    E_{\rm J}\mathcal J_\Sigma
    \left[
        \cos\phi_{\rm dc}\cos\hat\theta
        -
        d_{\rm eff}
        \sin\phi_{\rm dc}\sin\hat\theta
    \right].
    \label{eq:asym_static_deff}
\end{equation}
For a representation that remains regular even when
$\mathcal J_\Sigma=0$, we combine the two phase quadratures directly
into a shifted cosine,
\begin{equation}
    \hat H_{\rm static}
    =
    4E_{\rm C}\hat n^2
    -
    E_{\rm J}\mathcal R_{\rm static}
    \cos(\hat\theta+\theta_{\rm static}),
    \label{eq:asym_static_shifted}
\end{equation}
where
\begin{equation}
    \mathcal R_{\rm static}
    =
    \sqrt{
        \mathcal J_\Sigma^2\cos^2\phi_{\rm dc}
        +
        \mathcal J_\Delta^2\sin^2\phi_{\rm dc}
    },
    \label{eq:asym_static_amplitude}
\end{equation}
and the displacement $\theta_{\rm static}$ is defined by
\begin{align}
    \mathcal R_{\rm static}\cos\theta_{\rm static}
    &=
    \mathcal J_\Sigma\cos\phi_{\rm dc},
    \label{eq:asym_static_shift_cos}
    \\
    \mathcal R_{\rm static}\sin\theta_{\rm static}
    &=
    \mathcal J_\Delta\sin\phi_{\rm dc}.
    \label{eq:asym_static_shift_sin}
\end{align}

The remaining time-dependent part is separated into its even- and
odd-harmonic sectors,
\begin{equation}
    \hat H_{\rm pert}(t)
    =
    \hat H_{\rm pert}^{\rm e}(t)
    +
    \hat H_{\rm pert}^{\rm o}(t),
    \label{eq:asym_pert_exact}
\end{equation}
where
\begin{align}
    \hat H_{\rm pert}^{\rm e}(t)
    ={}&
    -2E_{\rm J}
    \sum_{\sigma=\pm}
    w_\sigma
    \sum_{m=1}^{\infty}
    J_{2m}(r_\sigma)
    \notag\\
    &\times
    \cos\left[
        2m(\xi+\delta_\sigma)
    \right]
    \cos(\hat\theta+\sigma\phi_{\rm dc}),
    \label{eq:asym_pert_even}
\end{align}
and
\begin{align}
    \hat H_{\rm pert}^{\rm o}(t)
    ={}&
    2E_{\rm J}
    \sum_{\sigma=\pm}
    w_\sigma
    \sum_{m=0}^{\infty}
    J_{2m+1}(r_\sigma)
    \notag\\
    &\times
    \sin\left[
        (2m+1)(\xi+\delta_\sigma)
    \right]
    \sin(\hat\theta+\sigma\phi_{\rm dc}),
    \label{eq:asym_pert_odd}
\end{align}
with
\begin{equation}
    w_+
    =
    \frac{1+d}{2},
    \qquad
    w_-
    =
    \frac{1-d}{2}.
\end{equation}
Equations~\eqref{eq:asym_static_exact} and
\eqref{eq:asym_pert_exact} give the exact decomposition
\begin{equation}
    \hat H'(t)
    =
    \hat H_{\rm static}
    +
    \hat H_{\rm pert}(t).
\end{equation}

For flux drive alone, $\phi_{\rm c}=0$, the two junctions acquire
the same Bessel renormalization,
\begin{equation}
    \mathcal J_\Sigma
    =
    J_0(\phi_{\rm ac}),
    \qquad
    \mathcal J_\Delta
    =
    dJ_0(\phi_{\rm ac}),
\end{equation}
and therefore $d_{\rm eff}=d$.
An instructive case occurs at the half-flux sweet spot,
$\phi_{\rm dc}=\pi/2$, where the Josephson confinement vanishes for a symmetric SQUID.
Junction asymmetry instead produces a finite Josephson confinement.
In the absence of drive, the Hamiltonian at this point is
\begin{equation}
    \left.
    \hat H
    \right|_{\phi_{\rm dc}=\pi/2,\,\phi_{\rm ac}=0}
    =
    4E_{\rm C}\hat n^2
    +
    dE_{\rm J}\sin\hat\theta,
\end{equation}
so junction asymmetry leaves a residual Josephson confinement of
magnitude $|d|E_{\rm J}$. Under flux modulation, the full Hamiltonian
becomes
\begin{align}
    \left.
    \hat H(t)
    \right|_{\phi_{\rm dc}=\pi/2}
    ={}&
    4E_{\rm C}\hat n^2
    +
    E_{\rm J}
    \sin\!\left[
        \phi_{\rm ac}\sin\xi
    \right]
    \cos\hat\theta
    \notag\\
    &+
    dE_{\rm J}
    \cos\!\left[
        \phi_{\rm ac}\sin\xi
    \right]
    \sin\hat\theta.
\end{align}
Its zero-frequency component follows directly from the
Jacobi--Anger expansion,
\begin{equation}
    \left.
    \hat H_{\rm static}
    \right|_{\phi_{\rm dc}=\pi/2}
    =
    4E_{\rm C}\hat n^2
    +
    dE_{\rm J}J_0(\phi_{\rm ac})\sin\hat\theta.
\end{equation}
Thus, the residual static Josephson confinement is renormalized from
$|d|E_{\rm J}$ to
$|dJ_0(\phi_{\rm ac})|E_{\rm J}$ and is suppressed with increasing
drive amplitude before the first zero of $J_0$.

This result cannot be inferred from the curvature of the undriven
energy spectrum around the half-flux sweet spot. Since the
undriven transition frequency has a local minimum at this point, a
quasistatic picture of sampling the spectrum on either side of the
minimum could instead suggest an increased characteristic energy
under flux modulation. Such a quasistatic construction does not capture the effective Hamiltonian of the periodically driven system. The direct
Jacobi--Anger decomposition shows that the static Josephson
component is instead reduced by the factor $J_0(\phi_{\rm ac})$.

For charge drive alone, $\phi_{\rm ac}=0$, both junctions are
likewise renormalized by the same factor $J_0(\phi_{\rm c})$, and
$d_{\rm eff}=d$. 
At zero dc flux, pure charge drive has the additional
simplification that the two junctions then experience the same phase,
so their junction-asymmetry contributions cancel in the complete
Hamiltonian:
\begin{equation}
    \left.
    \hat H'
    \right|_{
        \phi_{\rm dc}=0,\,
        \phi_{\rm ac}=0
    }
    =
    4E_{\rm C}\hat n^2
    -
    E_{\rm J}
    \cos\left[
        \hat\theta
        +
        \phi_{\rm c}\sin(\xi+\varphi)
    \right].
\end{equation}
Thus, either drive applied alone renormalizes the two junctions equally and leaves $d_{\rm eff}=d$. For pure charge drive at zero dc flux, the junction asymmetry drops out of the full Hamiltonian altogether.

The situation changes when flux and charge drives are applied simultaneously.
Their opposite- and same-sign contributions to the two junction phases generally produce $r_+\neq r_-$, so the two junctions acquire different Bessel renormalizations.
The charge-drive component can either enhance or reduce the effective junction asymmetry.
Through this drive asymmetry, even a SQUID with $d=0$ can acquire a nonzero $d_{\rm eff}$.

\subsection{Weak charge-drive component and weak junction asymmetry}

We now assume that the charge component and junction asymmetry are
small,
\begin{equation}
    |\phi_{\rm c}|\ll1,
    \qquad
    |d|\ll1,
\end{equation}
while retaining the full dependence on the primary flux-drive
amplitude $\phi_{\rm ac}$.
In this subsection, we write
\begin{equation}
    J_n
    \equiv
    J_n(\phi_{\rm ac}).
\end{equation}

To first order in $\phi_{\rm c}$,
\begin{equation}
    J_0(r_\sigma)
    =
    J_0
    -
    \sigma
    \phi_{\rm c}\cos\varphi\,J_1
    +
    \mathcal O(\phi_{\rm c}^2).
    \label{eq:asym_j0_small_charge}
\end{equation}
Equations~\eqref{eq:asym_jsigma} and
\eqref{eq:asym_jdelta} then give
\begin{align}
    \mathcal J_\Sigma
    ={}&
    J_0
    +
    \mathcal O(
        d\phi_{\rm c},
        \phi_{\rm c}^2
    ),
    \label{eq:asym_jsigma_weak}
    \\
    \mathcal J_\Delta
    ={}&
    dJ_0
    -
    \phi_{\rm c}\cos\varphi\,J_1
    +
    \mathcal O(
        \phi_{\rm c}^2
    ).
    \label{eq:asym_jdelta_weak}
\end{align}
Thus, to first order, the charge component leaves the
even-in-$\hat\theta$ static coefficient unchanged while modifying the
odd-in-$\hat\theta$ coefficient linearly.
Away from zeros of $J_0(\phi_{\rm ac})$, the corresponding effective
asymmetry is
\begin{equation}
    d_{\rm eff}
    \simeq
    d
    -
    \phi_{\rm c}\cos\varphi
    \frac{
        J_1
    }{
        J_0
    }.
    \label{eq:asym_deff_weak}
\end{equation}

At zero dc flux bias, the odd-in-$\hat\theta$ static component vanishes,
but junction asymmetry and charge drive still modify the time-dependent
Hamiltonian.
Retaining the dominant even-in-$\hat\theta$ flux-drive term at
$2\omega_{\rm d}$ and the fundamental-frequency terms at
$\omega_{\rm d}$ gives
\begin{align}
    \hat H_{\rm pert}(t)
    \simeq{}&
    -2E_{\rm J}J_2
    \cos(2\xi)\cos\hat\theta
    \notag\\
    &+
    E_{\rm J}
    \Big\{
        \Big[
            2dJ_1
            +
            \phi_{\rm c}(J_0-J_2)\cos\varphi
        \Big]
        \sin\xi
    \notag\\
    &\hspace{1.2cm}
        +
        \phi_{\rm c}(J_0+J_2)
        \sin\varphi\cos\xi
    \Big\}
    \sin\hat\theta
    +
    \cdots.
    \label{eq:asym_leading_zero_dc}
\end{align}
The first line is the $J_2$ channel of an ideal flux-driven SQUID
with identical junctions.
Junction asymmetry adds the fundamental-frequency contribution
$2dJ_1\sin\xi\sin\hat\theta$, while the charge drive contributes to
both temporal quadratures of the same odd-in-$\hat\theta$ channel.
The relative importance of these additional fundamental-frequency
contributions is set by their full Bessel-function coefficients and
can become appreciable even for small $d$ or $\phi_{\rm c}$.
Such additional $\omega_{\rm d}$ channels could contribute to the
discrepancy between theory and experiment at zero dc flux bias in
Fig.~\ref{fig:dc_ac_exp}.
Since the asymmetry- and charge-induced contributions act through the
same odd-in-$\hat\theta$ channel, they can interfere constructively or
destructively depending on their relative phase.
This provides a route to compensating the asymmetry-induced
fundamental-frequency channel, which we consider next.

\subsection{Compensation using a phase-controlled charge component}

\begin{figure}[t]
    \centering
    \includegraphics[width=\columnwidth]{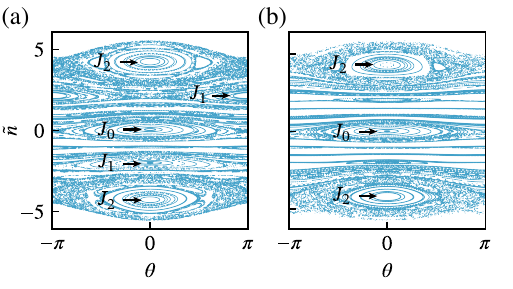}
    \caption{
    \textbf{Phase-space illustration of charge-drive compensation of junction asymmetry.}
    Poincar\'e sections at zero dc flux bias for (a) a junction-asymmetric SQUID and (b) the same asymmetric SQUID with an additional charge component chosen according to Eq.~\eqref{eq:asym_odd_cancel_weak}. The corresponding symmetric-SQUID case with $d=0$ is shown in Fig.~\ref{fig:bessel}(c). Junction asymmetry introduces the additional $J_1$-associated resonant tori in (a), corresponding to the leading $\omega_{\rm d}$ channel. With the compensating charge component, these resonant tori disappear in (b), consistent with cancellation of this channel. Parameters: $E_{\rm C}/2\pi=0.13$~GHz, $E_{\rm J}/2\pi=50$~GHz, $d=0.1$, $\omega_{\rm d}/2\pi=15$~GHz, $\phi_{\rm ac}=0.7\pi$, $n_{\rm g}=0$.
    }
\label{fig:asymmetry}
\end{figure}

We next consider how the effective charge component can be chosen to compensate for the junction asymmetry.
This common-mode component can be supplied by an additional charge-drive line or arise from the electromotive force (EMF) generated by the flux modulation itself~\cite{You2019TimeDependentFlux,Lu2025SystematicConstruction}.
In the latter case, its relative phase is fixed by the circuit response, while its amplitude is set by the circuit geometry.

There are two distinct compensation targets: the asymmetry of the
effective static Hamiltonian and the additional fundamental-frequency
drive generated by junction asymmetry.
A charge component can compensate either target, although the required
conditions generally differ.
We consider these two cases separately below.

\subsubsection{Compensation of the effective static asymmetry}

At finite dc flux, the effective static Hamiltonian contains the
odd-in-$\hat\theta$ component
\begin{equation}
    E_{\rm J}\mathcal J_\Delta
    \sin\phi_{\rm dc}\sin\hat\theta.
\end{equation}
This term can be removed by choosing the charge component such that
$\mathcal J_\Delta=0$, or, equivalently, $d_{\rm eff}=0$ away from
points at which $\mathcal J_\Sigma=0$. Using
Eq.~\eqref{eq:asym_jdelta}, the exact compensation condition is
\begin{equation}
    (1+d)J_0(r_+)
    =
    (1-d)J_0(r_-).
    \label{eq:asym_static_cancel_exact}
\end{equation}
Under this condition, the zero-frequency Hamiltonian reduces to
\begin{equation}
    \hat H_{\rm static}
    =
    4E_{\rm C}\hat n^2
    -
    E_{\rm J}\mathcal J_\Sigma
    \cos\phi_{\rm dc}\cos\hat\theta.
\end{equation}
The compensation therefore removes the odd-in-$\hat\theta$
component of the effective static Hamiltonian.

The coefficient $\mathcal J_\Sigma$ of the even-in-$\hat\theta$
component is not generally identical to $J_0(\phi_{\rm ac})$. For a
weak charge component and weak junction asymmetry, however,
Eq.~\eqref{eq:asym_jsigma_weak} gives
\begin{equation}
    \mathcal J_\Sigma
    =
    J_0(\phi_{\rm ac})
    +
    \mathcal O
    \left(
        d\phi_{\rm c},
        \phi_{\rm c}^2
    \right).
    \label{eq:asym_jsigma_compensated}
\end{equation}
Thus, to first order in the small imperfections, static compensation
removes the odd-in-$\hat\theta$ component while leaving the
even-in-$\hat\theta$ potential term
$-E_{\rm J}J_0(\phi_{\rm ac})\cos\phi_{\rm dc}\cos\hat\theta$
unchanged. Higher-order corrections can modify this term and are
captured by the exact coefficient $\mathcal J_\Sigma$ when the charge
component is not perturbatively small.

For a weak charge component and weak junction asymmetry,
Eq.~\eqref{eq:asym_jdelta_weak} gives the condition
\begin{equation}
    \phi_{\rm c}\cos\varphi
    \simeq
    d
    \frac{
        J_0(\phi_{\rm ac})
    }{
        J_1(\phi_{\rm ac})
    }.
    \label{eq:asym_static_cancel_weak}
\end{equation}
In the weak-flux-drive limit,
$J_0(\phi_{\rm ac})\simeq1$ and
$J_1(\phi_{\rm ac})\simeq\phi_{\rm ac}/2$, so the compensation
condition becomes
\begin{equation}
    \phi_{\rm c}\cos\varphi
    \simeq
    \frac{2d}{\phi_{\rm ac}}.
    \label{eq:asym_static_cancel_small_flux}
\end{equation}
The required charge amplitude therefore increases as
$\phi_{\rm ac}$ decreases. When this required amplitude is no longer
small, the perturbative condition above is not self-consistent and
the exact condition in
Eq.~\eqref{eq:asym_static_cancel_exact} should instead be used.

At $\phi_{\rm dc}=0$, the odd-in-$\hat\theta$ static component
vanishes identically since it is proportional to
$\sin\phi_{\rm dc}$. Consequently, no compensation of
$d_{\rm eff}$ is required for the effective static Hamiltonian at
this bias point. Junction asymmetry
can nevertheless generate additional time-dependent drive channels,
whose cancellation is discussed below.

\subsubsection{Compensation of the asymmetry-induced fundamental-frequency channel}

At a finite dc flux bias, the fundamental-frequency contribution
contains two independent operator-parity channels. Expanding
Eq.~\eqref{eq:asym_pert_odd} gives
\begin{equation}
    \hat H_{\omega_{\rm d}}(t)
    =
    E_{\rm J}
    \left[
        \cos\phi_{\rm dc}\,
        \mathcal F_{\rm o}(t)\sin\hat\theta
        +
        \sin\phi_{\rm dc}\,
        \mathcal F_{\rm e}(t)\cos\hat\theta
    \right],
    \label{eq:asym_fundamental_general}
\end{equation}
where
\begin{align}
    \mathcal F_{\rm o}(t)
    ={}&
    (1+d)J_1(r_+)\sin(\xi+\delta_+)
    \notag\\
    &+
    (1-d)J_1(r_-)\sin(\xi+\delta_-),
    \label{eq:asym_fundamental_odd}
\end{align}
and
\begin{align}
    \mathcal F_{\rm e}(t)
    ={}&
    (1+d)J_1(r_+)\sin(\xi+\delta_+)
    \notag\\
    &-
    (1-d)J_1(r_-)\sin(\xi+\delta_-).
    \label{eq:asym_fundamental_even}
\end{align}

The compensation target is mainly the odd-in-$\hat\theta$ channel, which is
absent for an ideal symmetric flux-driven SQUID.
To determine whether this channel can be suppressed by a weak charge
component, we consider weak junction asymmetry and weak charge drive
while retaining the full dependence on $\phi_{\rm ac}$.
Using $J_n\equiv J_n(\phi_{\rm ac})$, the leading forms are
\begin{align}
    \mathcal F_{\rm o}(t)
    \simeq{}&
    \left[
        2dJ_1
        +
        \phi_{\rm c}
        (J_0-J_2)\cos\varphi
    \right]\sin\xi
    \notag\\
    &+
    \phi_{\rm c}
    (J_0+J_2)\sin\varphi\cos\xi,
    \label{eq:asym_odd_weak}
\end{align}
and
\begin{align}
    \mathcal F_{\rm e}(t)
    \simeq{}&
    \left[
        2J_1
        +
        d\phi_{\rm c}
        (J_0-J_2)\cos\varphi
    \right]\sin\xi
    \notag\\
    &+
    d\phi_{\rm c}
    (J_0+J_2)\sin\varphi\cos\xi.
    \label{eq:asym_even_weak}
\end{align}

For $J_0(\phi_{\rm ac})+J_2(\phi_{\rm ac})\neq0$, the
odd-in-$\hat\theta$ channel generically vanishes when
\begin{equation}
    \sin\varphi=0,
\end{equation}
and, provided
$J_0(\phi_{\rm ac})-J_2(\phi_{\rm ac})\neq0$,
\begin{equation}
    \phi_{\rm c}\cos\varphi
    \simeq
    -
    \frac{
        2dJ_1(\phi_{\rm ac})
    }{
        J_0(\phi_{\rm ac})-J_2(\phi_{\rm ac})
    }.
    \label{eq:asym_odd_cancel_weak}
\end{equation}
In the weak-flux-drive limit, the second condition becomes
\begin{equation}
    \phi_{\rm c}\cos\varphi
    \simeq
    -d\phi_{\rm ac}.
    \label{eq:asym_odd_cancel_small_flux}
\end{equation}
The required charge component is therefore of order
$d\phi_{\rm ac}$ and remains perturbatively small for a weakly
asymmetric SQUID.

For an EMF-induced charge component, the corresponding phase has $\varphi=0$ or $\pi$, consistent with the phase condition above.
Its amplitude and sign are set by the circuit geometry~\cite{Lu2025SystematicConstruction}, which can in principle be designed so that the induced component satisfies the compensation condition.
Alternatively, an independently controlled charge tone allows the required amplitude and phase to be set directly.

By contrast, the even-in-$\hat\theta$ channel is already present at
finite dc flux bias for a SQUID with identical junctions.
The charge drive modifies this channel only through the product
$d\phi_{\rm c}$.
Under the odd-channel compensation condition,
$\phi_{\rm c}$ is of order $d$, so the resulting correction to
$\mathcal F_{\rm e}(t)$ is of order $d^2$.
The compensating charge tone therefore suppresses the
symmetry-breaking odd-in-$\hat\theta$ channel while leaving the
pre-existing even-in-$\hat\theta$ fundamental-frequency channel
unchanged to first order.

The zero-dc-flux operating point provides a particularly favorable
special case.
At $\phi_{\rm dc}=0$, the odd-in-$\hat\theta$ component of the
effective static Hamiltonian vanishes since it is proportional to
$\sin\phi_{\rm dc}$.
The even-in-$\hat\theta$ fundamental-frequency channel also vanishes
for the same reason, leaving
\begin{equation}
    \left.
    \hat H_{\omega_{\rm d}}(t)
    \right|_{\phi_{\rm dc}=0}
    =
    E_{\rm J}
    \mathcal F_{\rm o}(t)\sin\hat\theta.
\end{equation}
Moreover, for weak $d$ and $\phi_{\rm c}$, the coefficient of the
even-in-$\hat\theta$ static component remains
\begin{equation}
    \mathcal J_\Sigma
    =
    J_0(\phi_{\rm ac})
    +
    \mathcal O
    \left(
        d\phi_{\rm c},
        \phi_{\rm c}^2
    \right).
\end{equation}
Consequently, at zero dc flux bias, a weak charge tone chosen
according to Eq.~\eqref{eq:asym_odd_cancel_weak} cancels the
fundamental-frequency term to the order retained while leaving the
effective static potential unchanged to first order.
The leading even-harmonic flux-drive channel is likewise unchanged
to first order.
Higher odd temporal harmonics, corrections of order
$d\phi_{\rm c}$, and higher-order Floquet processes remain.

We numerically verify this compensation by calculating the Poincar\'e section at zero dc flux bias in the high-frequency regime. 
The compensation condition itself is not restricted to high frequency, and here we use the high-frequency regime only as a representative example.
The drive amplitude and frequency follow those used in Fig.~\ref{fig:bessel}(c), which also serves as the symmetric-SQUID baseline with $d=0$. In the presence of junction asymmetry $d=0.1$, two additional resonant tori appear, marked by $J_1$ in Fig.~\ref{fig:asymmetry}(a), corresponding to the leading $\omega_{\rm d}$ term induced by the junction asymmetry. 
By adding a compensating charge component according to Eq.~\eqref{eq:asym_odd_cancel_weak} with the relative phase $\varphi=\pi$, the $J_1$-associated resonant tori disappear in Fig.~\ref{fig:asymmetry}(b). This phase-space comparison provides a direct illustration of the compensation scheme.

\section{Multiple stationary states in Floquet--Markov dynamics}
\label{app:bistability}

Multiple stationary states appear in the high-frequency flux-driven regime near the first zero of $J_0(\phi_{\rm ac})$, shown in Fig.~\ref{fig:bessel}(b). We use this regime as a representative example since its underlying phase-space structure is particularly transparent. Similar behavior also occurs in other parts of the parameter space, including the charge-driven case near the second zero of $J_1(\phi_{\rm ac})$, shown in Fig.~\ref{fig:bessel_charge}(b).

This regime coincides with the appearance of three well-separated regular regions in the classical Poincar\'e section, with a central bound-state region and two outer resonant tori located at
$n_\pm=\pm\omega_{\rm d}/(4E_{\rm C})$.
The central region is controlled by the effective static potential proportional to $J_0(\phi_{\rm ac})$, whereas the two outer tori arise from resonances of above-barrier running trajectories.
As the drive frequency increases, the outer tori move farther apart in phase space.
Consistent with the phenomenological phase-space picture developed in Sec.~\ref{subsec:transfertime}, larger separations among the relevant phase-space regions are associated with smaller effective inter-region relaxation rates and hence slower population redistribution.

Near the first zero of $J_0(\phi_{\rm ac})$, the slow-mode analysis of Sec.~\ref{subsec:transfertime} provides a more direct description of the increasingly slow equilibration between the two outer regions.
In the coarse-grained three-region model, relaxation of a population imbalance between the two outer regions is governed by the outer-antisymmetric mode, with decay rate
$\gamma_{\mathrm{a}}=\Gamma_{\mathrm{o}\rightarrow\mathrm{c}}+2\Gamma_{\mathrm{o}\leftrightarrow\mathrm{o}}$.
As the first zero of $J_0$ is approached, $\gamma_{\mathrm{a}}$ tends toward zero, corresponding within the coarse-grained description to the suppression of both $\Gamma_{\mathrm{o}\rightarrow\mathrm{c}}$ and $\Gamma_{\mathrm{o}\leftrightarrow\mathrm{o}}$.
At the high drive frequency considered here, the large phase-space separation between the two outer resonances is consistent with a small $\Gamma_{\mathrm{o}\leftrightarrow\mathrm{o}}$, while the collapse of the central $J_0$ confinement near its first zero is accompanied by suppression of $\Gamma_{\mathrm{o}\rightarrow\mathrm{c}}$.
These effects provide a phenomenological interpretation of the vanishing outer-antisymmetric decay rate, while the underlying rates are determined by the full Floquet--Markov transition network.
The outer-antisymmetric relaxation mode therefore becomes effectively stationary, leaving the two outer resonant regions as nearly isolated stationary sectors on the timescale resolved by the Floquet--Markov calculation.
This provides the dynamical origin of the multiple-stationary-state region shown in Fig.~\ref{fig:bessel}(b).

The same behavior is visible directly in the spectrum of the full Floquet--Markov generator.
Let $\mathcal{R}_{\mathrm{FM}}$ denote the generator acting on the Floquet-state population vector $\mathbf{p}$, such that a stationary distribution satisfies
$\mathcal{R}_{\mathrm{FM}}\mathbf{p}=0$.
At a generic parameter point, $\mathcal{R}_{\mathrm{FM}}$ has a single zero eigenvalue, and the corresponding normalized null vector gives the unique steady state.
Near the first Bessel zero, however, the smallest nonzero decay rate decreases to a numerical magnitude comparable to that of the exact zero mode.
We therefore identify a two-dimensional numerical null space,
\begin{equation}
    \ker \mathcal{R}_{\mathrm{FM}}
    =
    \mathrm{span}\{n_1,n_2\}.
    \label{eq:two_dim_kernel_copy}
\end{equation}
The appearance of the second zero mode is the numerical signature that the outer-antisymmetric relaxation mode has become effectively stationary, consistent with the strongly suppressed equilibration between the two outer regions.

The two vectors returned by a numerical null-space routine are not, by themselves, physical steady states. An eigensolver or singular-value decomposition returns an arbitrary basis of the vector space in Eq.~\eqref{eq:two_dim_kernel_copy}. A general stationary vector therefore has the form
\begin{equation}
    p(c_1,c_2)
    =
    c_1 n_1+c_2 n_2,
    \label{eq:linear_combination_copy}
\end{equation}
which automatically satisfies $\mathcal R_{\rm FM}p=0$. A population vector is physical, however, only if
\begin{equation}
    p_i\geq 0
    \quad \mathrm{for\ all}\ i,
    \qquad
    \sum_i p_i=1.
    \label{eq:positive_trace_copy}
\end{equation}
An arbitrary linear combination of $n_1$ and $n_2$ can contain negative components and therefore cannot be interpreted as a physical population distribution.

We consequently identify the non-negative part of the null space. Before imposing normalization, the physical stationary vectors form the cone
\begin{equation}
    \mathcal{P}
    =
    \left\{
    p\in\mathrm{span}\{n_1,n_2\}:
    p_i\geq 0\ \mathrm{for\ all}\ i
    \right\}.
    \label{eq:positive_cone_copy}
\end{equation}
Since the null space is two dimensional, the positivity constraints define a wedge in the coefficient plane $(c_1,c_2)$. Its two boundary rays correspond to the extremal non-negative stationary distributions.

Each boundary ray can be obtained by saturating one positivity constraint. For a given component $i$, the direction in coefficient space that sets the $i$th population to zero is
\begin{equation}
    c^{(i)}
    =
    \left(-n_{2,i},\,n_{1,i}\right),
    \label{eq:edge_direction_copy}
\end{equation}
up to an overall scale, since
\begin{equation}
    \left[c^{(i)}_1n_1+c^{(i)}_2n_2\right]_i=0.
\end{equation}
We scan all components $i$ and retain a candidate direction only when the corresponding vector, or its negative, is non-negative in every component. After clustering identical directions and normalizing the remaining vectors to unit trace, we obtain two extremal population distributions and construct the corresponding Floquet-diagonal density matrices, denoted $\rho_{\mathrm t}$ and $\rho_{\mathrm b}$. These states are unique up to relabeling. In the parameter regime considered here, their Husimi functions are localized on the upper and lower tori, respectively.

Once the numerical null space becomes two dimensional, the steady state is no longer unique. Any convex combination
\begin{equation}
    \rho(w)
    =
    w\rho_{\mathrm t}+(1-w)\rho_{\mathrm b},
    \qquad
    0\leq w\leq 1,
    \label{eq:steady_manifold_copy}
\end{equation}
also satisfies the Floquet--Markov equation. The value of $w$ is determined by the initial condition. An initial distribution localized near the upper torus relaxes to a different stationary state from one localized near the lower torus.

This nonuniqueness has an important consequence for the impurity. When $\ker\mathcal R_{\rm FM}$ is one dimensional, the normalized steady state is unique and the impurity is a well-defined scalar diagnostic. In the multiple-stationary-state regime, however, the generator no longer selects a unique stationary distribution, so the impurity also depends on the initial state or preparation protocol. This loss of uniqueness reflects the strong-drive dynamics itself rather than a limitation of the impurity diagnostic. We therefore identify these regions separately rather than assigning them a unique impurity value.

\section{Offset-charge dependence}
\label{app:offset_charge}

\begin{figure}[t]
    \centering
    \includegraphics[width=\columnwidth]{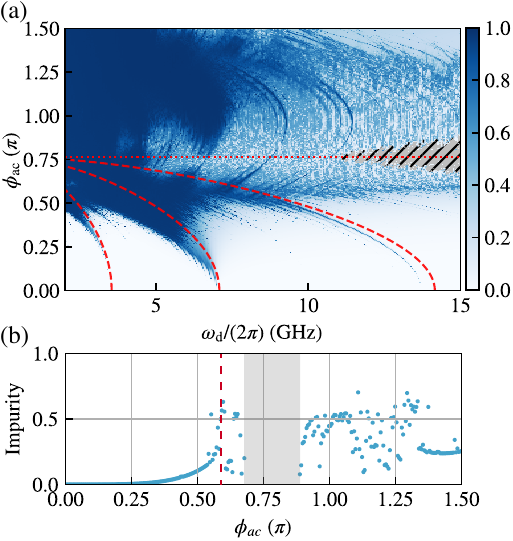}
    \caption{\textbf{Offset-charge dependence of the steady-state impurity.}
    (a) Steady-state impurity as a function of ac flux-drive amplitude
    and frequency for $n_{\rm g}=0.25$.
    Red dashed curves indicate resonance conditions extracted at
    $n_{\rm g}=0.25$ from the eigenenergies of the reduced static
    Hamiltonian in Eq.~\eqref{eq:offset_charge_static_hamiltonian}.
    The red dotted line marks the first zero of $J_0(\phi_{\rm ac})$.
    The hatched region denotes parameters for which the Floquet--Markov
    equation supports multiple stationary states.
    (b) Impurity as a function of ac amplitude at
    $\omega_{\rm d}/2\pi=15$~GHz.
    The shaded region indicates multiple stationary states.
    Parameters: $E_{\rm C}/2\pi=0.13$~GHz,
    $E_{\rm J}/2\pi=50$~GHz, $n_{\rm g}=0.25$.}
    \label{fig:offset}
\end{figure}

In the theoretical calculations throughout most of the main text, we fix the offset charge to $n_{\rm g}=0$. Only in the modeling of the experimental data do we explicitly average over $n_{\rm g}$. Here, we examine the effect of offset charge more systematically in both the low- and high-frequency regimes. As a representative example, we repeat the impurity calculation shown in Fig.~\ref{fig:impurity} at $n_{\rm g}=0.25$.
The flux-driven Hamiltonian at zero dc flux offset is
\begin{equation}
    \hat H(t)
    =
    4E_{\rm C}(\hat n-n_{\rm g})^2
    -
    E_{\rm J}
    \cos\!\left[
        \phi_{\rm ac}\sin(\omega_{\rm d}t)
    \right]
    \cos\hat\theta.
    \label{eq:offset_charge_hamiltonian}
\end{equation}
The resulting impurity map and a high-frequency line cut at $\omega_\text{d}/2\pi=15$~GHz are shown in Fig.~\ref{fig:offset}.

At the classical level, the offset charge can be absorbed by shifting the charge coordinate as $n\rightarrow n-n_{\rm g}$ and therefore does not change the phase-space structure. Quantum mechanically, the spectrum and dissipative transition rates can nevertheless depend on $n_{\rm g}$.

The low-frequency structure in Fig.~\ref{fig:offset}(a) is qualitatively the same as that obtained at $n_{\rm g}=0$ in Fig.~\ref{fig:impurity}. In particular, the narrow impurity features continue to follow resonances involving the low-lying bound states of the reduced static Hamiltonian
\begin{equation}
    \hat H_0
    =
    4E_{\rm C}(\hat n-n_{\rm g})^2
    -
    E_{\rm J}J_0(\phi_{\rm ac})\cos\hat\theta .
    \label{eq:offset_charge_static_hamiltonian}
\end{equation}
For the parameters considered here, these low-frequency resonance features occur predominantly at $\phi_{\rm ac}\lesssim0.4\pi$. Although the drive reduces the effective Josephson energy to
$E_{\rm J}J_0(\phi_{\rm ac})$, it is still much larger than
$E_{\rm C}$ over this range. The low-lying states therefore remain in the transmon regime, with strongly suppressed charge dispersion, and the corresponding resonance conditions depend only weakly on $n_{\rm g}$. The broader separatrix-chaos region is also nearly unchanged since its classical phase-space structure is invariant under the shift of the charge coordinate. At larger drive amplitudes, $J_0(\phi_{\rm ac})$ can become small or vanish, and the suppression of quantum charge dispersion no longer applies. In the low-frequency regime, however, these amplitudes already lie within the broadly chaotic region, whose qualitative behavior is not determined by the offset-charge dependence of individual low-lying levels.

The onset of the high-frequency transition is similarly insensitive to offset charge. The two unbound resonances are centered at
\begin{equation}
    n_{\pm}
    =
    n_{\rm g}
    \pm
    \frac{\omega_{\rm d}}{4E_{\rm C}}.
\end{equation}
Offset charge therefore translates the relevant phase-space structures along the charge axis without changing their classical local dynamics. In particular, the central and outer local confinement scales remain $2E_{\rm J}J_0(\phi_{\rm ac})$ and $2E_{\rm J}J_2(\phi_{\rm ac})$, respectively. Their local-confinement crossover is therefore unchanged and remains determined by
\begin{equation}
    J_0(\phi_{\rm ac}^{*})
    =
    J_2(\phi_{\rm ac}^{*}),
\end{equation}
giving $\phi_{\rm ac}^{*}=0.59\pi$. Consistent with this argument, Fig.~\ref{fig:offset}(b) shows that the sharp rise in impurity remains near $\phi_{\rm ac}=0.59\pi$ (marked by the red line), in agreement with the zero-offset result in Fig.~\ref{fig:impurity}.

Although the transition threshold is nearly unchanged, the impurity after the transition depends nontrivially on $n_{\rm g}$. At $n_{\rm g}=0$ and $1/2$, the Hamiltonian has the charge-reflection symmetry
\begin{equation}
    n\longrightarrow 2n_{\rm g}-n,
    \qquad
    \theta\longrightarrow-\theta,
\end{equation}
which exchanges the upper and lower outer tori. Since the charge eigenvalues are integers, this transformation preserves the charge lattice only when $2n_{\rm g}\in\mathbb{Z}$. At a generic offset charge, the two outer tori are therefore not related by an exact quantum symmetry, and their steady-state populations need not be equal.

At the symmetry points, the two outer sectors have identical quasienergy structures and symmetry-related bath-induced transition rates. Provided that the system--bath coupling preserves this symmetry, a unique Floquet--Markov steady state has equal total populations in the two outer sectors. When each outer sector is dominated by a single Floquet state and the central-sector population is negligible, this gives the impurity near $1/2$ observed in Fig.~\ref{fig:impurity}.

For a generic offset charge, including $n_{\rm g}=0.25$, no exact quantum symmetry exchanges the two outer tori. Their classical positions and local confinement scales remain equivalent after shifting the charge coordinate, but their dissipation-induced transition rates need not be identical. Consequently, when the steady state is dominated by the two outer sectors, its impurity is not constrained to $1/2$, consistent with Fig.~\ref{fig:offset}(b).

For the parameters considered here, offset charge therefore has little effect on either the low-frequency impurity structure or the high-frequency unbound-resonance threshold. The low-frequency bound-state resonances shift only weakly, while the broader chaos-induced impurity and the high-frequency threshold are governed primarily by classical phase-space structures that are unchanged by the shift of the charge coordinate. Offset charge can nevertheless modify the relative populations of the two outer sectors and therefore the impurity reached beyond the high-frequency threshold.

\section{Phase-space-area estimate for the separatrix-chaos threshold}
\label{app:chaos_area}

Here we provide the phase-space-area estimate underlying the low-frequency separatrix-chaos criterion used in Sec.~\ref{subsec:low_freq_dc}. The Melnikov amplitudes $M_1$ and $M_2$ in Eqs.~\eqref{eq:M1_dc} and \eqref{eq:M2_dc} characterize the separatrix-energy splitting produced by the leading $J_1$ and $J_2$ drive channels, normalized by the effective Josephson energy
\begin{equation}
    E_{\rm J}^{\rm eff}
    =
    E_{\rm J}
    J_0(\phi_{\rm ac})
    \cos\phi_{\rm dc}.
\end{equation}
Defining
\begin{equation}
    M
    =
    \max\{|M_1|,|M_2|\},
\end{equation}
we estimate the stochastic-layer energy width as
\begin{equation}
    \frac{\Delta E_{\rm ch}}{E_{\rm J}^{\rm eff}}
    \sim M.
    \label{eq:chaotic_energy_width}
\end{equation}

To convert this energy width into a phase-space area, we use the action-area relation for a one-degree-of-freedom Hamiltonian,
\begin{equation}
    \frac{dA}{dE}=T(E),
\end{equation}
where $A(E)$ is the phase-space area enclosed by an energy contour and $T(E)$ is the corresponding classical period. Near the pendulum separatrix, the period diverges logarithmically~\cite{Zaslavsky1991WeakChaos}. For an energy displacement
$\delta E=E-E_{\rm sep}$ from the separatrix energy $E_{\rm sep}$, its leading behavior can be written as
\begin{equation}
    T(\delta E)
    \sim
    \frac{1}{\omega_0}
    \ln\left(
        \frac{C E_{\rm J}^{\rm eff}}{|\delta E|}
    \right),
    \label{eq:sep_period_log}
\end{equation}
where
\begin{equation}
    \omega_0
    =
    \sqrt{
    8E_{\rm C}E_{\rm J}
    J_0(\phi_{\rm ac})
    \cos\phi_{\rm dc}
    }
\end{equation}
is the characteristic frequency of the effective static potential. Here, $C$ is a numerical constant whose precise value depends on the separatrix and energy-width convention and is subleading at logarithmic accuracy.

Integrating Eq.~\eqref{eq:sep_period_log} over a stochastic-layer energy width $\Delta E_{\rm ch}$ on both sides of the separatrix gives, to logarithmic accuracy,
\begin{equation}
    A_{\rm ch}
    \sim
    \frac{\Delta E_{\rm ch}}{\omega_0}
    \ln\left(
        \frac{C E_{\rm J}^{\rm eff}}
        {\Delta E_{\rm ch}}
    \right),
    \label{eq:chaotic_layer_area}
\end{equation}
where a convention-dependent numerical prefactor is omitted since the boundary of the stochastic layer is not sharply defined within the Melnikov treatment.

For the effective static pendulum, the phase-space area enclosed by the unperturbed separatrix is
\begin{equation}
    A_{\rm sep}
    =
    \frac{16E_{\rm J}^{\rm eff}}{\omega_0}.
    \label{eq:static_sep_area}
\end{equation}
Combining Eqs.~\eqref{eq:chaotic_energy_width}, \eqref{eq:chaotic_layer_area}, and \eqref{eq:static_sep_area} gives, up to a convention-dependent prefactor,
\begin{equation}
    \frac{A_{\rm ch}}{A_{\rm sep}}
    \sim
    M
    \ln\left(\frac{C}{M}\right).
    \label{eq:chaotic_area_ratio}
\end{equation}

Equation~\eqref{eq:chaotic_area_ratio} shows that the stochastic-layer area becomes a nonperturbative fraction of the bound-region area once the normalized Melnikov splitting $M$ is no longer small. Since the leading-log expression is controlled only for a thin stochastic layer, it does not reliably fix the numerical value at which broad separatrix chaos sets in. This motivates parameterizing the onset by the criterion
\begin{equation}
    \max\{|M_1|,|M_2|\}=M_{\rm c},
\end{equation}
where $M_{\rm c}$ is a numerical constant. We determine $M_{\rm c}=1$ from comparison with the Floquet--Markov threshold in Fig.~\ref{fig:dc_ac}(a). Changing $M_{\rm c}$ shifts the absolute critical ac amplitude but leaves the characteristic dc-flux-bias dependence and the crossover between the $J_2$- and $J_1$-dominated regimes essentially unchanged.

\section{Additional details for parametric operation}
\label{app:parametric_details}

\subsection{Removal of linear drives in the unbound-resonance frame}
\label{app:parametric_displacement}

We first detail the displacement transformations used in
Sec.~\ref{subsec:outer_parametric} to remove the linear drives that appear
when the coupler dynamics are expanded around an outer resonant region.
We retain the leading $J_0$ and $J_2$ contributions. The $J_4$
correction retained in Sec.~\ref{subsec:numeric_parametric} near the first zero of
$J_0$ can be treated by the same procedure.

For the positive-$n$ outer resonance, the rotating-frame Hamiltonian
in Eq.~\eqref{eq:Hcos_flipped} is
\begin{align}
  \hat{H}_+
  ={}&
  4\EC(\hat n-n_+)^2
  -
  \EJ J_2(\phi_{\rm ac})\cos\hat\theta
  \notag\\
  &-
  \EJ J_0(\phi_{\rm ac})
  \cos(\hat\theta+2\wdrv t)
  \notag\\
  &-
  \EJ J_2(\phi_{\rm ac})
  \cos(\hat\theta+4\wdrv t),
\end{align}
with $n_+=\wdrv/(4\EC)$.
Expanding around the local minimum $\theta=0$ gives the quadratic
static Hamiltonian
\begin{equation}
  \hat{H}_{+,\text{static}}
  =
  4\EC(\hat n-n_+)^2
  +
  \frac{1}{2}
  \EJ J_2(\phi_{\rm ac})\hat\theta^2,
  \label{eq:outer_harmonic_app}
\end{equation}
with frequency
\begin{equation}
  \bar\omega_{\mathrm q}^{\rm o}
  =
  \sqrt{
  8\EC\EJ J_2(\phi_{\rm ac})
  }.
\end{equation}
The terms linear in $\hat\theta$ are
\begin{equation}
  \hat{V}_{\rm lin}(t)
  =
  \EJ
  \left[
  J_0(\phi_{\rm ac})\sin(2\wdrv t)
  +
  J_2(\phi_{\rm ac})\sin(4\wdrv t)
  \right]
  \hat\theta.
  \label{eq:linear_drives_app}
\end{equation}
These terms drive a periodic classical orbit around the center of the
outer resonance. We remove this orbit by a time-dependent displacement
of the local coupler mode.

Writing $\hat\theta=\theta_{\rm zpf}^\text{o}
  (\hat{c}_++\hat{c}_+^\dagger)$,
and $\hat n-n_+ =-i n_{\rm zpf}^\text{o}(\hat{c}_+-\hat{c}_+^\dagger)$,
we introduce
\begin{equation}
  \hat D_+(t)
  =
  \exp\left[
  \alpha(t)\hat{c}_+^\dagger
  -
  \alpha^*(t)\hat{c}_+
  \right].
  \label{eq:Dc_app}
\end{equation}
The displacement is chosen to follow the classical trajectory
$\theta_{\rm cl}(t)$ and $\delta n_{\rm cl}(t)$ generated by
Eq.~\eqref{eq:linear_drives_app}. Hamilton's equations give
\begin{equation}
  \dot\theta_{\rm cl}
  =
  8\EC\,\delta n_{\rm cl},
\end{equation}
and
\begin{equation}
\begin{aligned}
  \ddot\theta_{\rm cl}
  +
  (\bar\omega_{\mathrm q}^{\rm o})^2\theta_{\rm cl}
  =
  -8\EC\EJ
  \big[
  &J_0(\phi_{\rm ac})\sin(2\wdrv t)
  \\
  &+
  J_2(\phi_{\rm ac})\sin(4\wdrv t)
  \big].
\end{aligned}
\label{eq:theta_cl_eom_app}
\end{equation}
Provided that neither $2\wdrv$ nor $4\wdrv$ is resonant with
$\bar\omega_{\mathrm q}^{\rm o}$, a periodic particular solution is
\begin{align}
  \theta_{\rm cl}(t)
  ={}&
  -8\EC\EJ
  \frac{
  J_0(\phi_{\rm ac})
  }{
  (\bar\omega_{\mathrm q}^{\rm o})^2-(2\wdrv)^2
  }
  \sin(2\wdrv t)
  \notag\\
  &-
  8\EC\EJ
  \frac{
  J_2(\phi_{\rm ac})
  }{
  (\bar\omega_{\mathrm q}^{\rm o})^2-(4\wdrv)^2
  }
  \sin(4\wdrv t),
  \label{eq:theta_cl_app}
\end{align}
with
\begin{equation}
  \delta n_{\rm cl}(t)
  =
  \frac{\dot\theta_{\rm cl}(t)}{8\EC}.
  \label{eq:n_cl_app}
\end{equation}
The corresponding displacement amplitude is
\begin{equation}
  \alpha(t)
  =
  \frac{\theta_{\rm cl}(t)}
  {2\theta_{\rm zpf}^\text{o}}
  +
  i
  \frac{\delta n_{\rm cl}(t)}
  {2 n_{\rm zpf}^\text{o}}.
  \label{eq:alpha_app}
\end{equation}
With this choice, the terms linear in the coupler fluctuation
operators vanish.

The coupler displacement also modifies the charge--charge interaction.
Writing the charge operator after displacement as
\begin{equation}
  \hat n
  =
  n_+
  +
  \delta n_{\rm cl}(t)
  +
  \delta\hat n,
\end{equation}
we obtain
\begin{equation}
\begin{aligned}
  g\hat n\hat n_\text{a}
  ={}&
  gn_+\hat n_\text{a}
  +
  g\delta n_{\rm cl}(t)\hat n_\text{a}
  +
  g\delta\hat n\hat n_\text{a}.
\end{aligned}
\label{eq:cavity_induced_drive_app}
\end{equation}
The first term acts as a static cavity drive and can be removed by a
static displacement. The second term drives periodic cavity
displacements at $2\wdrv$ and $4\wdrv$. These contributions can be
removed by a second displacement,
\begin{equation}
  \hat D_\text{a}(t)
  =
  \exp\left[
  \beta_\text{a}(t)\hat a^\dagger
  -
  \beta_\text{a}^*(t)\hat a
  \right].
  \label{eq:Da_app}
\end{equation}
For a Fourier component of $\delta n_{\rm cl}(t)$ at frequency
$\Omega$, the corresponding cavity displacement scales as
\begin{equation}
  |\beta_\text{a}^{(\Omega)}|
  \sim
  \frac{
  g\,n_{\text{a},\rm zpf}
  |\delta n_{\rm cl}^{(\Omega)}|
  }{
  |\omega_\text{a}-\Omega|
  },
  \qquad
  \Omega=2\wdrv,4\wdrv,
  \label{eq:beta_a_app}
\end{equation}
up to the corresponding counter-rotating contribution. When these
frequencies are sufficiently detuned from the cavity resonance, the
induced cavity displacement remains small.

After the two displacements, the leading interaction between the
quantum fluctuations retains the form
\begin{equation}
  g\,\delta\hat n\,\delta\hat n_\text{a},
\end{equation}
while the quadratic time-dependent part of the Josephson potential
generates the frequency modulation used in
Sec.~\ref{subsec:outer_parametric}. To leading order in the classical
displacement, the two modulation components are therefore
\begin{align}
  \delta\omega_{\mathrm q}^{\text{o},2\wdrv}
  &=
  \sqrt{2\EC\EJ}
  \frac{J_0(\phi_{\rm ac})}
  {\sqrt{J_2(\phi_{\rm ac})}},
  \\
  \delta\omega_{\mathrm q}^{\text{o},4\wdrv}
  &=
  \sqrt{2\EC\EJ J_2(\phi_{\rm ac})},
\end{align}
which reproduce Eq.~\eqref{eq:delta_w_flipped}.

Near the first zero of $J_0$, the $J_4(\phi_{\rm ac})$ contribution
retained in Sec.~\ref{subsec:numeric_parametric} generates an additional
$2\wdrv$ component in the outer-resonance frame. Its linear
contribution can be removed by the same displacement procedure, while
its quadratic contribution produces the replacement
\begin{equation}
  J_0(\phi_{\rm ac})
  \rightarrow
  J_0(\phi_{\rm ac})+J_4(\phi_{\rm ac})
\end{equation}
in the $2\wdrv$ modulation index used in
Eq.~\eqref{eq:beta1_corr}.

\subsection{Finite-dc-flux-bias extension of the sideband rates}
\label{app:parametric_finite_dc}

The zero-dc-flux-bias analysis extends to finite dc flux bias by
retaining the odd harmonics in the Jacobi--Anger expansion. Keeping
the leading three terms gives
\begin{equation}
\begin{aligned}
    \cos\!\left[
        \phi_{\rm dc}
        +
        \phi_{\rm ac}\sin(\wdrv t)
    \right]
    \simeq{}&
    A_0
    -
    2A_1\sin(\wdrv t)
    \\
    &+
    2A_2\cos(2\wdrv t),
\end{aligned}
\label{eq:finite_dc_expansion}
\end{equation}
where
\begin{align}
    A_0
    &=
    J_0(\phi_{\rm ac})\cos\phi_{\rm dc},
    \nonumber\\
    A_1
    &=
    J_1(\phi_{\rm ac})\sin\phi_{\rm dc},
    \nonumber\\
    A_2
    &=
    J_2(\phi_{\rm ac})\cos\phi_{\rm dc}.
    \label{eq:A012}
\end{align}
These terms generate three families of localized regions.
The central region is centered at $n=0$, the $J_1$-generated resonances are centered at
$n_{\pm,1}=\pm\wdrv/(8\EC)$, and the $J_2$-generated resonances are centered at
$n_{\pm,2}=\pm\wdrv/(4\EC)$.
The outer resonances belong to the above-barrier sector when their
centers lie outside the separatrix of the central effective
potential.

The magnitudes of the effective Josephson energies associated with
the three families are $\EJ|A_s|$, where $s=0,1,2$ labels the central,
$J_1$, and $J_2$ regions. Their local harmonic frequencies are
\begin{equation}
    \bar{\omega}_{\mathrm q}^s
    =
    \omega_{\mathrm q}^{(0)}\sqrt{|A_s|},
    \qquad
    \omega_{\mathrm q}^{(0)}
    =
    \sqrt{8\EC\EJ},
    \label{eq:finite_dc_well_frequency}
\end{equation}
and their local couplings to the cavity scale as
\begin{equation}
    g_{\text{a}}^s
    =
    g_\text{a}^{(0)}|A_s|^{1/4}.
    \label{eq:finite_dc_zpf_prefactor}
\end{equation}
A sign change of $A_s$ shifts the corresponding phase minimum but
does not change these frequency and coupling magnitudes. These local
harmonic expressions apply away from zeros of $A_s$, where the
corresponding confinement collapses.

For each localized region, the remaining harmonics produce a periodic
modulation of the local frequency. For the central region, the
magnitudes of the modulation indices at $\wdrv$ and $2\wdrv$ are
\begin{align}
    \beta_{0,1}
    &=
    \frac{\omega_{\mathrm q}^{(0)}}{\wdrv}
    \frac{|A_1|}{\sqrt{|A_0|}},
    \nonumber\\
    \beta_{0,2}
    &=
    \frac{\omega_{\mathrm q}^{(0)}}{2\wdrv}
    \frac{|A_2|}{\sqrt{|A_0|}}.
    \label{eq:beta_center_finite_dc}
\end{align}

For a $J_1$-generated resonance, the local rotating frame contains
frequency-modulation components at $\wdrv$, $2\wdrv$, and $3\wdrv$.
Their modulation indices have magnitudes
\begin{align}
    \beta_{1,1}
    &=
    \frac{\omega_{\mathrm q}^{(0)}}{2\wdrv}
    \frac{|A_0-A_2|}{\sqrt{|A_1|}},
    \nonumber\\
    \beta_{1,2}
    &=
    \frac{\omega_{\mathrm q}^{(0)}}{4\wdrv}
    \sqrt{|A_1|},
    \nonumber\\
    \beta_{1,3}
    &=
    \frac{\omega_{\mathrm q}^{(0)}}{6\wdrv}
    \frac{|A_2|}{\sqrt{|A_1|}}.
    \label{eq:beta_J1_finite_dc}
\end{align}
The difference $A_0-A_2$ arises since the $J_0$ contribution and
the down-converted component of the $J_2$ contribution enter the quadratic term at $\wdrv$ with opposite signs in the $J_1$ rotating frame.

For a $J_2$-generated resonance, the local rotating frame contains
components at $\wdrv$, $2\wdrv$, $3\wdrv$, and $4\wdrv$. Their
modulation indices have magnitudes
\begin{align}
    \beta_{2,1}
    &=
    \frac{\omega_{\mathrm q}^{(0)}}{2\wdrv}
    \frac{|A_1|}{\sqrt{|A_2|}},
    \nonumber\\
    \beta_{2,2}
    &=
    \frac{\omega_{\mathrm q}^{(0)}}{4\wdrv}
    \frac{|A_0|}{\sqrt{|A_2|}},
    \nonumber\\
    \beta_{2,3}
    &=
    \frac{\omega_{\mathrm q}^{(0)}}{6\wdrv}
    \frac{|A_1|}{\sqrt{|A_2|}},
    \nonumber\\
    \beta_{2,4}
    &=
    \frac{\omega_{\mathrm q}^{(0)}}{8\wdrv}
    \sqrt{|A_2|}.
    \label{eq:beta_J2_finite_dc}
\end{align}
The associated sine and cosine phases depend on the chosen outer
branch and local phase minimum. These phases do not affect the
magnitude of an isolated direct sideband contribution.

We define the sets of modulation harmonics
\begin{equation}
    \mathcal K_0=\{1,2\},
    \qquad
    \mathcal K_1=\{1,2,3\},
    \qquad
    \mathcal K_2=\{1,2,3,4\}.
\end{equation}
When the $k$th harmonic directly satisfies
\begin{equation}
    k\wdrv
    =
    \omega_\text{a}-\bar{\omega}_{\mathrm q}^s,
\end{equation}
the corresponding direct-path coupling has magnitude
\begin{equation}
\begin{aligned}
    \left|g_{\text{eff}}^{s,k}\right|
    \simeq{}&
    g_\text{a}^{(0)}
    |A_s|^{1/4}
    \left|J_1(\beta_{s,k})\right|
    \prod_{\substack{j\in\mathcal K_s\\j\neq k}}
    \left|J_0(\beta_{s,j})\right|.
\end{aligned}
\label{eq:finite_dc_direct_path}
\end{equation}
This expression retains the leading direct path associated with the
harmonic that directly satisfies the resonance condition. More
generally, several combinations of modulation harmonics can
contribute coherently to the same sideband, as in
Eq.~\eqref{eq:Bcomb}.

At zero dc flux bias, $A_1=0$. The $J_1$-generated resonances then
disappear, and the odd-harmonic modulation indices vanish. The
central $k=2$ process and the $J_2$-region $k=2$ and $k=4$ processes
reduce to the zero-bias results derived in
Sec.~\ref{sec:parametric}.

Finite dc flux bias therefore activates $J_1$-generated resonant
regions and additional sidebands with fundamental spacing $\wdrv$.
The resulting exchange rates depend on both the local-confinement
factor $|A_s|^{1/4}$ and the multiharmonic sideband amplitudes.

\section{Experimental details}\label{app:exp}
The measurements presented in Fig.~\ref{fig:dc_ac_exp} were performed in a dilution refrigerator at Yale University using a nominally symmetric SQUID device based on the differential-drive architecture of Ref.~\onlinecite{Lu2023HighFidelityParametricBeamsplitting}. A $\lambda/4$ stub cavity served as a buffer mode. Following the design principle of Ref.~\onlinecite{Lu2023HighFidelityParametricBeamsplitting}, the SQUID was positioned near a magnetic-field antinode and electric-field node of the buffer mode, maximizing the magnetic flux through the SQUID loop while suppressing direct common-mode excitation. The resulting modulation was therefore designed to be predominantly differential. The buffer resonance frequency was $\omega_{\rm b}/2\pi=2.023~\mathrm{GHz}$, and the measurements reported here used a drive at $\omega_{\rm d}/2\pi=2.030~\mathrm{GHz}$.

The two Josephson junctions were designed to have nominally equal Josephson energies. The undriven dc-flux spectroscopy was fitted using a symmetric-SQUID circuit model, which also provided the dc-flux calibration. The resulting effective parameters were $\EC/2\pi=0.09~\mathrm{GHz}$, $\EJ/2\pi=66~\mathrm{GHz}$, $\omega_{\mathrm r}/2\pi=8.967~\mathrm{GHz}$, and $g/2\pi=39~\mathrm{MHz}$, where $\EJ$ is the total Josephson energy of the SQUID and $g$ is the capacitive coupling between the SQUID and readout modes.

At each dc-flux bias, the buffer-drive power was increased while the readout-resonator transmission spectrum was recorded. The resonance frequency was extracted from the dominant response maximum at each drive setting. The experimental threshold was defined as the first drive point for which the extracted resonance frequency remained below $8.9677~\mathrm{GHz}$ for four consecutive power settings. A single conversion between applied drive amplitude and $\phi_{\rm ac}$ was obtained by comparing the measured and simulated ac-Stark shifts at zero dc bias and was then held fixed for all dc-flux biases.

\bibliography{Bright_stating}
\end{document}